\documentclass{LMCS}

\def\dOi{9(3:15)2013}
\lmcsheading%
{\dOi}
{1--63}
{}
{}
{Mar.~30, 2010}
{Sep.~13, 2013}
{}

\usepackage{enumerate,hyperref}

\usepackage{ifthen}

\newboolean{FoundationsAndTrends}
\setboolean{FoundationsAndTrends}{false}

\usepackage{graphics}

\usepackage{subfigure}

\usepackage{bussproofs}
  
\usepackage{ifthen}

\usepackage{ifthen}

\newboolean{maybeSTOC}
\newboolean{maybeFOCS}
\newboolean{maybeElsevier}
\newboolean{maybeNOW}
\newboolean{maybeLMCS}
\newboolean{maybePoster}
\newboolean{maybeSIAM}
\newboolean{maybeIEEE}
\newboolean{maybeICS}
\newboolean{maybeLNCS}
\newboolean{maybeACM}
\newboolean{maybeToC}
\newboolean{maybeArticle}
\newboolean{maybeReport}
\newboolean{maybeThesis}

\ifthenelse{\isundefined{\affaddr}}        
{\setboolean{maybeSTOC}{false}}
{\setboolean{maybeSTOC}{true}}

\ifthenelse{\not \isundefined{\affiliation}
  \and \not \isundefined{\Section}}        
{\setboolean{maybeFOCS}{true}}
{\setboolean{maybeFOCS}{false}}

\ifthenelse{\not \isundefined{\tnotetext}
  \and \not \isundefined{\ead}}        
{\setboolean{maybeElsevier}{true}}
{\setboolean{maybeElsevier}{false}}

\ifthenelse{\not \isundefined{\journal} \and 
  \not \isundefined{\volume} \and 
  \not \isundefined{\issue} \and 
  \not \isundefined{\copyrightowner} \and 
  \not \isundefined{\articletitle} \and 
  \not \isundefined{\pubyear}}
{\setboolean{maybeNOW}{true}}
{\setboolean{maybeNOW}{false}}

\ifthenelse{\not \isundefined{\keywords} \and
  \not \isundefined{\subjclass} \and
  \not \isundefined{\titlecomment} \and
  \not \isundefined{\revisionname} \and
  \not \isundefined{\lmcsheading}}
{\setboolean{maybeLMCS}{true}}
{\setboolean{maybeLMCS}{false}}

\ifthenelse{\not \isundefined{\IEEEtransversionmajor} \and 
  \not \isundefined{\IEEEtransversionminor} \and 
  \not \isundefined{\IEEEitemize} \and 
  \not \isundefined{\IEEEenumerate} \and 
  \not \isundefined{\IEEEdescription}}
{\setboolean{maybeIEEE}{true}}
{\setboolean{maybeIEEE}{false}}

\ifthenelse{\isundefined{\sicomp}}
{\setboolean{maybeSIAM}{false}}
{\setboolean{maybeSIAM}{true}}

\ifthenelse{\not \isundefined{\address} \and 
  \not \isundefined{\email} \and 
  \not \isundefined{\keywords} \and 
  \not \isundefined{\sanhao} \and 
  \not \isundefined{\wuhao}}
{\setboolean{maybeICS}{true}}
{\setboolean{maybeICS}{false}}

\ifthenelse{\not \isundefined{\institute} \and 
  \not \isundefined{\institutename} \and 
  \not \isundefined{\email} \and 
  \not \isundefined{\fnmsep}}
{\setboolean{maybeLNCS}{true}}
{\setboolean{maybeLNCS}{false}}

\ifthenelse{\not \isundefined{\acmVolume} \and 
  \not \isundefined{\acmNumber} \and 
  \not \isundefined{\acmArticle} \and 
  \not \isundefined{\acmYear} \and 
  \not \isundefined{\acmMonth}}
{\setboolean{maybeACM}{true}}
{\setboolean{maybeACM}{false}}

\ifthenelse{\not \isundefined{\tocpdftitle} \and 
  \not \isundefined{\tocpdfauthor}}
{\setboolean{maybeToC}{true}}
{\setboolean{maybeToC}{false}}

\ifthenelse{\isundefined{\conference}}
{\setboolean{maybePoster}{false}}
{\setboolean{maybePoster}{true}}

\ifthenelse{\isundefined{\chapter}}
{\setboolean{maybeArticle}{true}}
{\setboolean{maybeArticle}{false}}

\ifthenelse{\not \isundefined{\examen} 
  \and \not \isundefined{\disputationsdatum} 
  \and \not \isundefined{\disputationslokal}}   
  {\setboolean{maybeThesis}{true}}
  {\setboolean{maybeThesis}{false}}
           
\ifthenelse{\boolean{maybeArticle} \or \boolean{maybeThesis}
  \or \boolean{maybeSTOC} \or \boolean{maybeFOCS}
  \or \boolean{maybeSIAM} \or \boolean{maybeIEEE}
  \or \boolean{maybeICS} \or \boolean{maybePoster}}
{\setboolean{maybeReport}{false}}
{\setboolean{maybeReport}{true}}

\usepackage[latin1]{inputenc}
\usepackage{amsmath}
\usepackage{amssymb}
\usepackage{amsfonts}

\usepackage{mathtools}

\ifthenelse
{\boolean{maybeElsevier}}
{}
{\usepackage{varioref}}

\usepackage{xspace}

\ifthenelse    
{\boolean{maybeSTOC} \or \boolean{maybeSIAM} \or \boolean{maybeLMCS}
 \or \boolean{maybeNOW} \or \boolean{maybeLNCS} \or \boolean{maybeACM}
 \or \boolean{maybeToC}}    
{}
{\usepackage{local-amsthm}}

\ifthenelse
{\boolean{maybeElsevier}    \or \boolean{maybeFOCS} \or \boolean{maybeSTOC}
  \or \boolean{maybePoster} \or \boolean{maybeSIAM} \or \boolean{maybeLMCS}
  \or \boolean{maybeIEEE}   \or \boolean{maybeNOW}  \or \boolean{maybeICS}
  \or \boolean{maybeThesis} \or \boolean{maybeLNCS} \or \boolean{maybeACM}
  \or \boolean{maybeToC}} 
{}
{\usepackage{nada-typography}}

\usepackage{ifthen}
\usepackage{xspace}
\ifthenelse
{\boolean{maybeElsevier}}
{}
{\usepackage{varioref}}

\DeclareMathAlphabet{\mathsfsl}{OT1}{cmss}{m}{sl}

\newcommand{\eqperiod}{\enspace .}
\newcommand{\eqcomma}{\enspace ,}

\newcommand{\italicitem}[1][]{\item[\textit{#1}]}

\newcommand{\formatfunctiontoset}[1]{\mathit{#1}}

\newcommand{\introduceterm}[1]{{\emph{#1}}}

\newcommand{\wrt}{with respect to\xspace}

\ifthenelse{\boolean{maybeToC}}{}
  {\newcommand{\eg}{for instance\xspace} 
    }

\ifthenelse{\boolean{maybeToC}}{}{\newcommand{\ie}{i.e.,\ }}

\newcommand{\st}{such that\xspace}

\newcommand{\etal}{et al.\@\xspace}

\newcommand{\ifaoif}{if and only if\xspace}
\newcommand{\wolog}{without loss of generality\xspace}

\ifthenelse{\boolean{maybeIEEE}}{}{}

\newcommand{\polysize}{polynomial-size\xspace}

\newcommand{\superpoly}{superpolynomial\xspace}

\newcommand{\Ordocompact}[1]{{\ensuremath{\mathrm{O} \bigl( #1 \bigr)}}}
\newcommand{\Ordosmall}[1]{{\ensuremath{\mathrm{O} ( #1 )}}}

\newcommand{\polyboundsmall}[1]{\ensuremath{\mathrm{poly} ( #1 )}}

\newcommand{\BIGOH}[1]{\mathrm{O} \left( #1 \right)}
\newcommand{\Bigoh}[1]{\mathrm{O} \bigl( #1 \bigr)}
\newcommand{\bigoh}[1]{\mathrm{O} ( #1 )}
\newcommand{\LITTLEOH}[1]{\mathrm{o} \left( #1 \right)}
\newcommand{\Littleoh}[1]{\mathrm{o} \bigl( #1 \bigr)}
\newcommand{\littleoh}[1]{\mathrm{o} ( #1 )}

\newcommand{\Bigtheta}[1]{\Theta \bigl( #1 \bigr)}
\newcommand{\bigtheta}[1]{\Theta ( #1 )}
\newcommand{\BIGOMEGA}[1]{\Omega \left( #1 \right)}
\newcommand{\Bigomega}[1]{\Omega \bigl( #1 \bigr)}
\newcommand{\bigomega}[1]{\Omega ( #1 )}

\newcommand{\Littleomega}[1]{\omega \bigl( #1 \bigr)}
\newcommand{\littleomega}[1]{\omega ( #1 )}

\newcommand{\polybound}[1]{\mathrm{poly} ( #1 )}

\newcommand{\problemlanguageformat}[1]{\textsc{#1}\xspace}
\newcommand{\langstd}{\ensuremath{L}}

\newcommand{\TAUTOLOGY}{\problemlanguageformat{tautology}}

\newcommand{\SATISFIABILITY}{\problemlanguageformat{satisfiability}}
\newcommand{\MINIMALUNSATISFIABILITY}%
  {\problemlanguageformat{minimal unsatisfiability}}

\newcommand{\complclassformat}[1]{\textrm{\upshape{\textsf{#1}}}\xspace}

\newcommand{\cocomplclass}[1]%
        {\mbox{\complclassformat{co}-\complclassformat{#1}}\xspace}

\newcommand{\Pclass}{\complclassformat{P}}
\newcommand{\NP}{\complclassformat{NP}}
\newcommand{\NPclass}{\NP}
\newcommand{\coNP}{\cocomplclass{NP}}

\newcommand{\CoNP}{\coNP}

\newcommand{\PSPACE}{\complclassformat{PSPACE}}

\newcommand{\EXPTIME}{\complclassformat{EXPTIME}}

\newcommand{\EXPSPACE}{\complclassformat{EXPSPACE}}

\newcommand{\refsec}[1]{Section~\ref{#1}}

\newcommand{\Refsec}[1]{Section~\ref{#1}}

\newcommand{\reftwosecs}[2]{Sections~\ref{#1} and~\ref{#2}}

\newcommand{\reffig}[1]{Figure~\ref{#1}}

\newcommand{\refth}[1]{Theorem~\ref{#1}}
\newcommand{\reftwoths}[2]{Theorems~\ref{#1} and~\ref{#2}}

\newcommand{\reflem}[1]{Lemma~\ref{#1}}

\newcommand{\refcor}[1]{Corollary~\ref{#1}}

\newcommand{\refdef}[1]{Definition~\ref{#1}}
\newcommand{\reftwodefs}[2]{Definitions~\ref{#1} and~\ref{#2}}

\newcommand{\Refth}[1]{Theorem~\ref{#1}}

\newcommand{\refpart}[1]{part~\ref{#1}}

\ifthenelse
{\isundefined{\refeq}}
{\newcommand{\refeq}[1]{\eqref{#1}}}
{\renewcommand{\refeq}[1]{\eqref{#1}}}

\newcommand{\ceiling}[1]{\lceil #1 \rceil}

\ifthenelse{\boolean{maybeToC}}{}
{

}

\newcommand{\maxofsmall}[2][]{\max_{#1} \{ #2 \}}

\newcommand{\maxofexpr}[2][]{\max_{#1} \{ #2 \}}
\newcommand{\minofexpr}[2][]{\min_{#1} \{ #2 \}}

\newcommand{\fieldstd}{\mathbb{F}}

\DeclareMathOperator{\Expop}{E}

\ifthenelse{\boolean{maybeLMCS}}
{}
{}

\newcommand{\twincommandJN}[6]%
    {#1#2#3\vphantom{#2#5}\mspace{-2.25mu}#4.#5#6}

\newcommand{\CondExp}[2]%
    {\Expop\twincommandJN{\bigl[}{#1}{\bigl|}{\bigr}{\,#2}{\bigr]}}
\newcommand{\CONDEXP}[2]%
     {\Expop\twincommandJN{\left[}{#1}{\left|}{\right}{\,#2}{\right]}}

\newcommand{\CondProb}[3][]%
    {\Pr_{#1}\twincommandJN{\bigl[}{#2}{\bigl|}{\bigr}{\,#3}{\bigr]}}
\newcommand{\CONDPROB}[3][]%
    {\Pr_{#1}\twincommandJN{\left[}{#2}{\left|}{\right}{\,#3}{\right]}}

\newcommand{\isdistras}[2]{\ensuremath{#1} \sim \ensuremath{#2}}

\newcommand{\funcdescr}[3]{\ensuremath{ #1 : #2 \to #3}}

\newcommand{\vertices}[1]{V( #1 )}

\newcommand{\vneighbour}[1]{\ensuremath{N(#1)}}

\newcommand{\prednode}[1]{\formatfunctiontoset{pred}(#1)}

\newcommand{\setcompact}[1]{{\ensuremath{\bigl\{ #1 \bigr\}}}}
\newcommand{\setsmall}[1]{{\ensuremath{\{ #1 \}}}}
\newcommand{\setdescrcompact}[3][\mid]{{\setcompact{ #2 #1 #3 }}}
\newcommand{\setdescrsmall}[3][\mid]{{\setsmall{ #2 #1 #3 }}}

\newcommand{\set}[1]{\{ #1 \}}
\newcommand{\Set}[1]{\bigl\{ #1 \bigr\}}
\newcommand{\SET}[1]{\left\{ #1 \right\}}

\newcommand{\setdescr}[3][\mid]{\setsmall{ #2 #1 #3 }}
\newcommand{\Setdescr}[3][|]%
     {\twincommandJN{\bigl\{}{#2}{\bigl#1}{\bigr}{\,#3}{\bigr\}}}
\newcommand{\SETDESCR}[3][|]%
     {\twincommandJN{\left\{}{#2}{\left#1}{\right}{\,#3}{\right\}}}

\newcommand{\Setdescrbrackets}[3][|]%
     {\twincommandJN{\bigl[}{#2}{\bigl#1}{\bigr}{\,#3}{\bigr]}}
\newcommand{\SETDESCRBRACKETS}[3][|]%
     {\twincommandJN{\left[}{#2}{\left#1}{\right}{\,#3}{\right]}}

\newcommand{\Setsize}[1]{\bigl\lvert#1\bigr\rvert}
\newcommand{\setsize}[1]{\lvert#1\rvert}

\newcommand{\intersection}{\cap}

\newcommand{\union}{\cup}
\newcommand{\Union}{\bigcup}

\newcommand{\unionSP}{\, \union \, }
\newcommand{\intersectionSP}{\, \intersection \, }

\newcommand{\DisjointunionInText}%
    {{\smash{\overset{\mbox{\boldmath{.}}}{\bigcup}}}\vphantom{\bigcup}}

\newcommand{\intclcl}[2]{[ #1 , #2 ] }

\newcommand{\intnfirst}[1]{[{#1}]}

\newcommand{\Lor}{\bigvee}
\newcommand{\Land}{\bigwedge}

\newcommand{\Lornodisplay}{{\textstyle \bigvee}}

\newcommand{\xor}{\operatorname{\textsc{xor}}}

\newcommand{\olnot}[1]{\overline{#1}}
\newcommand{\stdnot}[1]{\olnot{#1}}
\newcommand{\zerooneset}{\ensuremath{\{0,1\}}}

\newcommand{\cnfform}{\cnfshort for\-mu\-la\xspace}

\newcommand{\dnfform}{DNF for\-mu\-la\xspace}
\newcommand{\cnfshort}{CNF\xspace}

\newcommand{\xdnf}[1]{\mbox{\ensuremath{#1}-DNF}\xspace}
\newcommand{\kdnf}{\xdnf{\clwidth}}
\newcommand{\xdnfform}[1]{\mbox{\ensuremath{#1}-}\dnfform}
\newcommand{\kdnfform}{\xdnfform{\clwidth}}

\newcommand{\xcnf}[1]{\mbox{\ensuremath{#1}-CNF}\xspace}
\newcommand{\kcnf}{\xcnf{\clwidth}}
\newcommand{\xcnfform}[1]{\mbox{\ensuremath{#1}-}\cnfform}
\newcommand{\kcnfform}{\xcnfform{\clwidth}}
\newcommand{\xclause}[1]{\mbox{\ensuremath{#1}-clause}\xspace}

\newcommand{\nvar}{n}
\newcommand{\nclause}{m}
\newcommand{\clwidth}{k}

\newcommand{\randkcnfnclwrepl}[3][\clwidth]%
        {\ensuremath{\mathcal{F}^{#2, #3}_{#1}}}
\newcommand{\randkcnfnclwreplstd}%
        {\randkcnfnclwrepl{\clwidth}{\nvar}{\nclause}}

\newcommand{\israndkcnfnclwrepl}[4]%
  {\isdistras{#1}{\randkcnfnclwrepl[#2]{#3}{#4}}}

\newcommand{\randkcnfprobcl}[3]%
        {\ensuremath{\mathcal{F}^{#2}_{#1} \bigl(#3 \bigr)}}

\newcommand{\pcfor}[4][to]{for #2 := #3 #1 #4 do}
\newcommand{\pcformath}[4][to]%
    {\pcfor[#1]{\ensuremath{#2}}{\ensuremath{#3}}{\ensuremath{#4}}}

\newcommand{\pcassigncompact}[2]{#1 := #2}
\newcommand{\pcassignmathcompact}[2]%
        {\pcassigncompact{\ensuremath{#1}}{\ensuremath{#2}}}

\newcommand{\inductionformat}[1]{\textit{#1}}

\newcommand{\BASE}[1][]
        {\inductionformat
                {%
                        \ifthenelse{\equal{#1}{}}%
                                {Base case: }%
                                {Base case (#1):}%
                }%
        }

\ifthenelse{\boolean{maybeSTOC}}                  
  {\input{definitions-STOC.tex}}
  {\ifthenelse{\boolean{maybeSIAM}}
    {\input{definitions-SIAM.tex}}
    {\ifthenelse{\boolean{maybeNOW}}
      {\input{definitions-NOW.tex}}
      {\ifthenelse{\boolean{maybeLMCS}}
        {%
\theoremstyle{plain}    

\newtheorem{theorem}[thm]{Theorem}
\newtheorem{lemma}[thm]{Lemma}

\newtheorem{corollary}[thm]{Corollary}

\newtheorem{openproblem}{Open Problem}

\theoremstyle{definition}

\newtheorem{definition}[thm]{Definition}

\newtheorem{remark}[thm]{Remark}
\newtheorem{example}[thm]{Example}

}
        {\ifthenelse{\boolean{maybeIEEE}}
          {\input{definitions-IEEE.tex}}
          {\ifthenelse{\boolean{maybeICS}}
            {\input{definitions-ICS.tex}}
            {\ifthenelse{\boolean{maybeLNCS}}
              {\input{definitions-LNCS.tex}}
              {\ifthenelse{\boolean{maybeACM}}
                {\input{definitions-ACM.tex}}
                {\ifthenelse{\boolean{maybeToC}}
                  {\input{definitions-ToC.tex}}
                  {\ifthenelse{\boolean{maybeArticle} \or \boolean{maybeElsevier}}
                    {\input{definitions-article.tex}}
                    {\input{definitions-report.tex}}}}}}}}}}}

\ifthenelse{\boolean{maybeThesis}}
{}
{\usepackage{ifpdf}}

\ifthenelse{\boolean{maybeToC}}
{}
{\usepackage{graphicx}}

\ifpdf         
\fi

\ifthenelse
{\boolean{maybeSTOC}  \or \boolean{maybeThesis} \or 
 \boolean{maybeLNCS}  \or \boolean{maybeToC}}    
{}
{
  \setcounter{secnumdepth}{3}
  \setcounter{tocdepth}{3}
}

\newcount\hours
\newcount\minutes
\def\SetTime{\hours=\time
\global\divide\hours by 60
\minutes=\hours
\multiply\minutes by 60
\advance\minutes by-\time
\global\multiply\minutes by-1 }
\SetTime
\def\now{\number\hours:\ifnum\minutes<10 0\fi\number\minutes}

\newcommand{\formuladots}{\cdots}

\newcommand{\impl}{\vDash}
\newcommand{\nimpl}{\nvDash}

\newcommand{\proplog}{propo\-sitional logic\xspace}
\newcommand{\tva}{truth value assignment\xspace}

\newcommand{\proofsystemformat}[1]{\ensuremath{\mathfrak{#1}}}

\newcommand{\psstd}{\proofsystemformat{P}}

\newcommand{\proofstd}{\ensuremath{\pi}}

\newcommand{\resnot}{\proofsystemformat{R}}
\newcommand{\treeresnot}{\proofsystemformat{T}}
\newcommand{\resknot}[1][k]{\proofsystemformat{R}({#1})}
\newcommand{\reskone}{\resknot[k+1]}

\newcommand{\pcrnot}{\proofsystemformat{PCR}}

\newcommand{\derivof}[4][\derives]
        {{\ensuremath{{#2} : {#3} \, {#1}\, {#4}}}}

\newcommand{\refof}[2]{\derivof{#1}{#2}{\falsenum}}
\renewcommand{\refof}[2]{\derivof{#1}{#2}{\emptycl}}

\newcommand{\deriveswithall}%
        {\vdash_{\!\!\!{\scriptscriptstyle \forall}}} 
\newcommand{\notderiveswithall}%
        {\nvdash_{\!\!\!{\scriptscriptstyle \forall}}} 

\newcommand{\clcfgtransition}[2]{\ensuremath{#1 \rightsquigarrow #2}}
\newcommand{\clcfgtransitioncrammed}[2]%
        {\ensuremath{#1 \!\rightsquigarrow\! #2}}

\newcommand{\truthval}{\nu}
\newcommand{\truthvalalt}{\mu}

\newcommand{\tvastd}{{\ensuremath{\alpha}}}
\newcommand{\tvaalt}{{\ensuremath{\beta}}}

\newcommand{\logeval}[2]{#2 ( #1 )}

\newcommand{\Logeval}[2]{#2 \bigl( #1 \bigr)}

\newcommand{\fstd}{{\ensuremath{F}}}
\newcommand{\fvar}{{\ensuremath{G}}}
\newcommand{\falt}{\fvar}

\newcommand{\emptycl}{\bot}

\newcommand{\varx}{\ensuremath{x}}
\newcommand{\vary}{\ensuremath{y}}
\newcommand{\varz}{\ensuremath{z}}
\newcommand{\varu}{\ensuremath{u}}
\newcommand{\varv}{\ensuremath{v}}

\newcommand{\lita}{\ensuremath{a}}
\newcommand{\litb}{\ensuremath{b}}
\newcommand{\litc}{\ensuremath{c}}

\newcommand{\cla}{\ensuremath{A}}
\newcommand{\clb}{\ensuremath{B}}
\newcommand{\clc}{\ensuremath{C}}
\newcommand{\cld}{\ensuremath{D}}

\newcommand{\clausesetformat}[1]{\ensuremath{\mathbb{#1}}}
\newcommand{\clsc}{\clausesetformat{C}}
\newcommand{\clsd}{\clausesetformat{D}}

\newcommand{\setsofvarsorlitlarge}[2]%
        {\mathit{#1}\left({#2}\right)}
\newcommand{\setsofvarsorlit}[2]%
        {\mathit{#1}({#2})}
\newcommand{\setsofvarsorlitcompact}[2]%
        {\mathit{#1}\bigl({#2}\bigr)}
\newcommand{\setsofvarsorlitsmall}[2]
        {\mathit{#1}({#2})}

\newcommand{\setsofvarsorlitsup}[3]%
        {\mathit{#1}^{#2}({#3})}
\newcommand{\setsofvarsorlitsuplarge}[3]%
        {\mathit{#1}^{#2}\left({#3}\right)}
\newcommand{\setsofvarsorlitsupcompact}[3]%
        {\mathit{#1}^{#2}\bigl({#3}\bigr)}

\newcommand{\vars}[1]{\setsofvarsorlitsmall{Vars}{#1}}
\newcommand{\lit}[1]{\setsofvarsorlitsmall{Lit}{#1}}

\newcommand{\Vars}[1]{\setsofvarsorlitcompact{Vars}{#1}}

\newcommand{\restr}{\ensuremath{\rho}}
\newcommand{\rstd}{\restr}

\newcommand{\measstd}[1][]{\genericmeasure{M}{#1}}
\newcommand{\measstdof}[2][]{\genericform{{M}_{#1}}{#2}}

\newcommand{\derivabbrev}[2]{\bigl( #1 \vdash #2 \bigr)}
\newcommand{\derivabbrevsmall}[2]{( #1 \vdash #2 )}
\newcommand{\derivabbrevcompact}[2]{\bigl( #1 \vdash #2 \bigr)}

\newcommand{\refutabbrevsmall}[1]{\derivabbrevsmall{#1}{\falsenum}}
\newcommand{\refutabbrevcompact}[1]{\derivabbrevcompact{#1}{\falsenum}}

\renewcommand{\refutabbrevsmall}[1]{\derivabbrevsmall{#1}{\!\emptycl}}
\renewcommand{\refutabbrevcompact}[1]{\derivabbrevcompact{#1}{\!\emptycl}}

\newcommand{\genericmeasure}[2]{{\mathit{#1}}_{#2}}
\newcommand{\genericform}[2]{\mathit{#1}\bigl( #2 \bigr)}
\newcommand{\genericformsmall}[2]{\mathit{#1}( #2 )}
\newcommand{\genericformcompact}[2]{\mathit{#1}\bigl( #2 \bigr)}

\newcommand{\genericrefsmall}[3]%
    {{\mathit{#1}}_{#2}\refutabbrevsmall{#3}}
\newcommand{\genericrefcompact}[3]%
    {{\mathit{#1}}_{#2}\refutabbrevcompact{#3}}
\newcommand{\genericderiv}[4]%
    {{\mathit{#1}}_{#2}\derivabbrev{#3}{#4}}
\newcommand{\genericderivsmall}[4]%
    {{\mathit{#1}}_{#2}\derivabbrevsmall{#3}{#4}}
\newcommand{\genericderivcompact}[4]%
    {{\mathit{#1}}_{#2}\derivabbrevcompact{#3}{#4}}
\newcommand{\generictaut}[3]%
    {{\mathit{#1}}_{#2}\derivabbrev{}{#3}}
\newcommand{\generictautcompact}[3]%
    {{\mathit{#1}}_{#2}\derivabbrevcompact{}{#3}}
\newcommand{\generictautsmall}[3]%
    {{\mathit{#1}}_{#2}\derivabbrevsmall{}{#3}}
\newcommand{\size}[1][]{\genericmeasure{S}{#1}}

\newcommand{\sizestd}{{\size}}
\newcommand{\sizeofarg}[1]{\genericformsmall{S}{#1}}

\newcommand{\sizeref}[2][]{\genericrefsmall{S}{#1}{#2}}

\newcommand{\length}[1][]{\genericmeasure{L}{#1}}

\newcommand{\lengthofsmall}[1]{\genericformsmall{L}{#1}}
\newcommand{\lengthrefsmall}[2][]{\genericrefsmall{L}{#1}{#2}}

\newcommand{\lengthstd}{\length}
\newcommand{\lengthofarg}[1]{\genericformsmall{L}{#1}}
\newcommand{\Lengthofarg}[1]{\genericformcompact{L}{#1}}
\newcommand{\lengthref}[2][]{\genericrefsmall{L}{#1}{#2}}
\newcommand{\Lengthref}[2][]{\genericrefcompact{L}{#1}{#2}}

\newcommand{\widthofsmall}[2][]{\genericformsmall{W_{#1}}{#2}}

\newcommand{\widthstd}{\ensuremath{w}} 

\newcommand{\widthofarg}[2][]{\genericformsmall{W_{#1}}{#2}}
\newcommand{\Widthofarg}[2][]{\genericformcompact{W_{#1}}{#2}}
\newcommand{\widthref}[2][]{\genericrefsmall{W}{#1}{#2}}
\newcommand{\Widthref}[2][]{\genericrefcompact{W}{#1}{#2}}

\newcommand{\totspaceof}[2][]{\genericformsmall{TotSp_{#1}}{#2}}

\newcommand{\totspaceref}[2][]{\genericrefsmall{TotSp}{#1}{#2}}

\newcommand{\clspaceofsmall}[1]{\genericformsmall{Sp}{#1}}

\newcommand{\clspaceof}[2][]{\genericformsmall{Sp_{#1}}{#2}}
\newcommand{\Clspaceof}[2][]{\genericformcompact{Sp_{#1}}{#2}}
\newcommand{\clspaceref}[2][]{\genericrefsmall{Sp}{#1}{#2}}
\newcommand{\Clspaceref}[2][]{\genericrefcompact{Sp}{#1}{#2}}

\newcommand{\varspaceof}[2][]{\genericformsmall{VarSp_{#1}}{#2}}

\newcommand{\varspaceref}[2][]{\genericrefsmall{VarSp}{#1}{#2}}

\newcommand{\formulaformat}[1]{\ensuremath{\mathit{#1}}}
\renewcommand{\formulaformat}[1]{\mathit{#1}}

\newcommand{\formatconfiguration}[1]{\mathbb{#1}}

\newcommand{\formatpebblingstrategy}[1]{\mathcal{#1}}

\newcommand{\transitionarrow}{\rightsquigarrow}
\newcommand{\pebcfgtransition}[2]%
    {\ensuremath{#1 \transitionarrow #2}}
\newcommand{\pebcfgtransitionsqueeze}[2]%
    {#1 \! \transitionarrow \! #2}

\newcommand{\pebbling}[1][P]{\formatpebblingstrategy{#1}}

\newcommand{\pconf}[1][P]{\formatconfiguration{#1}}

\newcommand{\formatpebblingprice}[1]{\text{\textsl{\textsf{#1}}}}

\newcommand{\pebblingprice}[1]{\formatpebblingprice{Peb}(#1)}
\newcommand{\Pebblingprice}[1]%
    {\formatpebblingprice{Peb}\bigl(#1\bigr)}
\newcommand{\pebblingpricecompact}[1]
    {\formatpebblingprice{Peb}\bigl(#1\bigr)}
\newcommand{\bwpebblingprice}[1]{\formatpebblingprice{BW-Peb}(#1)}
\newcommand{\Bwpebblingprice}[1]%
    {\formatpebblingprice{BW-Peb}\bigl(#1\bigr)}
\newcommand{\bwpebblingpricecompact}[1]
    {\formatpebblingprice{BW-Peb}\bigl(#1\bigr)}

\newcommand{\pebpersistentsymbol}{\bullet}
\newcommand{\pebvisitingsymbol}{\emptyset}

\newcommand{\bwpebpricepersistent}[1]%
    {\formatpebblingprice{BW-Peb}^{\pebpersistentsymbol}(#1)}
\newcommand{\Bwpebpricepersistent}[1]%
    {\formatpebblingprice{BW-Peb}^{\pebpersistentsymbol}\bigl(#1\bigr)}
\newcommand{\bwpebpricevisiting}[1]%
    {\formatpebblingprice{BW-Peb}^{\pebvisitingsymbol}(#1)}
\newcommand{\Bwpebpricevisiting}[1]%
    {\formatpebblingprice{BW-Peb}^{\pebvisitingsymbol}\bigl(#1\bigr)}

\newcommand{\pebpricepersistent}[1]%
    {\formatpebblingprice{Peb}^{\pebpersistentsymbol}(#1)}
\newcommand{\Pebpricepersistent}[1]%
    {\formatpebblingprice{Peb}^{\pebpersistentsymbol}\bigl(#1\bigr)}
\newcommand{\pebpricevisiting}[1]%
    {\formatpebblingprice{Peb}^{\pebvisitingsymbol}(#1)}
\newcommand{\Pebpricevisiting}[1]%
    {\formatpebblingprice{Peb}^{\pebvisitingsymbol}\bigl(#1\bigr)}

\newcommand{\bwpebblingpriceempty}[1]%
    {\formatpebblingprice{BW-Peb}^{\pebvisitingsymbol}(#1)}
\newcommand{\bwpebblingpriceemptycompact}[1]%
    {\formatpebblingprice{BW-Peb}^{\pebvisitingsymbol}\bigl(#1\bigr)}

\newcommand{\stoptime}{\tau}

\newcommand{\pebspace}[1]{\formatpebblingprice{space} ( #1 )}

\newcommand{\pebtime}[1]{\formatpebblingprice{time} ( #1 )}

\newcommand{\pebcontr}[2][G]{\ensuremath{\formulaformat{Peb}^{#2}_{#1}}}

\newcommand{\pebdeg}{\ensuremath{d}}

\newcommand{\pebaxcompact}[2]%
        [\pebdeg]{\ensuremath{\formulaformat{Ax}^{#1} \bigl(#2 \bigr)}}

\newcommand{\pqrxvar}[6]%
    {\ensuremath{\stdnot{\varx({#1})}_{#2} \lor \stdnot{\varx({#3})}_{#4} \lor %
    \sourceclausexvar[#6]{#5}}}

\newcommand{\pqr}[6]%
    {\ensuremath{\stdnot{#1}_{#2} \lor \stdnot{#3}_{#4} \lor %
    \sourceclausenodisplay[#6]{#5}}}
\newcommand{\pqrstd}{\pqr{p}{i}{q}{j}{r}{l}}
\newcommand{\pqrall}[6]%
        {\setdescrcompact
        {\pqr{#1}{#2}{#3}{#4}{#5}{#6}}{#2,#4 \in \intnfirst{\pebdeg}}}
\newcommand{\pqrallstd}%
        {\setdescrcompact{\pqrstd}{i,j \in \intnfirst{\pebdeg}}}

\newcommand{\sourceclausexvar}[2][n]%
        {\Lor_{#1 = 1}^{\pebdeg} \varx({#2})_{#1}}
\newcommand{\subsourceclausexvar}[3][n]%
        {\Lor_{#1 = {#2}}^{\pebdeg} \varx({#3})_{#1}}

\newcommand{\sourceclausexvarnodisplay}[2][n]%
        {\textstyle \Lor_{#1 = 1}^{\pebdeg} \varx({#2})_{#1}}

\newcommand{\sourceclausenodisplay}[2][n]%
        {\textstyle \Lor_{#1 = 1}^{\pebdeg} #2_{#1}}

\newcommand{\relativisation}[1]%
    {\ensuremath{\formulaformat{Rel}\bigl(#1 \bigr)}}

\newcommand{\extendedversion}[1]{\widetilde{#1}}

\usepackage{stmaryrd} 

\newcommand{\nclausesof}[1]{\setsize{#1}}

\newcommand{\formatfunctiontosubconfiguration}[1]{\mathsf{#1}}

\newcommand{\formatfunctiontomulti}[1]{\mathcal{#1}}

\newcommand{\pebcomplete}{complete\xspace}

\DeclareMathOperator{\dummystar}{*}
\newcommand{\pebblingcontrNT}[2][G]%
 {\ensuremath{\dummystar\!\!\formulaformat{Peb}^{#2}_{#1}}}

\newcommand{\somenodetrueclausedeg}[2]{\formulaformat{All}_{#1}^{+}({#2})}

\newcommand{\slashedstrickenletter}[1]{{\backslash\mkern-9mu #1}}
            
\newcommand{\strikethroughcommand}[1]{\slashedstrickenletter{#1}}

\newcommand{\abovevertices}[2][G]%
    {{#1}_{#2}^{\hspace{-0.2 pt}\triangledown}}
\newcommand{\aboveverticesNR}[2][G]%
    {{#1}_{\strikethroughcommand{#2}}^{\hspace{-0.3 pt}\triangledown}}
\newcommand{\belowvertices}[2][G]%
    {{#1}^{#2}_{\hspace{-0.6 pt}\vartriangle}}
\newcommand{\belowverticesNR}[2][G]%
    {{#1}^{\strikethroughcommand{#2}}_{\hspace{-0.6 pt}\vartriangle}}

\newcommand{\lpebblingtext}{\mbox{L-peb}\-bling\xspace}

\newcommand{\lpebblegame}{\mbox{L-peb}\-ble game\xspace}

\newcommand{\subconftext}{sub\-con\-figu\-ration\xspace}

\newcommand{\blunconditional}{independent\xspace}
\newcommand{\blconditional}{dependent\xspace}

\newcommand{\lpebbling}[1][L]{\pebbling[#1]}
\newcommand{\lconf}[1][L]{\pconf[#1]}

\newcommand{\lpebblingpricecompact}[1]%
    {\formatpebblingprice{L-Peb}\bigl(#1\bigr)}

\newcommand{\scnot}[2]{#1 \langle #2 \rangle}
\newcommand{\scnotcompact}[2]{#1 \bigl\langle #2 \bigr\rangle}

\newcommand{\introsubconfnot}[1]{\scnot{#1}{\prednode{#1}}}

\newcommand{\unconditionalblacknot}[1]{\scnot{#1}{\emptyset}}

\newcommand{\spmerge}[2]
        {\formatfunctiontosubconfiguration{merge}(#1, #2 )}

\newcommand{\spcanonconfcompact}[1]%
        {\formatfunctiontosubconfiguration{canon}\bigl({#1}\bigr)}

\newcommand{\spprojsubsub}[4]%
    {\formatfunctiontosubconfiguration{proj}_{\scnot{#1}{#2}}(\scnot{#3}{#4})}
\newcommand{\spprojsubsubcompact}[4]%
    {\formatfunctiontosubconfiguration{proj}_{\scnot{#1}{#2}}%
    \bigl(\scnot{#3}{#4}\bigr)}
\newcommand{\spprojsubconf}[3]%
    {\formatfunctiontosubconfiguration{proj}_{\scnot{#1}{#2}}({#3})}
\newcommand{\spprojsubconfcompact}[3]%
    {\formatfunctiontosubconfiguration{proj}_{\scnot{#1}{#2}}\bigl({#3}\bigr)}
\newcommand{\spprojconfsub}[3]%
    {\formatfunctiontosubconfiguration{proj}_{#1}(\scnot{#2}{#3})}
\newcommand{\spprojconfsubcompact}[3]%
    {\formatfunctiontosubconfiguration{proj}_{#1}\bigl(\scnot{#2}{#3}\bigr)}
\newcommand{\spprojconfconf}[2]%
    {\formatfunctiontosubconfiguration{proj}_{#1}({#2})}
\newcommand{\spprojconfconfcompact}[2]%
    {\formatfunctiontosubconfiguration{proj}_{#1}\bigl({#2}\bigr)}

\newcommand{\spclossubcompact}[2]%
        {\formatfunctiontoset{cl}\bigl(\scnotcompact{#1}{#2}\bigr)}

\newcommand{\spintersubcompact}[2]%
        {\formatfunctiontoset{int}\bigl(\scnotcompact{#1}{#2}\bigr)}

\newcommand{\spcoversubcompact}[2]%
        {\formatfunctiontoset{cover}\bigl(\scnotcompact{#1}{#2}\bigr)}

\newcommand{\spcoverconfcompact}[1]%
        {\formatfunctiontoset{cover}\bigl({#1}\bigr)}

\newcommand{\spinducedblack}[1]%
    {\formatfunctiontoset{Bl} (#1)}
\newcommand{\spinducedwhite}[1]%
    {\formatfunctiontoset{Wh} (#1)}
\newcommand{\spinducedblackcompact}[1]%
    {\formatfunctiontoset{Bl} \bigl(#1 \bigr)}
\newcommand{\spinducedwhitecompact}[1]%
    {\formatfunctiontoset{Wh} \bigl(#1 \bigr)}

\newcommand{\pathclausedeg}[2][\pebdeg]%
    {\somenodetrueclausedeg[#1]{\vertexpath{#2}}}
\newcommand{\pathclauseNRdeg}[2][\pebdeg]%
    {\somenodetrueclausedeg[#1]{\vertexpathNR{#2}}}

\newcommand{\blacktruthdegexplicit}[4]%
        {\setdescrcompact
        {{\textstyle \Lor_{#2 = 1}^{#3} {#1}_{#2}}}
        {{#1} \in {#4}}}

\newcommand{\binsubtree}[1]{T^{#1}}

\newcommand{\vertexpath}[1]{{P}^{#1}}
\newcommand{\vertexpathNR}[1]{{P}_{*}^{#1}}

\newcommand{\unrelatedNP}[1]%
        {T \setminus \bigl(\binsubtree{#1} \unionSP \vertexpath{#1} \bigr)}
\newcommand{\unrelatedsmallNP}[1]%
        {T \setminus (\binsubtree{#1} \unionSP \vertexpath{#1} )}

\newcommand{\abovelevelblockerminsizecompact}%
    [2]{L_{\succeq{#1}}\bigl({#2}\bigr)}

\newcommand{\necessaryhidingvert}[2]%
{{#1}{\scriptstyle{\llfloor {#2} \rrfloor}}}

\newcommand{\Klawepropertyprefix}{Limited hiding-cardinality\xspace}

\newcommand{\klawepropacronym}{LHC property\xspace}

\newcommand{\nongenklaweprop}%
{non-generalized \Klawepropertyprefix property\xspace}
\newcommand{\nongenklawepropacronym}%
{non-generalized \klawepropacronym}
\newcommand{\nongenklawepropacronymWithParam}%
{(non-generalized) \klawepropacronym}

\newcommand{\siblingnonreachabiblitypropertynoref}%
{Sibling non-reachability property\xspace}
\newcommand{\Siblingnonreachabiblitypropertynoref}%
{Sibling non-reachability property\xspace}
\newcommand{\siblingnonreachabiblityproperty}%
{\siblingnonreachabiblitypropertynoref~%
\ref{property:sibling-non-reachability-property}\xspace}
\newcommand{\Siblingnonreachabiblityproperty}%
{\Siblingnonreachabiblitypropertynoref~%
\ref{property:sibling-non-reachability-property}\xspace}

\newcommand{\introducetermanmpctext}%
    {a \introduceterm{\mpctext{}}\xspace}

\newcommand{\introducetermamultipebblingtext}%
  {a \introduceterm{\multipebblingtext{}}\xspace}

\newcommand{\blobpebblingtext}{blob-pebbling\xspace}
\newcommand{\Blobpebblingtext}{Blob-pebbling\xspace}

\newcommand{\multipebblingtext}{\blobpebblingtext}

\newcommand{\multipebblegame}{blob-pebble game\xspace}

\newcommand{\blob}{blob\xspace}

\newcommand{\mpcostblack}[1]%
        {\formatpebblingprice{cost}_{\mpcblacks}( #1 )}
\newcommand{\mpcostwhite}[1]%
        {\formatpebblingprice{cost}_{\mpcwhites}( #1 )}

\newcommand{\blobpebblingprice}[1]{\formatpebblingprice{Blob-Peb}(#1)}
\newcommand{\blobpebblingpricecompact}[1]%
    {\formatpebblingprice{Blob-Peb}\bigl(#1\bigr)}

\newcommand{\multipebblingpricecompact}[1]%
    {\formatpebblingprice{Blob-Peb}\bigl(#1\bigr)}

\newcommand{\mpcblacks}{\formatfunctiontomulti{B}}
\newcommand{\mpcwhites}{\formatfunctiontomulti{W}}

\newcommand{\mpscnot}[2]{[ {#1} ] \langle {#2} \rangle}
\newcommand{\mpscnotcompact}[2]%
        {\big[ {#1} \big] \bigl\langle {#2} \bigr\rangle}

\newcommand{\mpscnotstd}[1][]{\mpscnot{B_{#1}}{W_{#1}}}

\newcommand{\mpctext}{\blobpebblingtext con\-fig\-u\-ra\-tion\xspace}

\newcommand{\Mpsctext}{Sub\-con\-figu\-ration\xspace}
\newcommand{\mpsctext}{sub\-con\-figu\-ration\xspace}

\newcommand{\preciseimpl}{\vartriangleright}

\newcommand{\preciseimpladv}{precisely\xspace}

\newcommand{\Preciseimplsubst}{Precise implication\xspace}

\newcommand{\chargeablevertices}[1]%
{\formatfunctiontoset{chargeable}({#1}) }
\newcommand{\chargeableverticescompact}[1]%
{\formatfunctiontoset{chargeable}\bigl({#1}\bigr) }

\newcommand{\blackschargedfor}[1][]%
    {\mpcblacks_{#1}}
\newcommand{\whiteschargedfor}[1][]%
    {\mpcwhites_{#1}^{\hspace{-0.3 pt}\vartriangle}}

\newcommand{\whitesbelowjustblocked}%
    {\mpcwhites_{B}^{\hspace{-0.3 pt}\vartriangle}}
\newcommand{\whitesbelowhidden}%
    {\mpcwhites_{H}^{\hspace{-0.3 pt}\vartriangle}}

\newcommand{\whitestight}%
    {\mpcwhites_{T}^{\hspace{-0.3 pt}\vartriangle}}

\newcommand{\subsectionNOW}[1]{\section{#1}}
\newcommand{\subsubsectionNOW}[1]{\subsection{#1}}

\newcommand{\refsubsec}[1]{Section~\ref{#1}}

\newcommand{\reftwosubsecs}[2]{Sections~\ref{#1} and~\ref{#2}}

\usepackage{paralist} 

\newtheoremstyle{metacommenttheoremstyle}
  {3pt}
  {3pt}
  {\sffamily \itshape \scriptsize
  }
  {}
  {\bfseries \scshape \footnotesize }
  {:}
  { }
  {}
  
\theoremstyle{metacommenttheoremstyle}

\newcommand{\editcomment}[1]%
{}                                           

\newcommand{\proofcomment}[1]%
{}                                         

\newcommand{\trmt}{T}

\newcommand{\varsdary}[2]{\setsofvarsorlitsup{Vars}{#1}{#2}}

\newcommand{\farity}{d}
\newcommand{\ff}[1][]{f_{#1}}

\newcommand{\varvecx}{\varvec{\varx}}
\newcommand{\varvecy}{\varvec{\vary}}

\newcommand{\signedfunc}[3][\ff]{#1^{#2}({#3})}

\newcommand{\varwindex}[2]{\varx[#1]_{#2}}

\renewcommand{\lpebblingtext}{labelled pebbling\xspace}
\newcommand{\Lpebblingtext}{Labelled pebbling\xspace}

\newcommand{\alpebblingtext}{a labelled pebbling\xspace}

\renewcommand{\lpebblegame}{labelled pebble game\xspace}

\newcommand{\bpblackparam}{m}
\newcommand{\bpwhiteparam}{S}

\newcommand{\nsboundedpebblingfull}[2]{$({#1},{#2})$-bounded \lpebblingtext}
\newcommand{\nsboundedpebbling}[2]{$({#1},{#2})$-bounded pebbling\xspace}
\newcommand{\boundedpebblingtext}{bounded pebbling\xspace}

\newcommand{\boundedpebblingfulltext}{bounded \lpebblingtext}

\newcommand{\linestd}{L}

\newcommand{\largeconstant}{\kappa}
\newcommand{\kdnfboundsconst}{K}

\newcommand{\spacefuncupper}{s_{\mathrm{hi}}}
\newcommand{\spacefunclower}{s_{\mathrm{lo}}}

\renewcommand{\proofsystemformat}[1]{\ensuremath{\mathcal{#1}}}

\newcommand{\sourcestd}{s}
\newcommand{\sinkstd}{z}

\newcommand{\indegreedag}{\ell}

\newcommand{\spacestd}{s}

\newcommand{\graphcopyindex}[1]{(#1)}

\newcommand{\singlesinkdagtext}{single-sink DAG\xspace}

\newcommand{\supconcnot}[2][]%
    {\ifthenelse{\equal{#1}{}}
    {\mathit{SC}_{#2}}
    {\mathit{SC}^{\graphcopyindex{#1}}_{#2}}}

\newcommand{\gtdag}[2][]                
    {\ifthenelse{\equal{#1}{}}
    {\Xi({#2})}
    {\Xi^{\graphcopyindex{#1}}({#2})}}
\newcommand{\Gtdag}[2][]                
    {\ifthenelse{\equal{#1}{}}
    {\Xi \bigl({#2}\bigr)}
    {\Xi^{\graphcopyindex{#1}}\bigl({#2}\bigr)}}
\newcommand{\gtsupconc}[2][]
    {\ifthenelse{\equal{#1}{}}
    {C({#2})}
    {C^{\graphcopyindex{#1}}({#2})}}

\newcommand{\gtsource}[3][]%
    {\ifthenelse{\equal{#1}{}}
     {\sourcestd_{#2}[#3]}
     {\sourcestd^{\graphcopyindex{#1}}_{#2}[#3]}}
\newcommand{\gtsink}[3][]%
    {\ifthenelse{\equal{#1}{}}
     {\sinkstd_{#2}[#3]}
     {\sinkstd^{\graphcopyindex{#1}}_{#2}[#3]}}
\newcommand{\scsource}[3][]%
    {\ifthenelse{\equal{#1}{}}
      {x_{#2}[#3]}
      {x^{\graphcopyindex{#1}}_{#2}[#3]}}
\newcommand{\scsink}[3][]%
    {\ifthenelse{\equal{#1}{}}
      {y_{#2}[#3]}
      {y^{\graphcopyindex{#1}}_{#2}[#3]}}

\newcommand{\ksncols}{p}
\newcommand{\ksreclevel}{q}   
\newcommand{\ksnrgraphs}{k}
\newcommand{\ksnblocks}{m}

\newcommand{\kscold}{d}

\newcommand{\ksquadletter}{K}

\newcommand{\ksquadnot}[3][]%
  {\ifthenelse{\equal{#1}{}}
    {\ksquadletter}
    {\ksquadletter^{\graphcopyindex{#1}}}
    ({#2}, {#3})}
  
\newcommand{\kspolyletter}{\Lambda}

\newcommand{\kspolynot}[4][]%
  {\ifthenelse{\equal{#1}{}}
    {\kspolyletter}
    {\kspolyletter^{\graphcopyindex{#1}}}
    ({#2}, {#3}, {#4})}

\newcommand{\ksquadblockletter}{M}

\newcommand{\ksquadblock}[3][]%
  {\ifthenelse{\equal{#1}{}}
    {\ksquadblockletter}
    {\ksquadblockletter^{\graphcopyindex{#1}}}
    ({#2}, {#3})}
  
\newcommand{\kspolyblockletter}{N}

\newcommand{\kspolyblock}[4][]%
  {\ifthenelse{\equal{#1}{}}
    {\kspolyblockletter}
    {\kspolyblockletter^{\graphcopyindex{#1}}}
    ({#2}, {#3}, {#4})}

\newcommand{\ksquadnotnblocks}[4][]{\ksquadnot[#1]{#2}{#3}[#4]}

\newcommand{\ksquadnotnblocksext}[3]%
    {\ksquadletter^{+\ksbeforegraphletter} ({#1}, {#2})[#3]}
  
\newcommand{\kspolynotnblocks}[5][]{\kspolynot[#1]{#2}{#3}{#4}[#5]}

\newcommand{\kspolynotnblocksext}[4]%
    {\kspolyletter^{+\ksbeforegraphletter} ({#1}, {#2}, {#3})[#4]}

\newcommand{\ksquadnotnblocksstd}%
    {\ksquadnotnblocks{\ksncols}{\ksreclevel}{\ksnblocks}}
\newcommand{\ksquadnotnblocksextstd}%
    {\ksquadnotnblocksext{\ksncols}{\ksreclevel}{\ksnblocks}}

\newcommand{\kspolynotnblocksstd}%
    {\kspolynotnblocks{\ksncols}{\ksreclevel}{\ksnrgraphs}{\ksnblocks}}
\newcommand{\kspolynotnblocksextstd}%
    {\kspolynotnblocksext{\ksncols}{\ksreclevel}{\ksnrgraphs}{\ksnblocks}}

\newcommand{\ksbeforev}[1][]%
    {\ifthenelse{\equal{#1}{}}{b}{b^{\graphcopyindex{#1}}}}
\newcommand{\ksafterv}[1][]%
    {\ifthenelse{\equal{#1}{}}{a}{a^{\graphcopyindex{#1}}}}
\newcommand{\ksfirstrowv}[1][]%
    {\ifthenelse{\equal{#1}{}}{f}{f^{\graphcopyindex{#1}}}}
\newcommand{\kslastrowv}[1][]%
    {\ifthenelse{\equal{#1}{}}{l}{l^{\graphcopyindex{#1}}}}
\newcommand{\ksrcolv}[1][]%
    {\ifthenelse{\equal{#1}{}}{r}{r^{\graphcopyindex{#1}}}}

\newcommand{\ksaftervindicescopy}[3]{a^{\graphcopyindex{#1}}_{#2, #3}}

\newcommand{\ksbeforegraphletter}{B}
\newcommand{\ksbeforegraph}[1][]%
    {\ifthenelse{\equal{#1}{}}{B}{B^{\graphcopyindex{#1}}}}
\newcommand{\ksaftergraph}[1][]%
  {\ifthenelse{\equal{#1}{}}{A}{A^{\graphcopyindex{#1}}}}
\newcommand{\ksrcolgraph}[1][]%
  {\ifthenelse{\equal{#1}{}}{R}{R^{\graphcopyindex{#1}}}}
\newcommand{\ksrcolgraphindex}[2][]%
    {\ifthenelse{\equal{#1}{}}{R_{#2}}{R^{\graphcopyindex{#1}}_{#2}}}

\newcommand{\ksnewgoalvertex}%
    {\ksaftervindicescopy{\ksnblocks+1}{n'}{\kscold}}

\newcommand{\anfprojsubst}[1][\funcpebc]{an \mbox{${#1}$-pro}\-jec\-tion\xspace}
\newcommand{\fprojsubst}[1][\funcpebc]{\mbox{${#1}$-pro}\-jec\-tion\xspace}

\newcommand{\spacerespecting}{space-faithful\xspace}
\newcommand{\Spacerespecting}{Space-faithful\xspace}
\newcommand{\SPACERESPECTING}{Space-Faithful\xspace}

\newcommand{\sprdeg}{K}

\newcommand{\projectionsubst}{projection\xspace}

\newcommand{\projectedverb}{projected\xspace}

\newcommand{\Projectionsubst}{Projection\xspace}

\newcommand{\projclnot}[1][\funcpebc]{{\formatfunctiontoset{proj}\!}_{#1}}
\newcommand{\projcl}[2][\funcpebc]{{\formatfunctiontoset{proj}\!}_{#1}({#2})}
\newcommand{\localprojclnot}[1][\funcpebc]{{\formatfunctiontoset{proj}\!}^{\,L}_{#1}}
\newcommand{\localprojcl}[2][\funcpebc]{{\formatfunctiontoset{proj}\!}^{\,L}_{#1}({#2})}

\newcommand{\rprojclnot}[1][\funcpebc]{{\formatfunctiontoset{Rproj}\!}_{#1}}
\newcommand{\rprojcl}[2][\funcpebc]{{\formatfunctiontoset{Rproj}\!}_{#1}({#2})}
\newcommand{\localrprojclnot}[1][\funcpebc]{{\formatfunctiontoset{Rproj}\!}^{\,L}_{#1}}
\newcommand{\localrprojcl}[2][\funcpebc]{{\formatfunctiontoset{Rproj}\!}^{\,L}_{#1}({#2})}

\newcommand{\Lorspecfuncproj}[4][\pebdeg]%
    {\Lor_{{#2} \in \clpospart{#3}} \specfuncwithvecarg[#1]{#4}{#2}}    
\newcommand{\Lorspecnotfuncproj}[4][\pebdeg]%
    {\Lor_{\olnot{#2} \in \clnegpart{#3}} \specnotfuncwithvecarg[#1]{#4}{#2}}
\newcommand{\witnessproj}[4][\funcpebc]%
    {\Lorspecfuncproj{#2}{#4}{#1} \lor \Lorspecnotfuncproj{#3}{#4}{#1}}

\newcommand{\Lorspecfuncprojnodisplay}[4][\pebdeg]%
    {\Lornodisplay_{{#2} \in \clpospart{#3}} \specfuncwithvecarg[#1]{#4}{#2}}    
\newcommand{\Lorspecnotfuncprojnodisplay}[4][\pebdeg]%
    {\Lornodisplay_{\olnot{#2} \in \clnegpart{#3}} \specnotfuncwithvecarg[#1]{#4}{#2}}
\newcommand{\witnessprojnodisplay}[4][\funcpebc]%
    {\Lorspecfuncprojnodisplay{#2}{#4}{#1} \lor \Lorspecnotfuncprojnodisplay{#3}{#4}{#1}}

\newcommand{\clpospart}[1]{#1^{+}}
\newcommand{\clnegpart}[1]{#1^{-}}

\newcommand{\substform}[2]{{#1}[{#2}]}

\newcommand{\Substformtext}{Substitution formula\xspace}

\newcommand{\fsubsttext}[1][{\funcpebc[\pebdeg]}]%
    {${#1}$-substitution\xspace}

\newcommand{\pebcontrwithfunc}[3][G]{\formulaformat{Peb}^{#2}_{#1}[{#3}]}

\newcommand{\extendedpebcontrwithfunc}[3][G]%
    {{\vphantom{\formulaformat{Peb}}\smash{\extendedversion{\formulaformat{Peb}}}}^{#2}_{#1}[{#3}]}

\newcommand{\genericpebcontr}[2][G]{\pebcontrwithfunc[#1]{#2}{\funcpebc}}

\newcommand{\pebcontrwithfuncNT}[3][G]%
     {\dummystar\!\!\formulaformat{Peb}^{#2}_{#1}[{#3}]}
\newcommand{\genericpebcontrNT}[2][G]%
     {\pebcontrwithfuncNT[#1]{#2}{\funcpebc}}

\newcommand{\funcpebc}[1][]{f_{#1}}

\newcommand{\clausesreprfunction}{\formatfunctiontoset{Cl}}
\newcommand{\clfunc}[3]{\clausesreprfunction[{#1}_{{#3}}({\varvec{#2}})]}
\newcommand{\clnotfunc}[3]{\clausesreprfunction[\lnot{#1}_{{#3}}({\varvec{#2}})]}
\newcommand{\clfuncstd}[2][\funcpebc]{\clfunc{#1}{#2}{\pebdeg}}
\newcommand{\clnotfuncstd}[2][\funcpebc]{\clnotfunc{#1}{#2}{\pebdeg}}

\newcommand{\varvec}[1]{\vec{#1}}

\newcommand{\specfuncwithvecarg}[3][\pebdeg]%
	   {{#2}_{#1}(\varvec{#3})}
\newcommand{\specnotfuncwithvecarg}[3][\pebdeg]%
	   {\lnot {#2}_{#1}(\varvec{#3})}

\newcommand{\funcwithvecarg}[2][\pebdeg]%
           {\specfuncwithvecarg[{#1}]{\funcpebc}{#2}}
\newcommand{\notfuncwithvecarg}[2][\pebdeg]%
           {\specnotfuncwithvecarg[{#1}]{\funcpebc}{#2}}

\newcommand{\funcwithvecargstd}%
	   {\funcwithvecarg[\pebdeg]{v}}
\newcommand{\notfuncwithvecargstd}%
	   {\notfuncwithvecarg[\pebdeg]{v}}

\newcommand{\specfuncwithargspebcontr}[3][\pebdeg]%
	   {{#2}_{#1}({#3}_1, \ldots, {#3}_{\pebdeg})}
\newcommand{\specnotfuncwithargspebcontr}[3][\pebdeg]%
	   {\lnot {#2}_{#1}({#3}_1, \ldots, {#3}_{\pebdeg})}

\newcommand{\funcwithargspebcontr}[2][\pebdeg]%
	   {{\funcpebc}_{#1}({#2}_1, \ldots, {#2}_{\pebdeg})}
\newcommand{\notfuncwithargspebcontr}[2][\pebdeg]%
	   {\lnot {\funcpebc}_{#1}({#2}_1, \ldots, {#2}_{\pebdeg})}

\newcommand{\funcwithargspebcontrstd}%
	   {\funcwithargspebcontr[\pebdeg]{v}}
\newcommand{\notfuncwithargspebcontrstd}%
	   {\notfuncwithargspebcontr[\pebdeg]{v}}

\newcommand{\Lorspecfunc}[4][\pebdeg]%
    {\Lor_{{#2} \in {#3}} \specfuncwithargspebcontr[#1]{#4}{#2}}
\newcommand{\Lorspecnotfunc}[4][\pebdeg]%
    {\Lor_{{#2} \in {#3}} \specnotfuncwithargspebcontr[#1]{#4}{#2}}

\newcommand{\Lorspecfuncvec}[4][\pebdeg]%
    {\Lor_{{#2} \in {#3}} \specfuncwithvecarg[#1]{#4}{#2}}    
\newcommand{\Lorspecnotfuncvec}[4][\pebdeg]%
    {\Lor_{{#2} \in {#3}} \specnotfuncwithvecarg[#1]{#4}{#2}}
\newcommand{\Lorspecfuncvecnodisplay}[4][\pebdeg]%
    {\Lornodisplay_{{#2} \in {#3}} \specfuncwithvecarg[#1]{#4}{#2}}    
\newcommand{\Lorspecnotfuncvecnodisplay}[4][\pebdeg]%
    {\Lornodisplay_{{#2} \in {#3}} \specnotfuncwithvecarg[#1]{#4}{#2}}
    
\newcommand{\witnessclauserp}[5][\funcpebc]%
    {\Lorspecfuncvec{#2}{#3}{#1} \lor \Lorspecnotfuncvec{#4}{#5}{#1}}
\newcommand{\witnessclauserpnodisplay}[5][\funcpebc]%
    {\Lorspecfuncvecnodisplay{#2}{#3}{#1} \lor \Lorspecnotfuncvecnodisplay{#4}{#5}{#1}}

\newcommand{\witnessclauserpstd}[1][\funcpebc]%
    {\witnessclauserp[#1]{b}{B}{w}{W}}

\newcommand{\nonauthoritarian}{non-authoritarian\xspace}
\newcommand{\Nonauthoritarian}{Non-authoritarian\xspace}

\newcommand{\orpebcontrtext}{OR-pebbling contradiction\xspace}

\newcommand{\ntrues}{k}
\newcommand{\ktrue}[2][]{{\mathit{thr}}_{#1}^{#2}}
\newcommand{\ktruestd}[1][]{\ktrue[#1]{\ntrues}}
\newcommand{\ktruewithargs}[2]%
    {\ktrue[\pebdeg]{#1}({#2}_1, \ldots, {#2}_{\pebdeg})}

\newcommand{\ktruepebcontrtext}[1][\ntrues]%
    {${#1}$-true-pebbling contradiction\xspace}

\newcommand{\xorpebcontrtext}{XOR-pebbling contradiction\xspace}

\renewcommand{\xor}{\oplus}
\newcommand{\Xor}{{\textstyle \bigoplus}}

\newcommand{\xorvertex}[3][\pebdeg]{\Xor_{{#3}=1}^{#1} {#2}_{#3}}
\newcommand{\notxorvertex}[3][\pebdeg]{\lnot \Xor_{{#3}=1}^{#1} {#2}_{#3}}

\newcommand{\Lornotxor}[3][i]%
    {\Lor_{{#2} \in {#3}}\notxorvertex{#2}{#1}}
\newcommand{\Lorxor}[3][i]%
    {\Lor_{{#2} \in {#3}}\xorvertex{#2}{#1}}

\newcommand{\respebblingtext}{res-pebbling\xspace}
\newcommand{\Respebblingtext}{Res-pebbling\xspace}
\newcommand{\RESPEBBLINGTEXT}{Res-Pebbling\xspace}
\newcommand{\respebblinglongtext}{resolution-pebbling\xspace}
\newcommand{\Respebblinglongtext}{Resolution-pebbling\xspace}
\newcommand{\RESPEBBLINGLONGTEXT}{Resolution-Pebbling\xspace}

\newcommand{\respebbling}[1][R]{\pebbling[#1]}

\newcommand{\rpscnot}[2]{[ {#1} ] \langle {#2} \rangle}

\newcommand{\rpscnotstd}[1][]{\rpscnot{B_{#1}}{W_{#1}}}

\newcommand{\rpconftext}{\respebblingtext con\-fig\-u\-ra\-tion\xspace}

\newcommand{\rpconf}[1][R]{\pconf[#1]}

\newcommand{\rpscintro}[1]{\rpscnot{#1}{\prednode{#1}}}

\renewcommand{\bwpebpricepersistent}[1]%
    {\formatpebblingprice{BW-Peb}(#1)}
\renewcommand{\Bwpebpricepersistent}[1]%
    {\formatpebblingprice{BW-Peb}\bigl(#1\bigr)}
\renewcommand{\pebpricepersistent}[1]%
    {\formatpebblingprice{Peb}(#1)}
\renewcommand{\Pebpricepersistent}[1]%
    {\formatpebblingprice{Peb}\bigl(#1\bigr)}

\begin{document}

%
%

\title[Pebble Games, Proof Complexity, and Time-Space Trade-offs]
    {Pebble Games, Proof Complexity, \\ and Time-Space Trade-offs\rsuper*}

\author[J.~Nordström]{Jakob Nordström}

%
%

\address{School of Computer Science and Communication \\
  KTH Royal Institute of Technology \\
  SE-100 44 \mbox{Stockholm,} Sweden}

\email{jakobn@kth.se}



\ACMCCS{[{\bf Theory of computation}]: Models of
  computation---Abstract machines; Computational complexity and
  cryptography---Proof complexity\,/\,Complexity theory and logic;
  Logic---Logic and verification\,/\,Automated reasoning; [{\bf
      Mathematics of computing}]: Discrete
  mathematics---Combinatorics; Mathematical software---Solvers}

\titlecomment{{\lsuper*}This is a survey of the author's PhD thesis~%
  \cite{Nordstrom08Thesis}, presented with the Ackermann Award at 
  \emph{CSL~'09}, as well as of some subsequent developments.}
  


%
%

\begin{abstract}
  %
  %
  Pebble games were extensively studied in the 1970s and 1980s in a number
  of different contexts. The last decade has seen a revival of
  interest in pebble games coming from the field of proof
  complexity. Pebbling has proven to be a useful tool for studying
  resolution-based proof systems when comparing the strength of
  different subsystems, showing bounds on proof space, and
  establishing size-space trade-offs. This is a survey of research in
  proof complexity drawing on results and tools from pebbling, with a
  focus on proof space lower bounds and trade-offs between proof
  size and proof space.
\end{abstract}

\maketitle  

%
%
%
%
%
%

%
%

%
%

\ifthenelse{\boolean{FoundationsAndTrends}}{}{\theoremstyle{plain}}

\ifthenelse{\boolean{maybeLMCS}}
{
  \newtheorem{theoremtemplate}[thm]{Tentative Theorem}    
  \newtheorem{corollarytemplate}[thm]{Tentative Corollary}
}
{
  \newtheorem{theoremtemplate}[standardlocalcounter]{Tentative Theorem}    
  \newtheorem{corollarytemplate}[standardlocalcounter]{Tentative Corollary}
}

\newcommand{\theoremtemplatename}{tentative theorem\xspace}
\newcommand{\reftheoremtemplate}[1]{Tentative Theorem~\ref{#1}}
\newcommand{\Reftheoremtemplate}[1]{Tentative Theorem~\ref{#1}}
\newcommand{\reftwotheoremtemplates}[2]{Tentative Theorems~\ref{#1} and~\ref{#2}}

\newcommand{\corollarytemplatename}{tentative corollary\xspace}
\newcommand{\refcorollarytemplate}[1]{Tentative Corollary~\ref{#1}}
\newcommand{\Refcorollarytemplate}[1]{Tentative Corollary~\ref{#1}}

%
%
%

%
%

\ifthenelse{\boolean{maybeLMCS}}
{\subsectionNOW{Introduction}
  \label{sec:pc-intro}}
{\sectionNOW{Pebbling and Proof Complexity}
  \label{sec:proof-cplx}
  
  In this 
  \ifthenelse{\boolean{FoundationsAndTrends}}
  {chapter,\footnote{This chapter is adapted from 
      the paper~\cite{Nordstrom10SurveyLMCS}.}}
  {section,\footnote{This section is adapted from 
      the paper~\cite{Nordstrom10SurveyLMCS}.}}
  we describe the connections between pebbling and proof complexity that
  is the main motivation behind this survey. Our focus will be on how
  pebble games have been employed to study trade-offs between time and space
  in resolution-based proof systems, in particular in the line of work
  \cite{Nordstrom09NarrowProofsSICOMP,
    NH13TowardsOptimalSeparation,
    BN08ShortProofs,
    BN11UnderstandingSpace},
  but we will also discuss other usages of pebbling in proof
  complexity. Let us start, however, by giving a quick overview of proof
  complexity in general.
  
  \subsectionNOW{A Selective Introduction to Proof Complexity}
  \label{sec:pc-intro}
}

Ever since the  fundamental \NP-completeness result of 
Stephen Cook~\cite{Cook71CooksTheorem}, 
the problem of deciding 
whether a given \proplog formula in conjunctive normal form (CNF)
is satisfiable or not has been on center stage in Theoretical
Computer Science.  In more recent years, \SATISFIABILITY has gone from
a problem of mainly theoretical interest to a practical approach for
solving applied problems.  Although all known Boolean satisfiability
solvers (SAT solvers) have exponential running time in the worst case,
enormous progress in performance has led to satisfiability algorithms
becoming a standard tool for solving a large number of real-world
problems such as hardware and software verification, experiment
design, circuit diagnosis, and  scheduling.

Perhaps a somewhat surprising aspect of this development is that the
most successful SAT solvers to date are still variants of the
Davis-Putnam-Logemann-Loveland (DPLL) procedure
\mbox{\cite{DLL62MachineProgram,DP60ComputingProcedure}}
augmented with 
\introduceterm{clause learning}~\cite{BS97UsingCSP,SS99Grasp}, 
which
can be seen to search for proofs
in the  resolution proof system.
For instance, the great majority of the best algorithms in recent
rounds of the international SAT competition~\cite{SATcompetitionShort} 
fit this description.

DPLL procedures perform a recursive backtrack search in the space of
partial truth value assignments. The idea behind clause learning
is that at each failure
(backtrack) point in the search tree, the system derives a reason for
the inconsistency in the form of a new clause and then adds this
clause to the original \cnfform (``learning'' the clause).  This can
save much work later on in the proof search, when some other
partial truth value assignment fails for similar reasons.

The main bottleneck for this approach, other than the obvious one that
the  running time is known to be exponential in the worst case, 
is the amount of space used   by the algorithms. 
Since there is only a fixed amount of memory, all clauses cannot be
stored. The difficulty lies in obtaining a highly selective and
efficient clause caching scheme that nevertheless keeps the
clauses needed. Thus, understanding time and memory requirements for
clause learning algorithms, and how these requirements are related to
each other, is a question of great practical importance.

Some good papers discussing clause learning (and
SAT solving in general) with examples of applications  are
\cite{BKS04TowardsUnderstanding,KS07StateofSAT,Marques08Practical}.
A more exhaustive general reference is the recently published 
\emph{Handbook of Satisfiability}~%
\cite{HandbookSAT}.
The first chapter of this handbook provides an excellent historical overview of
the satisfiability problem,  with \mbox{pages 19--25} focusing in
particular on DPLL and resolution.
%

The study of proof complexity originated with the
seminal paper of 
Cook and Reckhow~\cite{CR79Relative}. In its most general form,
a proof system for a language $\langstd$ is a
binary 
\mbox{predicate~$P(x, \proofstd)$,}  
which is
computable 
(deterministically) 
in time polynomial in 
the sizes $\setsize{x}$ and $\setsize{\proofstd}$ of the input
and has
the property that   
for all $x \in \langstd$ there is a string $\proofstd$ 
(a \introduceterm{proof})
for which
$P(x, \proofstd)$ evaluates to true, 
whereas for any
$x \not\in \langstd$
it holds for all strings $\proofstd$ that
$P(x, \proofstd)$ evaluates to false.
A proof system is said to be polynomially bounded if for every 
$x \in \langstd$
there exists a proof~$\proofstd_{x}$ for~$x$
that has size at most polynomial in~$\setsize{x}$.    
A \introduceterm{propositional proof system} is a proof system for
the language of tautologies in propositional logic. 

From a theoretical point of view, one important motivation for proof
complexity is the intimate connection with the fundamental question of
$\Pclass$ versus $\NPclass$.  Since $\NP$ is exactly the set of
languages with polynomially bounded proof systems, and since
\TAUTOLOGY 
can be seen to be the dual problem of
\SATISFIABILITY,
we have the famous theorem of~\cite{CR79Relative}
that \NP = \CoNP \ifaoif there exists a polynomially bounded
propositional proof system. 
Thus, if it could be shown that there are no
polynomially bounded proof systems for propositional tautologies,
\Pclass $\neq$ \NP 
would follow as a corollary since \Pclass is closed under complement.
One way of approaching this distant goal is to study stronger and
stronger proof systems and try to prove \superpoly lower bounds on
proof size.  However, although great progress has been made in the
last couple of decades for a variety of propositional proof systems,
it seems that we still do not fully understand the reasoning power of
even quite simple ones. 

Another important motivation for proof complexity is that, as was
mentioned above, designing efficient algorithms for proving
tautologies (or, equivalently, testing satisfiability) is a very
important problem not only in the theory of computation but also in
applied research and industry.  All automated theorem provers,
regardless of whether they produce a written proof or not,
explicitly or implicitly define a system in which proofs are searched
for and rules which determine what proofs in this system look like.
Proof complexity analyzes what it takes to simply write down and
verify the proofs that such an automated theorem prover might find,
ignoring the computational effort needed to actually find them.  Thus,
a lower bound for a proof system tells us that any algorithm, even an
optimal (non-deterministic) one making all the right choices, must
necessarily use at least the amount of a certain resource specified by
this bound.
In the other direction, theoretical upper bounds on some
proof complexity measure give us hope of finding good proof search
algorithms \wrt this measure, provided that we can design algorithms
that search for proofs in the system in an efficient manner.  For DPLL
procedures with clause learning, also known as conflict-driven clause learning 
(CDCL) solvers,
the time and memory resources used are measured by the
\introduceterm{length}
and
\introduceterm{space}
of proofs in the resolution proof system.

The field of proof complexity also has rich connections to
cryptography, artificial intelligence and mathematical logic.
\ifthenelse{\boolean{maybeLMCS}}
{Some good sources providing more details are~%
\cite{%
  B00ProofComplexity,
  BP98Propositional,
  CK02BooleanFunctions,
  Segerlind07Complexity,
  U95Complexity}.}
{We again refer the reader to, \eg, 
\cite{%
  B00ProofComplexity,
  BP98Propositional,
  CK02BooleanFunctions,
  Segerlind07Complexity,
  U95Complexity}
for more information.}

\subsubsectionNOW{Resolution-Based Proof Systems}
\label{sec:pc-intro-resolution}

Any formula in \proplog can be converted to a \cnfform that is only
linearly larger and is unsatisfiable \ifaoif the original formula is a
tautology.  Therefore, any sound and complete system 
that certifies the unsatisfiability of CNF formulas
can be considered as a general propositional proof system.

Arguably the single most studied proof system in propositional
proof complexity, \introduceterm{resolution}, is such a system
that produces proofs of the unsatisfiability of  \cnfform{}s. 
The resolution proof system appeared in~%
\cite{B37Canonical}
and began to be investigated in connection with automated theorem proving 
in the 1960s
\mbox{\cite{DLL62MachineProgram,
  DP60ComputingProcedure,
  R65Machine-oriented}}.
Because of its simplicity---there is only one derivation rule---and
because all lines in a proof are disjunctive clauses, this proof
system readily lends itself to proof search algorithms.

Being so simple and fundamental, resolution was also a natural target
to attack when developing methods for proving lower bounds in proof
complexity.  In this context, it is most straightforward to prove
bounds on the \introduceterm{length} of refutations, \ie the number of
clauses, rather than on the size of refutations, \ie the total number
of symbols. The length and
size measures are easily seen to be polynomially related.  The first
superpolynomial lower bound on resolution was presented by Tseitin
in the paper~\cite{T68ComplexityTranslated}, which is the published version of a
talk given in 1966, for a restricted form of the proof system called
\introduceterm{regular} resolution. It took almost an additional
20~years before Haken~\cite{H85Intractability} was able to establish
superpolynomial bounds without any restrictions, showing that CNF
encodings of the pigeonhole principle are intractable for general
resolution.  This weakly exponential bound of Haken has later been
followed by many other strong results, among others truly exponential%
\footnote{In this paper, an \introduceterm{exponential} lower bound is 
  any bound $\exp \bigl( \Bigomega{n^{\delta}} \bigr)$ for some 
  $\delta > 0$, where $n$ is the size of the formula in question.
  By a \introduceterm{truly expontential}
  lower bound we mean a bound $\exp ( \bigomega{n} )$.}
lower bounds on resolution refutation length for different formula
families in, \eg, 
\cite{BKPS02Efficiency,BW01ShortProofs,CS88ManyHard,U87HardExamples}.

A second complexity measure for resolution, first made explicit by
Galil~\cite{Galil77Resolution}, is the \introduceterm{width}, measured
as the maximal size of a clause in the refutation. Clearly, the
maximal width needed to refute any unsatisfiable CNF formula is at
most the
number of variables in it, which is upper-bounded by the formula size.
Hence, while refutation length can be exponential in the worst case,
the width ranges between constant and linear measured in the formula size.
Inspired by
previous work \cite{BP96Simplified,CEI96Groebner,IPS99LowerBounds}, 
Ben-Sasson and Wigderson~\cite{BW01ShortProofs} identified width as a 
crucial resource of resolution proofs by showing that the minimal width of any
resolution refutation of a \mbox{$\clwidth$-CNF} formula~$\fstd$ (\ie a
formula where all clauses have size at most some constant~$\clwidth$)
is bounded from above by the minimal refutation length by
\begin{equation}
  \label{eq:intro-Ben-Sasson-Wigderson-bound}
  \text{minimal width}
  \leq
  \Bigoh{\sqrt{(\text{size of formula}) \cdot \log (\text{minimal length})}}. 
\end{equation}
Since it is also easy to see that resolution refutations of \polysize
formulas in small width must necessarily be short---quite simply   for
the reason that  
$(2 \cdot \# \text{variables})^w$ 
is an upper bound on the total number of distinct clauses of
width at most~$w$---the result in~\cite{BW01ShortProofs} can be interpreted as
saying roughly  that there exists a short refutation of the \kcnfform~$\fstd$
\ifaoif there exists a (reasonably) narrow refutation of~$\fstd$.
This interpretation also gives rise to a natural proof search
heuristic: to find a short refutation, search for
refutations in small width. It was shown in~%
\cite{BIW00Near-optimalSeparation}
that there are formula families for which this heuristic exponentially
outperforms any DPLL procedure (without clause learning) regardless of
branching function. 

The idea to study \introduceterm{space} in the context of proof complexity
appears to have been raised for the first time%
\footnote{In fact, going through the literature one can find that
  somewhat similar concerns have been addressed in 
  \cite{Kozen77Lower}
  and
  \cite[Section~4.3.13]{KBL99PropositionalLogicFullRef}.
  However, the space measure defined there is too strict, 
  since it turns out that for all proof systems examined in the current
  paper, a CNF formula~$\fstd$ with $m$~clauses will always have space
  complexity in the interval $\intclcl{m}{2m}$ (provided the technical
  condition that $\fstd$ is minimally unsatisfiable as defined below), and
  this does not make for a very interesting measure.}
%
%
by Armin Haken during a workshop in Toronto in 1998, and the formal study of
space in resolution was initiated by Esteban and Torán in~%
\cite{ET01SpaceBounds}
and was later extended to a 
more general setting by Alekhnovich \etal in~%
\cite{ABRW02SpaceComplexity}.
Intuitively, we can view a  resolution refutation of a \cnfform $\fstd$ 
as a sequence of derivation steps on a blackboard, where in each step 
we may write a clause from~$\fstd$ on the blackboard,
erase a clause from the blackboard, or derive some new
clause implied by the clauses currently written on the blackboard,
and where the refutation ends when we reach the contradictory empty
clause.  The space of a refutation is then the 
maximal number of clauses one needs to keep on the blackboard
simultaneously at any time during the refutation, and the space of
refuting~$\fstd$ is defined as the minimal space of any resolution
refutation of~$\fstd$. 
A number of upper and lower bounds for refutation space in resolution
and other proof systems were subsequently presented 
in, for example,
\cite{ABRW02SpaceComplexity,
  BG03SpaceComplexity,
  EGM04Complexity,
  ET03CombinatorialCharacterization},
and to distinguish the space measure of 
\cite{ET01SpaceBounds}
from other measures introduced in these later papers we will sometimes refer
to it as  
\introduceterm{clause space}
below for extra clarity.

Just as is the case for width, the minimum clause space of refuting a
formula can be upper-bounded by the formula size.
Somewhat unexpectedly, 
it was discovered in a sequence of works 
that lower bounds on resolution refutation space for different formula
families turned out to match exactly previously known lower bounds on
refutation  width. In an elegant paper~%
\cite{AD08CombinatoricalCharacterization},
Atserias and Dalmau showed that this was not a coincidence, but that
the inequality 
\begin{equation}
  \label{eq:intro-Atserias-Dalmau-bound}
  \text{minimal width}
  \leq
  \text{minimal clause space} + \text{small constant}
\end{equation}
holds for refutations of  any \kcnfform~$\fstd$, where the constant
term depends only on~$\clwidth$.
Since clause space is an upper bound on width by
\refeq{eq:intro-Atserias-Dalmau-bound},
and since width upper-bounds length by the counting argument discussed
above, it
%
%
follows that upper bounds on clause space imply upper bounds on length. 
Esteban and Tor\'{a}n \cite{ET01SpaceBounds} showed the converse that
length upper bounds imply clause space upper bounds for the restricted case of
\introduceterm{tree-like}
resolution (where every clause can only be used once
in the derivation and has to be rederived again from scratch if it is
needed again at some later stage in the proof).
Thus, clause space is an interesting complexity measure with
nontrivial relations to proof length and width. We note that apart
from being of theoretical interest, clause space has also been
proposed in~\cite{ABLM08Measuring} as a relevant measure of the
hardness in practice of CNF formulas for SAT solvers,
and such possible connections have been further investigated in~%
\cite{JMNZ12RelatingProofCplx}.

%
%

The resolution proof system was generalized by
Kraj{\'\i}{\v{c}}ek~\cite{K01OnTheWeak}, who introduced the the family
of \introduceterm{$k$-DNF resolution} proof systems as an intermediate
step between resolution and depth-$2$ Frege systems.  Roughly
speaking, for positive integers $k$ the $k$th member of this family, which we
denote $\resknot$, is allowed to reason in terms of formulas in
disjunctive normal form (\introduceterm{DNF formulas}) with the added
restriction that any conjunction in any formula is over at most $k$
literals. For $k=1$, the lines in the proof are hence disjunctions of
literals, and the system
$\resknot[1] = \resnot$ 
is standard resolution. At the other extreme, 
$\resknot[\infty]$ 
is equivalent to depth-$2$ Frege.

The original motivation to study 
the family of $k$-DNF resolution proof systems,
as stated in~\cite{K01OnTheWeak}, was to better understand the complexity
of counting in weak models of bounded arithmetic, and it was later
observed that these systems are also related to SAT solvers that
reason using multi-valued logic (see \cite{JN02OptimalLowerBound} for
a discussion of this point). A number of subsequent works have shown
superpolynomial lower bounds on the length of \mbox{$\resknot$-refutations},
most notably for (various formulations of) the pigeonhole principle
and for random CNF formulas
\cite{AB04automatizabilityOfResolutionAndRelated,
  ABE02LowerBoundsWPHPandRandom,
  Alekhnovich11LowerBoundsk-DNF3-CNF,
  JN02OptimalLowerBound,
  Razborov03PseudorandomGeneratorsHard,
  SBI04SwitchingLemma,
  Segerlind05Exponential}.
Of particular interest in the current context are the results of Segerlind
\etal~\cite{SBI04SwitchingLemma} and of Segerlind~\cite{Segerlind05Exponential} 
showing that the family of 
\mbox{$\resknot$-systems} form a strict hierarchy with respect to proof
length. More precisely, they prove that for every~$k$ there exists a
family of formulas 
$\set{F_n}_{n=1}^{\infty}$ 
of arbitrarily large size $n$ such that
$F_n$ has an $\reskone$-refutation of polynomial length 
but any $\resknot$-refutation of $F_n$ requires exponential length.

With regard to space, 
Esteban et al.~\cite{EGM04Complexity}
established essentially optimal linear lower bounds in  $\resknot$ on
\introduceterm{formula space},
extending the clause space measure for resolution in the natural way
by counting the number of \kdnfform{}s. They also proved that the family
of \emph{tree-like} $\resknot$ systems form a strict hierarchy with
respect to formula space in the sense that there are arbitrarily large
formulas $F_n$ of size $n$ that can be refuted in tree-like $\reskone$ in
constant space but require space $\Omega(n/\log^2 n)$ to be refuted in
tree-like~$\resknot$. It should be pointed out, however, that as observed in
\cite{K01OnTheWeak,EGM04Complexity} the family of tree-like
$\resknot$ systems for all $k>0$ are strictly weaker than standard
resolution. As was mentioned above, the family of general, unrestricted
 $\resknot$ systems are strictly stronger than resolution, so the results
in \cite{EGM04Complexity} left 
open the question of
whether there is a strict formula space hierarchy for
(non-tree-like) $\resknot$ or not.

\subsubsectionNOW{Three Questions Regarding Space}
\label{sec:pc-questions-relations-time-space}

Although resolution is simple and by now very well-studied, the
research surveyed above left open a few fundamental questions about
this proof system. In what follows, our main focus will be on the three
questions considered below.%
\footnote{In the interest of full disclosure, it should perhaps be
  noted that these questions also happened to be the focus of the
  author's PhD thesis~\cite{Nordstrom08Thesis}.}
\begin{enumerate}[(1)]
\item
  \label{question-space-width}
  What is the relation between clause space and width?  The
  inequality~\refeq{eq:intro-Atserias-Dalmau-bound} says that clause
  space $\gtrapprox$ width, but it leaves open whether this
  relationship is tight or not.  Do the clause space and width
  measures always coincide, or is there a formula family that
  separates the two measures asymptotically?
  
\item
  \label{question-space-length-separation}
  What is the relation between clause space and length?
  Is there some nontrivial correlation between the two in the sense that
  formulas refutable in short length must also be refutable in small
  space, or can ``easy'' formulas \wrt length be ``arbitrarily complex''
  \wrt space? (We will make these notions more precise shortly.)
  
\item
  \label{question-space-length-trade-offs}
  Can the length and space of refutations be optimized simultaneously,
  or are there trade-offs in the sense that any refutation that
  optimizes one of the two measures must suffer a blow-up in the other? 
%
%
\end{enumerate}

\noindent To put the questions about length versus space in perspective,
consider what has been known for length versus width.
It follows from the inequality
\refeq{eq:intro-Ben-Sasson-Wigderson-bound}
that if the width of refuting a \kcnf formula family
$\set{\fstd_{\nvar}}_{\nvar=1}^{\infty}$ of size $\nvar$
grows asymptotically faster than
$\sqrt{\nvar \log \nvar}$,
then the length of refuting $\fstd_{\nvar}$ must be superpolynomial.
This is known to be almost tight, since
Bonet and Galesi~\cite{BG01Optimality}
showed that there is a family of \kcnfform{}s of size~$\nvar$
with minimal refutation width growing like~$\sqrt[3]{\nvar}$, 
but which is nevertheless refutable in length linear in~$\nvar$.
Hence, formulas refutable in polynomial
length can have somewhat wide minimum-width refutations, but not
arbitrarily wide ones.

Turning to the relation between clause space and length, we
note that the inequality~\refeq{eq:intro-Atserias-Dalmau-bound}  
tells us that any correlation between length
and clause space cannot be tighter than the correlation between length and
width. In particular, we get from the previous paragraph 
that \kcnfform{}s refutable in polynomial length may have at least
``somewhat spacious'' minimum-space refutations. 
At the other end of the spectrum, given any resolution refutation
of~$\fstd$ in length~$\lengthstd$ it 
is a straightforward consequence of
\mbox{\cite{ET01SpaceBounds, HPV77TimeVsSpace}} 
that the space needed to refute $\fstd$ is at most on the order of
$\lengthstd / \log \lengthstd$.
This gives a trivial upper bound on any possible separation of the two
measures. 
Thus, what the question above is asking is whether it can be
that length and space are ``completely unrelated'' in
the sense that there exist \kcnfform{}s with refutation length $\lengthstd$
that need maximum possible space
$\Bigomega{L / \log L}$, or whether there is
a nontrivial upper bound on clause space in terms of length
analogous to 
the inequality
in~\refeq{eq:intro-Ben-Sasson-Wigderson-bound}, 
perhaps even stating that
$
\text{minimal clause space}
\leq
\Bigoh{\sqrt{(\text{size of formula}) \cdot \log (\text{minimal length})}} 
$
or similar.
Intriguingly, as we discussed above it was shown in~%
\cite{ET01SpaceBounds} 
that for the restricted case of so-called tree-like resolution there
is in fact a tight correspondence between length and clause space,
exactly as for length versus width.

\subsubsectionNOW{Pebble Games to the Rescue}
\label{sec:pc-pebbling-to-the-rescue}

Although the above questions have been around for a while, as
witnessed by discussions in, \eg, the papers  
\cite{ABRW02SpaceComplexity,
  Ben-Sasson09SizeSpaceTradeoffs,
  BG03SpaceComplexity,
  EGM04Complexity,
  ET03CombinatorialCharacterization,
  Segerlind07Complexity,
  Toran04Space}, 
there appears to have been no consensus on what the right answers
should be. However, what most of these papers did agree on was that a
plausible formula  family for answering these questions were so-called 
\ifthenelse{\boolean{maybeLMCS}}
{\introduceterm{pebbling contradictions}
  defined in terms of pebble games over directed acyclic graphs.}
{\introduceterm{pebbling contradictions}, 
  which are CNF formulas encoding pebble games played on graphs.}
Pebbling contradictions had already appeared in various disguises in
some of the papers 
mentioned in \refsubsec{sec:pc-intro-resolution},
and it had been noted 
that non-constant lower bounds on the clause space of
refuting pebbling contradictions would separate space and width 
and possibly also clarify the relation between space and length 
if the bounds were good enough. 
On the other hand, a constant upper bound on the refutation space
would improve the trade-off results for different proof complexity 
measures for resolution in~%
\cite{Ben-Sasson09SizeSpaceTradeoffs}.

And indeed, pebbling turned out to be just the right tool to
understand the interplay of length and space in resolution. 
The main purpose of this
\ifthenelse{\boolean{maybeLMCS}}          
{survey}
{\ifthenelse{\boolean{FoundationsAndTrends}}{chapter}{section}}
is to give an overview of the works establishing connections
between pebbling and proof complexity \wrt time-space trade-offs.
We will need to give some preliminaries in order to state the formal results,
but before we do so let us conclude this
\ifthenelse{\boolean{FoundationsAndTrends}}{introductory section}{introduction}
by giving a brief description of the relevant results.

The first progress was reported in 2006 (journal version in~%
\cite{Nordstrom09NarrowProofsSICOMP}),
where pebbling formulas of a very particular form, namely pebbling
contradictions defined over complete binary trees, were studied. 
This was sufficient to establish a logarithmic separation of
clause space and width, thus answering
question~\ref{question-space-width} 
above. This separation was 
improved from logarithmic
to polynomial in 2008
(journal version in~%
\cite{NH13TowardsOptimalSeparation}),
where a broader class of graphs were analyzed, but where unfortunately
a rather involved argument was required for this analysis to go through.
In~\cite{BN08ShortProofs}, a somewhat different approach was taken by
modifying the pebbling formulas slightly. This made the analysis both
much simpler and much stronger, and led to a resolution of
question~\ref{question-space-length-separation}
by establishing an optimal separation between clause space and length,
\ie 
showing
that there are formulas with refutation length~$\lengthstd$ that require
clause space~$\Bigomega{L / \log L}$.
In a further improvement,  the paper
\cite{BN11UnderstandingSpace}
used similar ideas to translate pebbling time-space trade-offs to
trade-offs between length and space in resolution, thus answering
question~\ref{question-space-length-trade-offs}.
In the  same paper these results were also extended 
to the \kdnf resolution proof systems, 
which also yielded as a corollary that the $\resknot$-systems indeed
form a strict hierarchy \wrt space.

\ifthenelse{\boolean{maybeLMCS}}
{\subsubsectionNOW{Outline of This Survey}

The rest of this survey is organized as follows.
\Refsec{sec:pc-preliminaries}
presents the necessary formal preliminaries,
and
\refsec{sec:pc-pebcontrs-and-proof-cplx}
gives a high-level overview of pebbling in proof complexity.
In 
\refsec{sec:pc-tradeoffs},
we describe in more detail how time-space separations and trade-offs
have been proven with the help of pebble games, and in 
\refsec{sec:pc-substitution-theorem}
we examine how this approach can be cast as a simple and generic
technique of proving lower bounds via variable substitutions. 
\Refsec{sec:pc-open-problems}
discusses a number of open problems. Finally, some concluding remarks are
given in
\refsec{sec:concluding-remarks}.

  \subsectionNOW{Preliminaries}
  \label{sec:pc-preliminaries}}
{\subsectionNOW{Proof Complexity Preliminaries}
  \label{sec:pc-preliminaries}
  
  Below we present the definitions, notation and terminology
  that we will need to make more precise the informal exposition
  in \refsubsec{sec:pc-intro}.
}

\subsubsectionNOW{Variables, Literals, Terms, Clauses,
  Formulas and Truth Value Assignments}
\label{sec:pc-variables-literals}

For $\varx$ a Boolean variable, a 
\introduceterm{literal over $\varx$} 
is either the variable $\varx$ itself, called a 
\introduceterm{positive literal over $\varx$}, 
or its  negation, denoted $\lnot \varx$ or $\olnot{\varx}$ and called a 
\introduceterm{negative literal over $\varx$}. 
Sometimes it will also be convenient to write $\varx^1$ for unnegated
variables and $\varx^0$ for negated ones.
We define $\lnot \lnot \varx$ to be $\varx$. 
A \introduceterm{clause} 
$\clc = \lita_1 \lor \formuladots \lor \lita_{\clwidth}$ 
is a disjunction of literals,
and a \introduceterm{term}
$\trmt = \lita_1 \land \formuladots \land \lita_{\clwidth}$ 
is a conjunction of literals. Below we will think of clauses and terms
as sets, so that the ordering of the literals is inconsequential and
that, in particular, no literals are repeated. We will also 
(\wolog) assume that all clauses and terms are nontrivial in the sense
that they do not contain both a literal and its complement.
A clause (term) containing at most $\clwidth$~literals is called a
\mbox{\introduceterm{$\clwidth$-clause}}
(\mbox{\introduceterm{$\clwidth$-term}}).
A \introduceterm{CNF formula} 
$\fstd = \clc_1 \land \formuladots \land \clc_m$ 
is a conjunction of clauses, and a \introduceterm{DNF formula} is a
disjunction of terms.  We will think of CNF and DNF formulas as sets
of clauses and terms, respectively.
A \introduceterm{\kcnfform{}} 
is a \cnfform consisting of
\xclause{\clwidth}{}s, and a \introduceterm{\kdnfform{}} consists of 
\mbox{$\clwidth$-terms}.

The \introduceterm{variable set} of a
clause~$\clc$, denoted $\vars{\clc}$, 
is the set of Boolean variables over
which there are literals 
in $\clc$,
%
and we write
$\lit{\clc}$ to denote the set of literals in~$\clc$. 
The variable and literal sets of a 
term
are  similarly defined and these definitions are extended to CNF and
DNF formulas by taking unions.%
\footnote{%
    Although the notation $\lit{\cdot}$ is slightly redundant for clauses
    and terms given that we consider them to be sets of literals, we find
    that it increases clarity to have a uniform notation for literals
    appearing in clauses or terms \emph{or formulas}. Note that 
    $\varx \in \fstd$ 
    means that that the unit clause $\varx$ appears in the CNF
    formula~$\fstd$, whereas 
    $\varx \in \lit{\fstd}$
    denotes that the positive literal~$\varx$ appears in some clause
    in~$\fstd$ and
    $\varx \in \vars{\fstd}$
    denotes that the variable~$\varx$ appears in~$\fstd$ (positively
    or negatively).} 
If $V$ is a set of Boolean variables and
$\vars{\clc}\subseteq V$, we say $\clc$ is a clause
\introduceterm{over $V$} and
similarly define 
terms,
CNF formulas, and DNF formulas over~$V$.

In what follows,  
we will usually write
$\lita, \litb, \litc$
to denote literals,
$\cla, \clb, \clc, \cld$
to denote clauses,
$\trmt$ 
to denote terms,
$\fstd, \falt$
to denote CNF formulas, 
and
$\clsc, \clsd$
to denote sets of clauses, \kdnfform{}s or sometimes other Boolean functions.
We will assume the existence of an arbitrary but fixed set
of variables
$V = \set{\varx, \vary, \varz, \ldots}$.
For a variable 
$\varx \in V$
we define
$\varsdary{d}{\varx} = \set{\varx_1, \ldots, \varx_d}$,
and we will tacitly assume that $V$ is such that 
the new set of variables
$
\varsdary{d}{V} =
\set{\varx_1, \ldots, \varx_d,
\vary_1, \ldots, \vary_d,
\varz_1, \ldots, \varz_d,
\ldots}
$
is disjoint from~$V$.
We will say that the variables $\varx_1, \ldots, \varx_{\farity}$, and
any literals over these variables, all \introduceterm{belong} to the 
variable~$x$. 

We write
$\tvastd, \tvaalt$
to denote truth value assignments, usually over
$\varsdary{d}{V}$ 
but sometimes over~$V$.
Partial truth value assignments, or \introduceterm{restrictions}, will
often be denoted~$\rstd$.
Truth value assignments are functions to 
$\set{0,1}$,
where we identify $0$ with false and $1$ with true.
We have the usual semantics that  a clause is true under~$\tvastd$, or 
\introduceterm{satisfied} 
by~$\tvastd$,  if at least one literal in it is
true, and a term is true if all literals evaluate to true.
We write
$\emptycl$
to denote the empty clause without literals that is false under all
truth value assignments. (The empty clause is also denoted, \eg,
$\lambda$,
$\Lambda$,
or
$\{\}$
in the literature.)
A CNF formula is satisfied if all clauses in it are satisfied, and for
a DNF formula we require that some term should be satisfied. 
In general, we will not distinguish between a formula and the Boolean
function computed by it.

If $\clsc$ is a set of Boolean functions we say that a restriction (or
assignment) satisfies $\clsc$ if and only if it satisfies every
function in $\clsc$.  For $\clsd, \clsc$ two sets of Boolean functions over a
set of variables $V$, we say that $\clsd$ \introduceterm{implies}~$\clsc$,
denoted $\clsd\impl\clsc$, if and only if every assignment 
$\funcdescr{\tvastd}{V}{\set{0,1}}$ 
that satisfies $\clsd$ also satisfies~$\clsc$. In particular,
$\clsd\impl\emptycl$ if and only if $\clsd$ is
\introduceterm{unsatisfiable}
or
\introduceterm{contradictory},
\ie if no assignment satisfies $\clsd$.
If a CNF formula~$\fstd$ is unsatisfiable but for any clause
$\clc \in \fstd$
it holds that the clause set
$\fstd \setminus \set{\clc}$
is satisfiable, we say that $\fstd$ is
\introduceterm{minimally unsatisfiable}.

\subsubsectionNOW{Proof Systems}
\label{sec:pc-proof-systems}

In this paper, we will focus on proof systems for refuting
unsatisfiable CNF formulas. (As was discussed in 
\refsubsec{sec:pc-intro-resolution}
this is essentially \wolog.) In this context it should be noted that, 
perhaps somewhat confusingly, a refutation of a
formula~$\fstd$ is often also referred to as a ``proof of~$\fstd$'' in
the literature.  We will try to be consistent and talk only about
``refutations of~$\fstd$,'' but will otherwise use the two terms
``proof'' and ``refutation'' interchangeably.

We say that a proof system is \introduceterm{sequential} if a proof
$\proofstd$ in the system is a \introduceterm{sequence} of lines
$\proofstd=\set{\linestd_1,\ldots, \linestd_\tau}$
of some prescribed syntactic form depending on the proof system in
question,  where each line is derived from
previous lines by one of a finite set of allowed 
\introduceterm{inference  rules}.
We say that a sequential proof system is
\introduceterm{implicational} if in addition
it holds for each inferred line 
that it is semantically implied by previous lines in the proof.
We remark that a proof system such as, for instance, extended Frege does \emph{not}
satisfy this property, since introducing a new extension variable as a
shorthand for a formula declares an equivalence that is
not the consequence of this formula.
All proof systems studied in this paper are implicational, however.

Following the exposition in~%
\cite{ET01SpaceBounds},
we view a proof  as similar to a non-deterministic
Turing machine computation, with a special read-only input tape from
which the clauses of the formula $\fstd$ being refuted
(the \introduceterm{axioms})
can be downloaded and a working memory where all
derivation steps are made. Then the length of a proof is essentially
the time of the computation and space measures memory
consumption.
The following definition is a straightforward generalization 
to arbitrary sequential proof systems
of the
definition in~%
\cite{ABRW02SpaceComplexity}
for the resolution proof system. 
We note that proofs defined in this way have been referred to as
\introduceterm{configuration-style}
proofs or 
\introduceterm{space-oriented} 
proofs in the literature.

\begin{definition}[Refutation]
  \label{def:sequential-refutation}
  For a sequential proof system
  $\psstd$
  with lines of the form
  $\linestd_i$,
  a 
  \introduceterm{\mbox{\introduceterm{$\psstd$-con}}\-fig\-u\-ration}~%
  $\clsd$, or, simply, a
  \introduceterm{configuration}, is a set of such lines. A sequence of
  configurations $\set{\clsd_0,\ldots, \clsd_\stoptime}$ is said to be
  a \introduceterm{$\psstd$-derivation} from a \cnfform $\fstd$ if
  $\clsd=\emptyset$ and for all $t\in [\stoptime]$, the set $\clsd_t$
  is obtained from $\clsd_{t-1}$ by one of the following
  \introduceterm{derivation steps}: 
  \begin{enumerate}[\hbox to8 pt{\hfill}]                                    
    \italicitem\noindent{\hskip-12 pt\bf\textit{Axiom Download:}}\ 
      $\clsd_t=\clsd_{t-1}\union\set{\linestd_C}$, where $\linestd_C$
    is 
    the encoding of a clause $C\in \fstd$ in the syntactic form
    prescribed by the proof system
    (an
    \introduceterm{axiom clause})
    or an axiom of the proof system.
    \italicitem\noindent{\hskip-12 pt\bf\textit{Inference:}}\ 
      $\clsd_t=\clsd_{t-1} \union \set{\linestd}$ for some $\linestd$ inferred by one
      of the inference rules for~$\psstd$ from a set of assumptions
      $\linestd_1, \ldots, \linestd_d \in \clsd_{t-1}$.
    \italicitem\noindent{\hskip-12 pt\bf\textit{Erasure:}}\ 
      $\clsd_t =  \clsd_{t-1} \setminus \set{\linestd}$ for some 
      $\linestd \in \clsd_{t-1}$.
  \end{enumerate}
  A $\psstd$-refutation
  $\derivof{\proofstd}{\fstd}{\emptycl}$ 
  of a CNF formula $F$ is a derivation
  $\proofstd=\set{\clsd_0,\ldots,\clsd_\stoptime}$ such that
  $\clsd_0=\emptyset$
  and
  $\emptycl \in \clsd_\stoptime$,
  where    
  $\emptycl$
  is the representation of contradiction (e.g.\ for resolution and $\resknot$-systems
  the empty clause without literals).

  If every line $\linestd$ in a derivation is used at most once before being
  erased (though it can possibly be rederived later), we say that the
  derivation is 
  \introduceterm{tree-like}.
  This corresponds to changing the inference rule so that
  $\linestd_1, \ldots, \linestd_d$
  must all be erased after they have been used to derive~$\linestd$.
\end{definition}

To every refutation $\proofstd$
\label{page:dag-representation-discussed-here}
we can associate  a DAG~$G_{\proofstd}$,
with the lines in $\proofstd$ labelling the vertices and
with edges from the assumptions  to the  consequence  for each
application of an inference rule.  
There might be several different derivations of a line~$\linestd$ during the
course of the refutation~$\proofstd$, but
if so we can label each occurrence of $\linestd$ with a time-stamp when it
was derived and keep track of which copy of $\linestd$ is used where. Using
this representation, a refutation~$\proofstd$ can be seen to be
tree-like if $G_{\proofstd}$ is a tree.

\begin{definition}[Refutation size, length and space]
  \label{def:length-and-space}
  Given a size measure
  $\sizeofarg{\linestd}$
  for lines~$\linestd$ in $\psstd$\nobreakdash-derivations
  (which we usually think of as the number of symbols in~$\linestd$, but
  other definitions can also be appropriate depending on the context), 
  the
  \introduceterm{size}
  of a $\psstd$-derivation~$\proofstd$ is the
  sum of the sizes of all lines in a derivation, where lines that
  appear multiple times are counted with repetitions (once for every
  vertex in~$G_\proofstd$).
  The \introduceterm{length} of a \mbox{$\psstd$-derivation}~$\proofstd$
  is the number of axiom downloads and inference steps in it,
  \ie the number of vertices 
  in~$G_\proofstd$.%
  \footnote{The reader who so prefers can instead 
    define the length of a derivation
    $\proofstd=\set{\clsd_0,\ldots,\clsd_\stoptime}$ 
    as the number of steps  $\stoptime$ in it, since the difference is
    at most a factor of~$2$. We have chosen the definition above
    for consistency with previous papers defining length as the number
    of lines in a listing of the derivation.}
  For a space measure
  $\clspaceof[\psstd]{\clsd}$
  defined for $\psstd$-configurations, 
  the \introduceterm{space} of a
  derivation~$\proofstd$ is defined as the maximal space 
  of a configuration in~$\proofstd$.
  
  If $\proofstd$ is a refutation of a formula $\fstd$ in 
  size $\sizestd$   and space $\spacestd$,  then we say that $\fstd$ can be
  refuted in size~$\sizestd$ and space $\spacestd$
  \emph{simultaneously}.  Similarly, $\fstd$ can be refuted in
  length~$\lengthstd$ and space $\spacestd$ simultaneously  
  if there is a $\psstd$-refutation $\psstd$ with
  $\lengthofarg{\proofstd} = \lengthstd$
  and 
  \mbox{$\clspaceof{\proofstd} = \spacestd$.}

  We define the \introduceterm{$\psstd$-refutation size} of a formula~$\fstd$,
  denoted
  $\sizeref[\psstd]{\fstd}$,
  to be the minimum size of any $\psstd$-refutation of it. The
  \introduceterm{$\psstd$-refutation length}
  $\lengthref[\psstd]{\fstd}$
  and 
  \introduceterm{$\psstd$-refutation space} 
  $\clspaceref[\psstd]{\fstd}$
  of $\fstd$ are analogously defined
  by taking the minimum \wrt length or space, respectively, over all
  $\psstd$-refutations of $\fstd$.  
\end{definition}

When the proof system in question is clear from context, we will drop
the subindex in the proof complexity measures.

Let us now show how some proof systems that will be of interest to us
can be defined in the framework of
\refdef{def:sequential-refutation}.
We remark that although we will not discuss this in any detail, all
proof systems below are sound and implicationally complete, \ie they
can refute a CNF formula~$\fstd$ \ifaoif $\fstd$ is unsatisfiable.
Below, the notation
\AxiomC{$G_1$}
\AxiomC{$\formuladots$}
\AxiomC{$G_m$}
\TrinaryInfC{$H$}
\DisplayProof
means that if
$G_1, \ldots, G_m$
have been derived previously in the proof (and are currently in
memory), then we can infer~$H$.

\begin{definition}[\kdnf-resolution]
  \label{def:resk}
  The \introduceterm{\kdnf-resolution} proof systems are a
  family of sequential proof systems $\resknot$ parameterized by 
  $k \in \mathbb{N}^+$. Lines in a \kdnf-resolution refutation are
  \kdnfform{}s and we have the following inference rules
  (where
  $G,H$ denote $k$-DNF formulas,
  $\trmt, \trmt'$ denote $k$-terms,
  and $\lita_1, \ldots, \lita_k$ denote literals):
\begin{description}
    \italicitem[$k$-cut]
%
%
    \AxiomC{$(\lita_1\land \formuladots \land
      \lita_{k'})\,\lor\, G$}
    \AxiomC{$\olnot{\lita}_1 \lor\formuladots \lor \olnot{\lita}_{k'} \,\lor\, H$}
    \BinaryInfC{$G\,\lor\, H$}
    \DisplayProof,
      where $k'\leq k$.
      \vspace{4pt}
    \italicitem[$\land$-introduction] 
%
%
      \AxiomC{$G \,\lor\, \trmt$}
      \AxiomC{$G \,\lor\, \trmt'$}
      \BinaryInfC{$G \,\lor\, (\trmt \land \trmt')$}
      \DisplayProof,
      as long as $\setsize{\trmt\union \trmt'} \leq k$.
      \vspace{4pt}
    \italicitem[$\land$-elimination] 
%
%
    \AxiomC{$G\,\lor\, \trmt$}
    \UnaryInfC{$G\,\lor\, \trmt'$}
    \DisplayProof
    for any $\trmt'\subseteq \trmt.$
      \vspace{4pt}
    \italicitem[Weakening] 
%
%
    \AxiomC{$G$}
    \UnaryInfC{$G\,\lor\, H$} 
    \DisplayProof
    for any $k$-DNF formula $H$.

  \end{description}
\end{definition}

\noindent For standard resolution, \ie $\resknot[1]$,
the \mbox{$k$-cut rule}
simplifies to the 
\introduceterm{resolution rule}
\begin{equation}
  \label{eq:resolution-rule} 
  \AxiomC{$\clb \lor \varx$}
  \AxiomC{$\clc \lor \stdnot{\varx}$}
  \BinaryInfC{$\clb \lor \clc$}
  \DisplayProof
\end{equation}
for clauses $\clb$ and~$\clc$.
We refer to~\eqref{eq:resolution-rule} as 
\introduceterm{resolution on the variable}~$\varx$ 
and to $\clb \lor \clc$ as the
\introduceterm{resolvent} of $\clb \lor \varx$ and $\clc \lor
\stdnot{\varx}$ on~$\varx$.
Clearly, in resolution the
$\land$-introduction and $\land$-elimination rules do not apply. In
fact, it can also be shown that the weakening rule never needs to be
used in resolution refutations, but it is convenient to allow it in
order to simplify some technical arguments in proofs.

For $\resknot$-systems, the length measure is as defined in
\refdef{def:length-and-space},
and for space we get the two measures
\introduceterm{formula space}
and
\introduceterm{total space}
depending on whether we consider the number of \kdnfform{}s in a
configuration or all literals in it, counted with repetitions. For
standard resolution there are two more space-related measures that
will be relevant, namely
\introduceterm{width}
and
\introduceterm{variable space}. 
For clarity, let us give an explicit definition of all space-related
measures for resolution that will be of interest.

\begin{definition}[Width and space in resolution]
  \label{def:width-space}
  The
  \introduceterm{width}
  $\widthofsmall{\clc}$
  of a clause $\clc$ is the number of literals in it, and the width of
  a CNF formula or clause configuration is the size of a widest clause
  in it.
  The
  \introduceterm{clause space}
  (as the formula space measure is known in resolution)
  $\clspaceof{\clsc}$
  of a clause configuration $\clsc$
  is
  $\setsize{\clsc}$, \ie the number of clauses in~$\clsc$,
  the
  \introduceterm{variable space}%
  \footnote{%
    We remark that there is some terminological confusion in the
    literature here. The term ``variable space'' has also been used in
    previous papers (including by the current author) to refer to
    what is here called ``total space.''
    The terminology adopted in this paper is due to Alex Hertel and Alasdair Urquhart
    (see~\cite{AlexHertel08Thesis}),
    and we feel  that although their naming convention is as of yet
    less well-established, it feels much more natural than other alternatives.}
  $\varspaceof{\clsc}$ 
  is
  $\setsize{\vars{\clc}}$,
  \ie the number of distinct variables mentioned in~$\clsc$,
  and the
  \introduceterm{total space}
  $\totspaceof{\clsc}$ 
  is
  $\sum_{\clc \in \clsc} \setsize{\clc}$,
  \ie the total number of literals in
  $\clsc$
  counted with repetitions.

  The width or space of a 
  \ifthenelse{\boolean{maybeLMCS}}
  {resolution proof~$\proofstd$}
  {resolution refutation~$\proofstd$} 
  is the maximum that the corresponding measures attains over any
  configuration $\clsc \in \proofstd$, and taking the minimum
  over all refutations of a
  formula~$\fstd$, 
  \ifthenelse{\boolean{maybeLMCS}}
  {we define
    $\mbox{$\widthref{\fstd}$} =
    \minofexpr[\refof{\proofstd}{\fstd}]{\widthofsmall{\proofstd}}$,
    $\mbox{$\clspaceref{\fstd}$} =
    \minofexpr[\refof{\proofstd}{\fstd}]{\clspaceof{\proofstd}}$,
    $\mbox{$\varspaceref{\fstd}$} =
    \minofexpr[\refof{\proofstd}{\fstd}]{\varspaceof{\proofstd}}$,
    and
    $\mbox{$\totspaceref{\fstd}$} =
    \minofexpr[\refof{\proofstd}{\fstd}]{\totspaceof{\proofstd}}$
    as the width, clause space, variable space and total space, 
    respectively, of refuting~$\fstd$ in resolution.}
  {we can define the width
    $\mbox{$\widthref{\fstd}$} =
    \minofexpr[\refof{\proofstd}{\fstd}]{\widthofsmall{\proofstd}}$,
    clause space
    $\mbox{$\clspaceref{\fstd}$} =
    \minofexpr[\refof{\proofstd}{\fstd}]{\clspaceof{\proofstd}}$,
    variable space
    $\mbox{$\varspaceref{\fstd}$} =
    \minofexpr[\refof{\proofstd}{\fstd}]{\varspaceof{\proofstd}}$,
    and total space
    $\mbox{$\totspaceref{\fstd}$} =
    \minofexpr[\refof{\proofstd}{\fstd}]{\totspaceof{\proofstd}}$
    of refuting~$\fstd$ in resolution.}
\end{definition}

Restricting the refutations to tree-like resolution, we
can define the measures
$\lengthref[\treeresnot]{\fstd}$,
$\clspaceref[\treeresnot]{\fstd}$,
$\varspaceref[\treeresnot]{\fstd}$,
and
$\totspaceref[\treeresnot]{\fstd}$
(note that width in general and tree-like resolution are the same,
so defining tree-like width separately does not make much sense).
However, in this paper we will almost exclusively focus on general,
unrestricted versions of resolution and other proof systems.

\begin{remark}
  When studying and comparing the complexity measures for resolution in
  \refdef{def:width-space}, 
  as was noted in
  \cite{ABRW02SpaceComplexity}
  it is preferable to prove the results for 
  \kcnfform{}s, \ie formulas where all clauses have size upper-bounded by
  some constant. This is especially so  since the width and space
  measures can ``misbehave'' rather artificially for formula families
  of unbounded width (see \cite[Section~5]{Nordstrom09SimplifiedWay}
  for a discussion of this). Since every CNF formula can be rewritten
  as an equivalent formula of bounded width---in fact, as a $3$-CNF
  formula, by using extension variables as described on 
  page~\pageref{page:extended-version-discussed-here}---it therefore
  seems natural to insist that the formulas under study should have
  width bounded by some constant.%
  \footnote{We note that there have also been proposals to deal with
    unbounded-width CNF formulas by defining and studying other
    notions of width-restricted resolution, \eg, in
    \cite{Kullman99Investigating,Kullmann04UpperAndLowerBounds}.
    While this very conveniently eliminates the need to convert wide
    CNF formulas to $3$-CNFs, it leads to other problems, and on
    balance we find the established width measure in 
    \refdef{def:width-space}
    to be more natural and interesting.}
\end{remark}

Let us also give examples of some other propositional proof systems
that have been studied in the literature, and that will be of some
interest later in this survey. The first example is the 
\introduceterm{cutting planes}
proof system, or \introduceterm{CP} for short, which was introduced in
\cite{CCT87ComplexityCP}
based on ideas in
\cite{Chvatal73EdmondPolytopes,Gomory63AlgorithmIntegerSolutions}.
Here, clauses are translated to linear inequalities---for instance, 
$x \lor y \lor \olnot{z}$ gets translated to
$x + y + (1-z) \geq 1$, 
\ie
$x + y -z \geq 0$---and these linear inequalities are then manipulated
to derive a contradiction.

\begin{definition}[Cutting planes (CP)]
  \label{def:CP}
  Lines in a cutting planes proof are linear inequalities with integer
  coefficients. 
  The derivation rules are as follows:
  \begin{enumerate}[\hbox to8 pt{\hfill}]
    \vspace{2pt}
    \italicitem\noindent{\hskip-12 pt\bf\textit{Variable axioms:}}\  
    \AxiomC{\rule{0pt}{8pt}}
    \UnaryInfC{$x \geq 0$}
    \DisplayProof
    \t and 
    \AxiomC{\rule{0pt}{8pt}}
    \UnaryInfC{$-x \geq -1$}  
    \DisplayProof
    for all variables $x$.
    \vspace{4pt}
    \italicitem\noindent{\hskip-12 pt\bf\textit{Addition:}}\
    \AxiomC{$\sum a_i x_i \geq A$}
    \AxiomC{$\sum b_i x_i \geq B$}
    \BinaryInfC{$\sum (a_i +  b_i) x_i \geq A + B$}
    \DisplayProof
    \vspace{4pt}
    \italicitem\noindent{\hskip-12 pt\bf\textit{Multiplication:}}\
    \AxiomC{$\sum a_i x_i \geq A$}
    \UnaryInfC{$\sum c a_i x_i \geq  c A$}
    \DisplayProof
    for a positive integer~$c$.
    \vspace{4pt}
    \italicitem\noindent{\hskip-12 pt\bf\textit{Division:}}\
    \AxiomC{$\sum c a_i x_i \geq A$}
    \UnaryInfC{$\sum a_i x_i \geq \ceiling{A/c}$}
    \DisplayProof
    for a positive integer~$c$.
  \end{enumerate}
  A CP refutation ends when the inequality
  $0 \geq 1$
  has been derived.
\end{definition}

As shown in~\cite{CCT87ComplexityCP},
cutting planes is exponentially stronger than resolution \wrt length,
since a CP refutation can mimic any resolution refutation line by line
and furthermore CP can easily handle the pigeonhole principle which is
intractable for resolution. Exponential lower bounds on proof length
for cutting planes were first proven in~%
\cite{BPR95LowerBoundsCP}
for the restricted subsystem $\textrm{CP}^*$ where all coefficients in
the linear inequalities can be at most polynomial in the formula size, 
and were later extended to general CP in~%
\cite{Pudlak97}.
To the best of our knowledge, it is open whether CP is in fact
strictly stronger than $\textrm{CP}^*$ or not. We are not aware of any
papers studying CP space other than the early work by William Cook~%
\cite{Cook90CuttingPlaneProofs},
but this work uses the space concept in
\cite{Kozen77Lower}
that is not the right measure in this context, as was argued very
briefly in  
\refsubsec{sec:pc-intro}.
(It can be noted, though, that the study of space in cutting planes
was mentioned as an interesting open problem in~%
\cite{ABRW02SpaceComplexity}.)

The $\resknot$-systems are 
\introduceterm{logic-based} 
proof systems in the sense that they
manipulate logic formulas, and cutting planes is an example of a
\introduceterm{geometry-based}
proof systems where clauses are treated as geometric objects. Another
class of proof systems is
\introduceterm{algebraic}
systems. One such proof system is
\introduceterm{polynomial calculus (PC)},
which was introduced in~%
\cite{CEI96Groebner}
under the name of ``Gr{\"o}bner proof system.''
In a PC refutation, clauses are interpreted as multilinear polynomials.
For instance, the requirement that the clause 
$x \lor y \lor \olnot{z}$ 
should be satisfied
gets translated to the equation
$(1-x)(1-y)z = 0$
or
$xyz - xz - yz + z = 0$,
and we derive contradiction by showing that there is no common root
for the polynomial equations corresponding to all the clauses.%
\footnote{%
  In fact, from a mathematical point of view it seems more natural to
  think of $0$ as true and $1$ as false in polynomial calculus, so
  that the unit clause $\varx$ gets translated to  $\varx=0$. For
  simplicity and consistency in this survey, however, we stick to
  thinking about $\varx=1$ as meaning that $\varx$ is true and
  $\varx=0$ as meaning that $\varx$ is false.}

\begin{definition}[Polynomial calculus (PC)]
  \label{def:PC}
  Lines in a polynomial calculus proof are multivariate polynomial equations
  $p = 0$,
  where
  $p \in \fieldstd[x,y,z,\ldots]$
  for some (fixed) field~$\fieldstd$.
  It is customary to omit ``$=0$'' and only write $p$.
  The derivation rules are as follows,
  where $\alpha, \beta \in \fieldstd$,
  $p, q \in \fieldstd[x,y,z,\ldots]$,
  and $x$ is any variable:
  \begin{enumerate}[\hbox to8 pt{\hfill}]
    \vspace{2pt}
    \italicitem\noindent{\hskip-12 pt\bf\textit{Variable axioms:}}\
    \AxiomC{\rule{0pt}{8pt}}
    \UnaryInfC{$\ x^2 - x$}
    \DisplayProof
    for all variables~$x$ (forcing 0/1-solutions).
    \vspace{4pt}
    \italicitem\italicitem\noindent{\hskip-12 pt\bf\textit{Linear combination:}}\
    \AxiomC{$p$}
    \AxiomC{$q$}
    \BinaryInfC{$\alpha p + \beta q$}
    \DisplayProof
    \vspace{4pt}
    \italicitem\italicitem\noindent{\hskip-12 pt\bf\textit{Multiplication:}}\
    \AxiomC{$p$}
    \UnaryInfC{$xp$}
    \DisplayProof
  \end{enumerate}
  A PC refutation ends when $1$ has been derived (\ie $1=0$). 
  The \introduceterm{size} of a PC refutation is defined as the total
  number of monomials in the refutation
  (counted with repetitions),
  the \introduceterm{length} of a refutation is the number 
  of polynomial equations, 
  and the \introduceterm{(monomial) space} 
  is the
  maximal number of monomials in any configuration (counted with repetitions).
  Another important measure is the
  \introduceterm{degree} of a refutation, which is the maximal (total) degree
  of any monomial
  (where we note that because of the variable axioms, all polynomials
  can be assumed to be multilinear without loss of generality).
\end{definition}

The minimal refutation degree for a 
$k$-CNF formula $\fstd$ is closely
related to the minimal refutation size. 
Impagliazzo \etal~\cite{IPS99LowerBounds}
showed that every PC proof of size
$\sizestd$
can be transformed into another PC proof of degree
$\Bigoh{\sqrt{n \log \sizestd}}$.
A number of strong lower bounds on proof size have been obtained by proving
degree lower bounds in, \eg,
\cite{AR03LowerBounds,
  BI10Random,
  BGIP01LinearGaps,
  IPS99LowerBounds,
  Razborov98LowerBound}.

The representation of a clause
$\Lor_{i=1}^{n} x_i$
as a PC polynomial is
$\prod_{i=1}^{n} (1- x_i)$,
which means that the number of monomials is exponential in the clause
width. This problem arises only for positive literals, however---a
large clause with only negative literals is translated to a single monomial.
This is a weakness of monomial space in polynomial calculus
when compared to clause space in resolution.
To get a cleaner and more symmetric treatment of proof space, 
in~\cite{ABRW02SpaceComplexity} the proof system
\introduceterm{polynomial calculus (with) resolution},
or \introduceterm{PCR} for short, 
was introduced as a common extension of polynomial calculus and
resolution. 
The idea is to add an extra set of parallell formal variables
$x', y', z', \ldots$ 
so that positive and negative literals can both be represented in a
space-efficient fashion.

\begin{definition}[Polynomial calculus resolution (PCR)]
  \label{def:PCR}
  Lines in a PCR proof are polynomials over the ring
  $\fieldstd[x,x',y,y',z,z',\ldots]$,
  where as before $\fieldstd$ is some field.
  We have all the axioms and rules of PC plus the following axioms:
  \begin{enumerate}[\hbox to8 pt{\hfill}]
    \italicitem\italicitem\noindent{\hskip-12 pt\bf\textit{Complementarity:}}\
    \AxiomC{\rule{0pt}{8pt}}
    \UnaryInfC{$x + x' - 1$}
    \DisplayProof
    for all pairs of variables $(x, x')$.
  \end{enumerate}
  Size, length, and degree are defined as for polynomial calculus, and
  the   (monomial) space
  of a PCR refutation is again the maximal number of monomials in any
  configuration counted with repetitions.%
  \footnote{%
    We remark that in 
    \cite{ABRW02SpaceComplexity}
    monomial space was defined to be the  number of \emph{distinct}
    monomials in a configuration (\ie not counted with
    repetitions), but we find this restriction to be somewhat artificial.}
\end{definition}

The point of the complementarity rule is to force
$x$ and~$x'$ to have opposite values in~$\set{0,1}$,
so that they encode complementary literals.
One gets the same degree bounds for PCR as in PC (just substitute
$1-x$ for $x'$), but one can potentially avoid an exponential blow-up
in size measured in the number of monomials (and thus also for space).
Our running example clause
$x \lor y \lor \olnot{z}$ 
is rendered as
$x' y' z$ 
in PCR.
In PCR, monomial space is a natural generalization of clause space
since every clause translates into a monomial as just explained.
%
%

It was observed in
\cite{ABRW02SpaceComplexity}
that the tight relation between degree and size in PC carries over to PCR.
In a recent paper~\cite{GL10Optimality}, 
Galesi and Lauria
showed that this trade-off is essentially optimal, and also studied a
generalization of PCR that unifies polynomial calculus and \kdnf resolution.

In general, the admissible inferences in a proof system according to 
\refdef{def:sequential-refutation}
are defined by a set of syntactic inference rules. In what follows, we
will also be interested in a strengthened version of this concept,
which was made explicit in~\cite{ABRW02SpaceComplexity}.

\begin{definition}[Syntactic and semantic derivations]
  \label{def:semantic-syntactic-derivations}
  We refer to derivations according to 
  \refdef{def:sequential-refutation},
  where each new line $\linestd$  has to be inferred by one
  of the inference rules for~$\psstd$,
  as \introduceterm{syntactic} derivations.
  If instead \introduceterm{any line} $\linestd$ that is semantically implied
  by the current configuration can be derived in one atomic step, we
  talk about a \introduceterm{semantic}%
  \footnote{It should be noted here that the term
    \introduceterm{semantic resolution} 
    is also used in the literature to refer to  something very
    different, namely a restricted subsystem of (syntactic)
    resolution. In this paper, however, semantic proofs will always be
    proofs in the sense of
    \refdef{def:semantic-syntactic-derivations}.}
  derivation.
\end{definition}

Clearly, semantic derivations are at least as strong as syntactic
ones, and they are easily seen to be superpolynomially stronger \wrt
length for any proof system where superpolynomial lower bounds are
known. This is so since a semantic proof system can download all
axioms in the formula one by one, and then deduce contradiction in one
step since the formula is unsatisfiable. Therefore, semantic versions of
proof systems are mainly interesting when we want to reason about
space or the relationship between space and length.

This concludes our presentation of proof systems, and we next turn to
the connection between proof complexity and pebble games.


\ifthenelse{\boolean{maybeLMCS}}
{\subsubsectionNOW{Pebble Games and Pebbling Contradictions}
  \label{sec:pc-pebcontr}
  Pebbling is a tool for studying time-space relationships by means of a
game played on directed acyclic graphs.  This game models computations
where the execution is independent of the input and can be performed
by straight-line programs.  Each such program is encoded as a graph,
and a pebble on a vertex in the graph indicates that the corresponding
value is currently kept in memory. The goal is to pebble the output
vertex of the graph with minimal number of pebbles (amount of
memory) and steps (amount of time).

Pebble games were originally devised for studying programming
languages and  compiler construction, but have later found a broad range of
applications in computational complexity theory. 
The pebble game model seems to have appeared for the first time
(implicitly) in~%
\cite{PH70Comparative},
where it was used to study flowcharts and recursive schemata, and it
was later employed  to model register allocation~%
\cite{S75CompleteRegisterAllocation},
and analyze the relative power of time and space as Turing-machine
resources~%
\cite{C74ObservationTimeStorageTradeOff,
  HPV77TimeVsSpace}.
Moreover, pebbling has been used 
to  derive time-space trade-offs for algorithmic concepts such as 
linear recursion~%
\cite{Chandra73Efficient, SS83SpaceTimeLinear},
fast Fourier transform~%
\cite{SS77SpaceTimeFFT, Tompa78TimeSpaceComputing},
matrix multiplication~%
\cite{Tompa78TimeSpaceComputing}, 
and integer multiplication~%
\cite{SS79SpaceTimeOblivious}.
An excellent survey of pebbling up to ca 1980 is~\cite{P80Pebbling},
and
another in-depth treatment of some pebbling-related questions can be
found in chapter~10 of~%
\cite{Savage98Models}.
Some more recent developments are covered in the author's upcoming survey~%
\cite{Nordstrom09PebblingSurveyFTTCS}.

The \introduceterm{pebbling price} of a 
directed acyclic graph 
$G$ in the black pebble game captures the memory space, or number of
registers, required to perform the deterministic computation described
by~$G$.  We will mainly be interested in the the more general 
\introduceterm{black-white pebble game} 
modelling nondeterministic computation,
which was introduced in~\cite{CS76Storage}
and has been studied in
\cite{%
  GT78VariationsPebbleGameVerboseURL,
  Klawe85TightBoundPebblesPyramid,
  LT82AsymptoticallyTightBounds,
  MadH81ComparisonOfTwoVariationsOfPebbleGame,
  KS91OnThePowerOfWhitePebbles,
  Wilber88WhitePebblesHelp}
and other papers.
Let us refer to vertices of a directed graph having indegree~$0$ as
\introduceterm{sources} 
and vertices having outdegree~$0$ as
\introduceterm{sinks}.

\begin{definition}[Pebble game]
  \label{def:bw-pebble-game}
  Let $G$ be a directed acyclic graph (DAG) with a 
  unique sink vertex~$\sinkstd$. The
  \introduceterm{black-white pebble game}
  on~$G$  is the following \mbox{one-player} game.
  At any time~$t$, we have a configuration 
  $\pconf_t = (B_t, W_t)$
  of black pebbles $B_t$ and white pebbles $W_t$ on 
  the vertices of~$G$, 
  at most one pebble per vertex. 
  The rules of the  game are as
  follows:
  \begin{enumerate}[(1)]
    
  \item
    \label{pebrule:black-placement} If all immediate predecessors of an
    empty vertex~$v$ have pebbles on them, a black pebble may be placed
    on~$v$. In particular, a black pebble can always be placed on
    a source vertex.
    
  \item
    \label{pebrule:black-removal} A black pebble may be removed from any
    vertex  at any time.
    
  \item
    \label{pebrule:white-placement} A white pebble may be placed on any
    empty vertex at any time.
    
  \item
    \label{pebrule:white-removal} If all immediate predecessors of a
    white-pebbled vertex~$v$ have pebbles on them, the white pebble on~$v$  
    may be removed. In particular, a white pebble can always be
    removed from a source vertex.
    
  \end{enumerate}
  A \introduceterm{(\pebcomplete{}) black-white pebbling of $G$}, 
  also called a
  \introduceterm{pebbling strategy for~$G$},
  is a sequence of pebble configurations
  $\pebbling = \setsmall{\pconf_0, \ldots, \pconf_{\stoptime}}$ \st
  $\pconf_0 = (\emptyset, \emptyset)$,
  $\pconf_{\stoptime} = (\set{\sinkstd}, \emptyset)$,
  and for all $t \in
  \intnfirst{\stoptime}$,
  $\pconf_{t}$ follows from $\pconf_{t-1}$ by
  one of the rules above.
  The \introduceterm{time} of a pebbling
  $\pebbling = \setsmall{\pconf_0, \ldots, \pconf_{\stoptime}}$
  is simply
  $\pebtime{\pebbling} = \stoptime$
  and the
  \introduceterm{space}
  is
  $\pebspace{\pebbling} =
  \maxofexpr[0 \leq t \leq {\stoptime}]{\setsize{B_t \unionSP W_t}}$.
  The
  \introduceterm{black-white pebbling price}
  (also known as the
  \introduceterm{pebbling measure}
  or
  \introduceterm{pebbling number})
  of~$G$, denoted $\bwpebblingprice{G}$, is the minimum
  space of any \pebcomplete 
  pebbling of~$G$.

  A \introduceterm{black pebbling} is a pebbling using black pebbles
  only, \ie having $W_t = \emptyset$ for all~$t$. The
  \introduceterm{(black) pebbling price}
  of~$G$, denoted
  $\pebblingprice{G}$,
  is the minimum space of any \pebcomplete black pebbling of~$G$.
\end{definition}

In the last decade, there has been renewed interest in pebbling in
the context of proof complexity.
A (non-exhaustive)  list of proof complexity papers 
using pebbling in one way or another is 
\cite{%
  AJPU07ExponentialSeparation,
  BEGJ00RelativeComplexity,
  BIPS10FormulaCaching,
  Ben-Sasson09SizeSpaceTradeoffs,
  BIW00Near-optimalSeparation,
  BN08ShortProofs,
  BN11UnderstandingSpace,
  BW01ShortProofs,
  EGM04Complexity,
  ET03CombinatorialCharacterization,
  HN11Communicating,
  HU07Game,
  Nordstrom09NarrowProofsSICOMP,
  NH13TowardsOptimalSeparation,
  Nordstrom11RelativeStrength,
  SBK03UsingProblemStructure}.
The way pebbling results have been used in proof complexity has mainly
been by studying so-called
\introduceterm{pebbling contradictions}
(also known as
\introduceterm{pebbling formulas}
or
\introduceterm{pebbling tautologies}).
These are \cnfform{}s encoding the pebble game played on a DAG~$G$ by
postulating  the sources to be true and the sink to be false, and
specifying that truth propagates through the graph according to the
pebbling rules. The idea to use such formulas seems to 
have appeared for the first time in
Kozen~\cite{Kozen77Lower},
and they were also studied in
\cite{RM99Separation, BEGJ00RelativeComplexity}
before being defined in full generality by Ben-Sasson and Wigderson in~%
\cite{BW01ShortProofs}.
}
{\subsubsectionNOW{Pebbling Contradictions and Substitution Formulas}
  \label{sec:pc-pebcontr}
  \label{sec:pc-subst-form}
  
  The way pebbling results have been used in proof complexity has mainly
  been by studying so-called
  \introduceterm{pebbling contradictions}
  (also known as
  \introduceterm{pebbling formulas}
  or
  \introduceterm{pebbling tautologies}).
  These are \cnfform{}s encoding the pebble game played on a DAG~$G$ by
  postulating  the sources to be true and the sink to be false, and
  specifying that truth propagates through the graph according to the
  pebbling rules. The idea to use such formulas seems to 
  have appeared for the first time in
  Kozen~\cite{Kozen77Lower},
  and they were also studied in
  \cite{RM99Separation, BEGJ00RelativeComplexity}
  before being defined in full generality by Ben-Sasson and Wigderson in~%
  \cite{BW01ShortProofs}.}


\begin{definition}[Pebbling contradiction]
  \label{def:pebbling-contradiction}
  Suppose that $G$ is
  a DAG with sources~$S$ and a unique sink~$z$.
  Identify every vertex
  $v \in \vertices{G}$
  with a propositional logic variable~$v$. The
  \introduceterm{pebbling contradiction} over~$G$,
  denoted~$\pebcontr{}$,
  is the conjunction of the following clauses:
  \begin{itemize}[(1)]
    
  \item
    for all
    $s \in S$,
    a unit clause
    $s$
    (\introduceterm{source axioms}),
    
  \item
    \ifthenelse{\boolean{maybeLMCS}}
    {For all non-sources $v$ with immediate predecessors}
    {For all non-source vertices $v$ with immediate predecessors}
    $\prednode{v}$,
    the clause
    $\Lor_{u \in \prednode{v}} \olnot{u} \lor v$
    (\introduceterm{pebbling axioms}),
    
  \item
    for the sink $z$,
    the unit clause $\olnot{z}$
    (\introduceterm{target} or \introduceterm{sink axiom}).
    
  \end{itemize}
\end{definition}

\noindent If $G$ has $n$ vertices and maximal indegree $\ell$,
the formula
$\pebcontr{}$
is a minimally unsatisfiable
\xcnfform{\text{($1+$$\ell$)}}{}
with
$n+1$ clauses over $n$~variables.
We will almost exclusively be interested in DAGs with
bounded indegree $\ell = \bigoh{1}$, usually 
$\ell = 2$. We note that DAGs with fan-in $2$ and a single sink
have sometimes been referred to as
\introduceterm{circuits} 
in the proof complexity literature, although we will not use that term
here.
For an example of a pebbling contradiction, see the CNF formula in
\reffig{fig:pebbling-contradiction-for-Pi-2-Peb1}
defined in terms of the graph in
\reffig{fig:pebbling-contradiction-for-Pi-2-graph}.

\begin{figure}[tp]
  \subfigure
  [Pyramid graph $\Pi_2$ of height 2.]	    
  {
    \label{fig:pebbling-contradiction-for-Pi-2-graph}
    \begin{minipage}[b]{0.40\linewidth}
      \centering
      \includegraphics{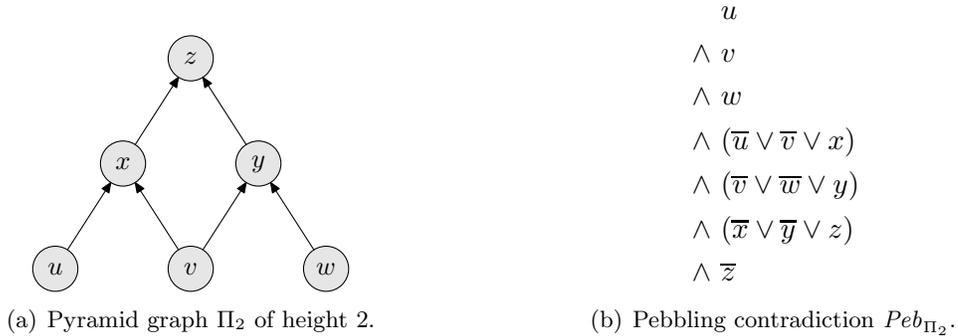}%
    \end{minipage}
  }
  \hfill
  \subfigure
  [Pebbling contradiction {$\pebcontr[\Pi_2]{}$}.]	    
  {     
    \label{fig:pebbling-contradiction-for-Pi-2-Peb1}
    \begin{minipage}[b]{0.55\linewidth}
      \centering
      \begin{gather*}
	\begin{aligned}
	  & 
	  u
	  \\
	  \land \ 
	  &v
	  \\
	  \land \ 
	  &w
	  \\
	  \land \ 
	  &(\olnot{u} \lor \olnot{v} \lor x)
	  \\
	  \land \ 
	  &(\olnot{v} \lor \olnot{w} \lor y)
	  \\
	  \land \ 
	  &(\olnot{x} \lor \olnot{y} \lor z)
	  \\
	  \land \ 
	  &\olnot{z}
	\end{aligned}
      \end{gather*}
    \end{minipage}
  }
  \caption{Pebbling contradiction for the pyramid graph $\Pi_2$.}
  \label{fig:pebbling-contradiction-for-Pi-2}
\end{figure}

\ifthenelse{\boolean{maybeLMCS}}
{\subsubsectionNOW{Substitution Formulas}
  \label{sec:pc-subst-form}}
{}

In many of the cases we will be interested in below, 
the formulas in
\refdef{def:pebbling-contradiction}
are not quite sufficient for our purposes since they are a bit too
easy to refute. We therefore want to make them (moderately)
harder, and it turns out that a good way of achieving this is to
substitute some suitable Boolean function
$\funcpebc(\varx_1, \ldots, \varx_\pebdeg)$
for each variable $\varx$ and expand to get a new CNF formula.

It will be useful to formalize this concept of substitution for any
CNF formula~$\fstd$ and any Boolean function~$\funcpebc$.
To this end, let
$\funcpebc[\pebdeg]$
denote any (non-constant) Boolean function
\mbox{$\funcdescr{\funcpebc[\pebdeg]}{\zerooneset^{\pebdeg}}{\zerooneset}$}
of arity~$\pebdeg$.
We use the shorthand
$\varvec{\varx} = (\varx_1, \ldots, \varx_{\pebdeg})$,
so that
$\funcwithvecarg[\pebdeg]{\varx}$
is just an equivalent way of writing
$\funcwithargspebcontr[\pebdeg]{\varx}$.
Every function $\funcwithargspebcontr[\pebdeg]{\varx}$ is equivalent
to a \cnfform over $\varx_1, \ldots, \varx_{\pebdeg}$ with at most
$2^\pebdeg$ clauses. 
Fix some canonical set of clauses $\clfuncstd{x}$
representing
$\funcwithvecarg[\pebdeg]{\varx}$
and let $\clnotfuncstd{x}$ denote the clauses in some chosen canonical
representation of the negation of $\funcpebc[\pebdeg]$
applied on $\varvec{\varx}$.

This canonical representation can be given by a formal
definition (in terms of min- and maxterms), but we do not want to get
too formal here and instead try to convey the intuition by providing a
few examples. For instance, we have
\begin{equation}
  \clfunc{\lor}{x}{2}  = \set{x_1 \lor x_2}
  \quad  \text{and}    \quad
  \clnotfunc{\lor}{x}{2}  = \set{\olnot{x}_1, \, \olnot{x}_2}
\end{equation}
for logical or of two variables and
\begin{equation}
  \clfunc{\xor}{x}{2}
  = \set{x_1 \lor x_2, \, \olnot{x}_1 \lor \olnot{x}_2}
  \quad  \text{and}    \quad
  \clnotfunc{\xor}{x}{2}
  =  \set{x_1 \lor \olnot{x}_2, \, \olnot{x}_1 \lor {x}_2}
\end{equation}
for exclusive or of two variables.
If we let
$\ktruestd[\pebdeg]$
denote the threshold function
saying that $\ntrues$ out of $\pebdeg$ variables are true, then for
$\ktrue[4]{2}$
we have
\begin{equation}
  \clfunc{\ktrue[4]{2}}{x}{} =
  \SET{%
    \begin{aligned}[l]
      x_1 \lor x_2 \lor x_3, \\
      x_1 \lor x_2 \lor x_4, \\
      x_1 \lor x_3 \lor x_4, \\
      x_2 \lor x_3 \lor x_4
    \end{aligned}%
  }
  \quad  \text{and}    \quad
  \clnotfunc{\ktrue[4]{2}}{x}{} =
  \SET{%
    \begin{aligned}[l]
      \olnot{x}_{1} \lor \olnot{x}_{2}, \\
      \olnot{x}_{1} \lor \olnot{x}_{3}, \\
      \olnot{x}_{1} \lor \olnot{x}_{4}, \\
      \olnot{x}_{2} \lor \olnot{x}_{3}, \\
      \olnot{x}_{2} \lor \olnot{x}_{4}, \\
      \olnot{x}_{3} \lor \olnot{x}_{4}
    \end{aligned}%
  }
  \eqperiod
\end{equation}

%
%
%
%

\noindent We want to define formally what it means to substitute
$\funcpebc[\pebdeg]$
for the variables $\vars{\fstd}$
in a \cnfform~$\fstd$.
For notational convenience, we
assume that
$\fstd$
only has variables
$\varx, \vary, \varz$, et cetera,
without subscripts, so that
$\varx_1, \ldots, \varx_{\pebdeg},$
$\vary_1, \ldots, \vary_{\pebdeg},$
$\varz_1, \ldots, \varz_{\pebdeg}, \ldots
$
are new variables not occurring in~$\fstd$.

\begin{definition}[\Substformtext{}] 
  \label{def:substitution-formula}
  For a positive literal $\varx$
  and a non-constant Boolean function
  $\funcpebc[\pebdeg]$,
  we define   the
  \introduceterm{\fsubsttext{}}
  of $\varx$ to be
  $\substform{\varx}{\funcpebc[\pebdeg]}
  =
  \clfuncstd{\varx}$, \ie the canonical representation of
  $\funcwithargspebcontr[\pebdeg]{\varx}$
  as a \cnfform.
  For a negative literal $\lnot y$,
  the \fsubsttext is
  $\substform{\lnot y}{\funcpebc[\pebdeg]}
  =
  \clnotfuncstd{\vary}$.
  The \fsubsttext of a clause
  $\clc = \lita_1 \lor \formuladots \lor \lita_{\clwidth}$
  is the \cnfform
  \begin{equation}
    \label{eq:f-substitution-for-clause}
    \substform{\clc}{\funcpebc[\pebdeg]}
    =
    \Land_{\clc_{1} \in \substform{\lita_{1}}{\funcpebc[\pebdeg]}}
    \ldots
    \Land_{\clc_{\clwidth} \in \substform{\lita_{\clwidth}}{\funcpebc[\pebdeg]}}
    \bigl( \clc_1 \lor \ldots \lor \clc_{\clwidth} \bigr)
\end{equation}
and the \fsubsttext of a \cnfform $\fstd$ is
$
\substform{\fstd}{\funcpebc[\pebdeg]}
=
\Land_{\clc \in \fstd} \substform{\clc}{\funcpebc[\pebdeg]}
$.
\end{definition}

For example, for the clause
$\clc = \varx \lor \olnot{\vary}$
and the exclusive or function
$\funcpebc[2] = \varx_1 \xor \varx_2$
we have
\begin{equation}
\label{eq:example-subst-clause}
\begin{split}
\substform{\clc}{\funcpebc[2]}
= \ \ \
&(\varx_{1} \lor \varx_{2} \lor \vary_{1} \lor \olnot{\vary}_{2})
\land
(\varx_{1} \lor \varx_{2} \lor \olnot{\vary}_{1} \lor \vary_{2})
\\
\land \
&(\olnot{\varx}_{1} \lor \olnot{\varx}_{2} \lor   \vary_{1} \lor \olnot{\vary}_{2})
\land
(\olnot{\varx}_{1} \lor \olnot{\varx}_{2} \lor \olnot{\vary}_{1} \lor \vary_{2})
\eqperiod
\end{split}
\end{equation}
Note that
$\substform{\fstd}{\funcpebc[\pebdeg]}$
is a \cnfform over
$\pebdeg \cdot \setsize{\vars{\fstd}}$ variables
containing strictly less than
$\nclausesof{\fstd} \cdot 2^{\pebdeg \cdot \widthofarg{\fstd}}$
clauses.
(Recall that we defined a \cnfform as a set of clauses,
which means that
$\nclausesof{\fstd}$
is  the number of clauses in~$\fstd$.)
It is easy to verify that
$\substform{\fstd}{\funcpebc[\pebdeg]}$
is unsatisfiable \ifaoif
$\fstd$ is unsatisfiable.

Two examples of substituted version of the pebbling formula in
\reffig{fig:pebbling-contradiction-for-Pi-2-Peb1}
are the substitution with logical or in
\reffig{fig:pebbling-contradiction-for-Pi-2-binary-or}
and with exclusive or in
\reffig{fig:pebbling-contradiction-for-Pi-2-exclusive-or}.    
As we shall see, these formulas have played an important role in the
line of research trying to understand proof space in resolution.
For our present purposes, there is an important difference between
logical or and exclusive or which is captured by the next definition.

\begin{figure}[tp]
  \subfigure
  [Substitution pebbling contradiction 
  {$\pebcontrwithfunc[{\Pi_2}]{}{\lor_2}$} 
  \wrt binary logical or.]
  {   
    \label{fig:pebbling-contradiction-for-Pi-2-binary-or}
    \label{fig:subst-peb-forms-for-Pi-2-binary-or}
    \begin{minipage}[b]{0.95\linewidth}
      \centering
      \begin{gather*}      
	\begin{aligned}
	  & 
	  ( u_1 \lor u_2 )
	  &     {\land \ }
	  &
	  ( \olnot{v}_2 \lor \olnot{w}_1 \lor y_1 \lor y_2 )  
	  \\
	  \land \ 
	  &
	  ( v_1 \lor v_2 ) 
	  &\land \ 
	  &
	  ( \olnot{v}_2 \lor \olnot{w}_2 \lor y_1 \lor y_2 )  
	  \\
	  \land \ 
	  &
	  ( w_1 \lor w_2 ) 
	  &\land \ 
	  &
	  ( \olnot{x}_1 \lor \olnot{y}_1 \lor z_1 \lor z_2 )  
	  \\
	  \land \ 
	  &
	  ( \olnot{u}_1 \lor \olnot{v}_1 \lor x_1 \lor x_2 )  
	  \ \ \ \ \ \ \ \ \ \  
	  &\land \ 
	  &
	  ( \olnot{x}_1 \lor \olnot{y}_2 \lor z_1 \lor z_2 )  
	  \\
	  \land \ 
	  &
	  ( \olnot{u}_1 \lor \olnot{v}_2 \lor x_1 \lor x_2 )  
	  &\land \ 
	  &
	  ( \olnot{x}_2 \lor \olnot{y}_1 \lor z_1 \lor z_2 )  
	  \\
	  \land \ 
	  &
	  ( \olnot{u}_2 \lor \olnot{v}_1 \lor x_1 \lor x_2 )  
	  &\land \ 
	  &
	  ( \olnot{x}_2 \lor \olnot{y}_2 \lor z_1 \lor z_2 )  
	  \\
	  \land \ 
	  &
	  ( \olnot{u}_2 \lor \olnot{v}_2 \lor x_1 \lor x_2 )  
	  &\land \ 
	  &
	  \olnot{z}_1
	  \\
	  \land \ 
	  &
	  ( \olnot{v}_1 \lor \olnot{w}_1 \lor y_1 \lor y_2 )  
	  &\land \ 
	  &
	  \olnot{z}_2
	  \\
	  \land \ 
	  &
	  ( \olnot{v}_1 \lor \olnot{w}_2 \lor y_1 \lor y_2 )  
	\end{aligned}
      \end{gather*}
    \end{minipage}
  }
  
  \subfigure
      [Substitution pebbling contradiction {$\pebcontrwithfunc[{\Pi_2}]{}{\xor_2}$} 
    \wrt binary exclusive or.]
  {   
    \label{fig:pebbling-contradiction-for-Pi-2-exclusive-or}    
    \label{fig:subst-peb-forms-for-Pi-2-exclusive-or}
    \begin{minipage}[b]{0.95\linewidth}
      \centering
      \begin{gather*}      
        \begin{aligned}
          & 
          ( u_1 \lor u_2 )
          &     \land \ 
          &
          ( v_1 \lor \olnot{v}_2 \lor \olnot{w}_1 \lor {w}_2 \lor y_1 \lor y_2 )  
          \\
          \land \ 
          &
          ( \olnot{u}_1 \lor \olnot{u}_2 ) 
          &\land \ 
          &
          ( v_1 \lor \olnot{v}_2 \lor \olnot{w}_1 \lor {w}_2 \lor \olnot{y}_1 \lor \olnot{y}_2 )  
          \\
          \land \ 
          &
          ( v_1 \lor v_2 ) 
          &\land \ 
          &
          ( \olnot{v}_1 \lor {v}_2 \lor w_1 \lor \olnot{w}_2 \lor y_1 \lor y_2 ) 
          \\
          \land \ 
          &
          ( \olnot{v}_1 \lor \olnot{v}_2 )  
          &\land \ 
          &
          ( \olnot{v}_1 \lor {v}_2 \lor w_1 \lor \olnot{w}_2 \lor \olnot{y}_1 \lor \olnot{y}_2 )
          \\
          \land \ 
          &
          ( w_1 \lor w_2 ) 
          &\land \ 
          &
          ( \olnot{v}_1 \lor {v}_2 \lor \olnot{w}_1 \lor {w}_2 \lor y_1 \lor y_2 ) 
          \\
          \land \ 
          &
          ( \olnot{w}_1 \lor \olnot{w}_2 )  
          &\land \ 
          &
          ( \olnot{v}_1 \lor {v}_2 \lor \olnot{w}_1 \lor {w}_2 \lor \olnot{y}_1 \lor \olnot{y}_2 )  
          \\
          \land \ 
          &
          ( u_1 \lor \olnot{u}_2 \lor v_1 \lor \olnot{v}_2 \lor x_1 \lor x_2 )  
          \ \ \ \ \ \ \ \ \ \  
          &\land \ 
          &
          ( x_1 \lor \olnot{x}_2 \lor y_1 \lor \olnot{y}_2 \lor z_1 \lor z_2 )  
          \\
          \land \ 
          &
          ( u_1 \lor \olnot{u}_2 \lor v_1 \lor \olnot{v}_2 \lor \olnot{x}_1 \lor \olnot{x}_2 )  
          &\land \ 
          &
          ( x_1 \lor \olnot{x}_2 \lor y_1 \lor \olnot{y}_2 \lor \olnot{z}_1 \lor \olnot{z}_2 )  
          \\
          \land \ 
          &
          ( u_1 \lor \olnot{u}_2 \lor \olnot{v}_1 \lor {v}_2 \lor x_1 \lor x_2 )  
          &\land \ 
          &
          ( x_1 \lor \olnot{x}_2 \lor \olnot{y}_1 \lor {y}_2 \lor z_1 \lor z_2 )  
          \\
          \land \ 
          &
          ( u_1 \lor \olnot{u}_2 \lor \olnot{v}_1 \lor {v}_2 \lor \olnot{x}_1 \lor \olnot{x}_2 )  
          &\land \ 
          &
          ( x_1 \lor \olnot{x}_2 \lor \olnot{y}_1 \lor {y}_2 \lor \olnot{z}_1 \lor \olnot{z}_2 )  
          \\
          \land \ 
          &
          ( \olnot{u}_1 \lor {u}_2 \lor v_1 \lor \olnot{v}_2 \lor x_1 \lor x_2 )  
          &\land \ 
          &
          ( \olnot{x}_1 \lor {x}_2 \lor y_1 \lor \olnot{y}_2 \lor z_1 \lor z_2 )  
          \\
          \land \ 
          &
          ( \olnot{u}_1 \lor {u}_2 \lor v_1 \lor \olnot{v}_2 \lor \olnot{x}_1 \lor \olnot{x}_2 )  
          &\land \ 
          &
          ( \olnot{x}_1 \lor {x}_2 \lor y_1 \lor \olnot{y}_2 \lor \olnot{z}_1 \lor \olnot{z}_2 )  
          \\
          \land \ 
          &
          ( \olnot{u}_1 \lor {u}_2 \lor \olnot{v}_1 \lor {v}_2 \lor x_1 \lor x_2 )  
          &\land \ 
          &
          ( \olnot{x}_1 \lor {x}_2 \lor \olnot{y}_1 \lor {y}_2 \lor z_1 \lor z_2 )  
          \\
          \land \ 
          &
          ( \olnot{u}_1 \lor {u}_2 \lor \olnot{v}_1 \lor {v}_2 \lor \olnot{x}_1 \lor \olnot{x}_2 )  
          &\land \ 
          &
          ( \olnot{x}_1 \lor {x}_2 \lor \olnot{y}_1 \lor {y}_2 \lor \olnot{z}_1 \lor \olnot{z}_2 )  
          \\
          \land \ 
          &
          ( v_1 \lor \olnot{v}_2 \lor w_1 \lor \olnot{w}_2 \lor y_1 \lor y_2 )  
          &\land \ 
          &
          ({z}_1 \lor \olnot{z}_2)
          \\
          \land \ 
          &
          ( v_1 \lor \olnot{v}_2 \lor w_1 \lor \olnot{w}_2 \lor \olnot{y}_1 \lor \olnot{y}_2 )  
          &\land \ 
          &
          (\olnot{z}_1 \lor {z}_2)
        \end{aligned}
      \end{gather*}
    \end{minipage}
  }
  \caption{Examples of substitution pebbling formulas 
    for the pyramid graph $\Pi_2$.}
  \label{fig:subst-peb-forms-for-Pi-2}
\end{figure}

\begin{definition}[\Nonauthoritarian function \cite{BN11UnderstandingSpace}]
  \label{def:non-authoritarian-func}
  We say that a Boolean function 
  $\funcwithargspebcontr[]{\varx}$
  is
  \introduceterm{\mbox{$k$-non}-authoritarian}%
  \footnote{Such functions have previously also been referred to as
    \introduceterm{$(k+$$1)$-robust} functions in~%
    \cite{ABRW04Pseudorandom}.}
  if no restriction to $\set{x_1,\ldots, x_d}$ of size $k$ can fix the
  value of $f$. In other words, for every restriction $\restr$ to
  $\set{x_1,\ldots, x_d}$ with $\setsize{\restr}\leq k$ 
  there exist two assignments $\tvastd_0, \tvastd_1\supset \restr$ such
  that $f(\tvastd_0)=0$ and $f(\tvastd_1)=1$. 
  If this does not hold, 
  $f$ is \introduceterm{$k$-authoritarian}.
  A   \mbox{$1$-(non-)}authoritarian function is called just
  \introduceterm{(non-)authoritarian}.
\end{definition}

Observe that a function on $d$ variables can be $k$-non-authoritarian
only if $k<d$. Two natural examples of 
$d$-non-authoritarian functions are exclusive or 
$\xor$ of $d+1$ variables and majority of  $2d+1$ variables,
\mbox{\ie
$\ktrue[2 \pebdeg + 1]{\pebdeg+1}$.}
Non-exclusive or of any arity is easily seen to be an authoritarian
function, however, since setting any variable $x_i$ to true forces the
whole disjunction to true.

Concluding our presentation of preliminaries, 
we remark that the idea of  combining
\refdef{def:pebbling-contradiction}
with
\refdef{def:substitution-formula}
was not a dramatic new insight originating with~%
\cite{BN11UnderstandingSpace},
but rather the natural generalization of ideas in many previous
articles. For instance, the papers
\cite{%
  BIW00Near-optimalSeparation,
  Ben-Sasson09SizeSpaceTradeoffs,
  BIPS10FormulaCaching,
  BW01ShortProofs,
  BOP07Complexity,
  ET03CombinatorialCharacterization,
  Nordstrom09NarrowProofsSICOMP,
  NH13TowardsOptimalSeparation}
all study formulas
$\pebcontrwithfunc[G]{}{\lor_2}$, 
and
\cite{EGM04Complexity}
considers formulas
$\pebcontrwithfunc[G]{}{\land_{l} \lor_{k}}$. 
And in fact, already back in 2006 Atserias~%
\cite{Atserias06personalcommunication}
proposed that \xorpebcontrtext{}s
$\pebcontrwithfunc[G]{}{\xor_2}$ 
could potentially be used to separate length and space in resolution,
as was later shown to be the case in~\cite{BN08ShortProofs}.

\subsectionNOW{Overview of Pebbling Contradictions in Proof Complexity}
\label{sec:pc-pebcontrs-and-proof-cplx}

Let us now give a general overview of how pebbling contradictions have been
used in proof complexity. While we have striven to give a reasonably
full picture below, we should add the caveat that our main focus is on
resolution-based proof systems, \ie standard resolution and
$\resknot[k]$ for $k>1$. Also, to avoid confusion it should be
\ifthenelse{\boolean{maybeLMCS}}{pointed out}{pointed out (again)}
that the pebble games examined here should not be mixed up with
the very different 
\introduceterm{existential pebble games} 
which have also proven to be a useful tool in proof complexity
in, \eg,
\cite{%
  Atserias04Sufficient,
  AKV04Constraint,
  BG03SpaceComplexity, 
  GT05Resolution}
and in this context perhaps most notably in the paper
\cite{AD08CombinatoricalCharacterization}
establishing the upper bound 
$\clspaceref{\fstd} \geq \widthref{\fstd} - \bigoh{1}$
on width in terms of clause space
for \kcnfform{}s~$\fstd$.

We have divided the  overview into four parts covering
(a) questions about time-space trade-offs and separations,
(b) comparisons of proof systems and subsystems of proof systems,
(c) formulas used as benchmarks for SAT solvers, and
(d) the computational complexity of various proof measures.
In what follows, our goal is to survey the results in fairly
non-technical terms. A more detailed discussion of the techniques used
to prove results on time and space will follow in  
\reftwosubsecs{sec:pc-tradeoffs}{sec:pc-substitution-theorem}.

\subsubsectionNOW{Time Versus Space}
\label{sec:pc-time-vs-space}

\ifthenelse{\boolean{maybeLMCS}}
{Pebble games have been used extensively
  as a tool to prove time and
  space lower bounds and trade-offs for computation. 
  Loosely put, a lower bound for the pebbling price of a graph says that
  although the computation that the graph describes can be performed
  quickly, it requires large space.} 
{As we have seen in this survey, pebble games have been used
  extensively as a tool to prove time and
  space lower bounds and trade-offs for computation.} 
Our hope is that when we encode pebble games in terms of \cnfform{}s,
these formulas should inherit the same properties as the underlying
graphs.  That is, if we start with a DAG~$G$ such that any pebbling
of~$G$ in short time must have large pebbling space, then we would
like to 
argue
that the corresponding pebbling contradiction
should have the property that any short resolution refutation of this
formula must also require large proof space.

In one direction the correspondence between pebbling and resolution
is straightforward. As was observed in~%
\cite{BIW00Near-optimalSeparation},
if there is a black pebbling strategy for~$G$ in time~$\stoptime$ and
space~$\spacestd$, then
$\pebcontr[G]{}$
can be refuted by resolution in length
$\bigoh{\stoptime}$
and space~%
$\bigoh{\spacestd}$.
Very briefly, the idea is that whenever the pebbling strategy places a
black pebble on $\varv$,
we derive in resolution the corresponding unit clause $\varv$.
This is possible since in the pebbling strategy all predecessors
$\varu$ of $\varv$ must be pebbled at this point, and so by induction
in the resolution derivation we have derived the unit clause
$\varu$
for all predecessors. But if so, it is easy to verify that 
the pebbling axiom for $\varv$ will allow us to derive
$\varv$.
When the pebbling ends, we have derived the unit clause $z$
corresponding to the unique sink of the DAG, at which point
we can download the sink axiom 
$\olnot{z}$
and derive a contradiction.

The other direction is much less obvious.  Our intuition is that the
resolution proof system should have to conform to the combinatorics of
the pebble game in the sense that from any resolution refutation of a
pebbling contradiction $\pebcontr[G]{}$ we should be able to extract a
pebbling of the DAG~$G$. To formalize this intuition, we would like to
prove something along the following lines:
\begin{enumerate}[(1)]
  
\item
  \label{item:tentative-proof-idea-part-one}
  First, 
  find a natural interpretation of sets of clauses   currently
  ``on the blackboard''   in a refutation of the formula 
  $\pebcontr[G]{}$
  in terms of 
  pebbles on the vertices of
  the DAG~$G$.
  
\item
  \label{item:tentative-proof-idea-part-two}
  Then,
  prove that this interpretation of clauses in terms of pebbles
  captures the pebble game in the following sense:
  for any resolution refutation of
  $\pebcontr[G]{}$,
  looking at consecutive sets of clauses on the blackboard
  and considering the corresponding sets of pebbles in the graph,
  we get a black-white pebbling of~$G$ 
  in accordance with the rules of the pebble game.
  
\item 
  \label{item:tentative-proof-idea-part-three}
  Finally, show that the interpretation captures 
  space in the sense that if the content of the blackboard induces
  $N$~pebbles on the graph, then there must be at least $N$~clauses on
  the   blackboard.
  
\end{enumerate}
Combining the above with known space lower bounds and time-space
trade-offs for pebble games, we would then be able to lift such bounds and
trade-offs to resolution.  For clarity, let us spell out what the
formal argument would look like.
Consider a resolution refutation~$\proofstd$ of the CNF
formula~$\pebcontr[G]{}$ defined over a graph~$G$ exhibiting a strong
time-space trade-off, and suppose this refutation has short
length. Using the approach outlined above, we extract a pebbling of
$G$ from~$\proofstd$. Since this is a pebbling in short time, because
of the time-space properties of~$G$ it follows that at some time~$t$
during this pebbling there must be many pebbles on the vertices of
$G$. But this means that at time~$t$, there are many clauses in the
corresponding configuration~$\clsc_t$  on the
blackboard. Since this holds for any refutation, we obtain
a length-space trade-off for resolution.

The first important step towards realizing the above program was taken
by Ben-Sasson in 2002 (journal version
in~\cite{Ben-Sasson09SizeSpaceTradeoffs}), who was the first
to prove trade-offs between proof complexity measures in resolution. 
The key insight in~\cite{Ben-Sasson09SizeSpaceTradeoffs}
is to interpret resolution refutations of
$\pebcontr[G]{}$
in terms of black-white pebblings of~$G$. The idea is to let
positive literals on the blackboard correspond to black pebbles and
negative literals to white pebbles. One can then show that using this
correspondence (and modulo some technicalities), any resolution refutation of 
$\pebcontr[G]{}$
results in a black-white pebbling of~$G$ in pebbling time
upper-bounded by the refutation length and pebbling space
upper-bounded by the refutation \emph{variable} space
(\refdef{def:width-space}). 

This translation of refutations to black-white pebblings was used  by
Ben-Sasson to establish strong trade-offs between clause space
and width in resolution. He showed that there are \kcnfform{}s
$\fstd_{\nvar}$ of size $\bigtheta{\nvar}$
which can be refuted both in constant clause space 
$\clspaceref{\fstd_n}$
and in constant width
$\widthref{\fstd_n}$,
but for which any refutation $\proofstd_n$ that tries to optimize both
measures simultaneously can never do better than
$\clspaceof{\proofstd_n} \cdot \widthofsmall{\proofstd_n}
= \bigomega{n / \log n}$.
This result was obtained by studying formulas
$\pebcontr[G_n]{}$
over the graphs~$G_n$ in~%
\cite{GT78VariationsPebbleGameVerboseURL}
with black-white pebbling price
$\bwpebblingprice{G_n} = \bigomega{n / \log n}$.
Since the upper bounds
$
\mbox{$\clspaceof{\proofstd} \cdot \widthofsmall{\proofstd}$} \geq
\totspaceof{\proofstd} \geq
\varspaceof{\proofstd}
$
are easily seen to hold for any resolution refutation~$\proofstd$, and since by
what was just said we must have  
$\varspaceof{\proofstd_n} = \bigomega{n / \log n}$, 
one gets the space-width trade-off stated above. In a separate
argument, one shows that 
$\Clspaceref{\pebcontr[G_n]{}} = \bigoh{1}$
and
$\Widthref{\pebcontr[G_n]{}} = \bigoh{1}$.
Using the same ideas plus upper bound on space in terms of size in~%
\cite{ET01SpaceBounds}, 
\cite{Ben-Sasson09SizeSpaceTradeoffs}~also proved that for tree-like
resolution it holds that
$\Lengthref[\treeresnot]{\pebcontr[G_n]{}} = \bigoh{n}$
but for any particular tree-like refutation $\proofstd_n$ there is a
length-width trade-off
$
\widthofsmall{\proofstd_n} \cdot 
\log \lengthofsmall{\proofstd_n} 
= \bigomega{n / \log n}
$.

Unfortunately, the results in~%
\cite{Ben-Sasson09SizeSpaceTradeoffs}
also show that the program outlined above for proving time-space
trade-offs will \emph{not} work for general resolution. This is so
since for any DAG~$G$ the formula $\pebcontr[G]{}$ is refutable in
linear length and constant clause space simultaneously.  What we have
to do instead is to look at substitution formulas
$\substform{\pebcontr[G]{}}{\funcpebc[]}$ 
for suitable Boolean functions~$\funcpebc$,
but this leads to a number of technical complications.
However, building on previous works
\cite{Nordstrom09NarrowProofsSICOMP,NH13TowardsOptimalSeparation},
a way was finally found to realize this program in~%
\cite{BN11UnderstandingSpace}.
We will give a more detailed exposition of the proof techniques in
\reftwosubsecs{sec:pc-tradeoffs}{sec:pc-substitution-theorem},
but let us conclude this discussion of time-space trade-offs by
describing the flavour of the results obtained in these latter papers.

Let 
$\set{G_n}_{n=1}^{\infty}$
be a family of \singlesinkdagtext{}s
of size
$\bigtheta{n}$
and with bounded fan-in. Suppose that there are functions
$\spacefunclower(n) \ll
\spacefuncupper(n) = 
\bigoh{n / \log \log n}$
such that
$G_n$
has black pebbling price
$\pebblingprice{G_n} = \spacefunclower(n)$
and there are 
black-only pebbling strategies for $G_n$ in time $\bigoh{n}$ and space
$\spacefuncupper(n)$,
but any black-white pebbling strategy in space
$\littleoh{\spacefuncupper(n)}$
must have superpolynomial or even exponential length.
Also, let
$\kdnfboundsconst$ be a fixed positive integer.
Then there are explicitly constructible CNF formulas
$\set{\fstd_n}_{n=1}^{\infty}$
of size $\bigoh{n}$ and width $\bigoh{1}$
(with constants depending on $\kdnfboundsconst$)
such that the following holds:
\begin{iteMize}{$\bullet$}
\item 
  The formulas $\fstd_n$ are refutable in syntactic resolution
  in  (small) total space
  $\bigoh{\spacefunclower(n)}$.
  
\item 
  There are also syntactic resolution refutations
  $\proofstd_n$ of~$\fstd_n$
  in simultaneous length $\bigoh{n}$ and (much larger) total space 
  $\bigoh{\spacefuncupper(n)}$.
  
\item 
  However, any resolution refutation, even semantic, in formula
  space
  $\littleoh{\spacefuncupper(n)}$
  must have superpolynomial or sometimes even exponential length.
  
\item 
  Even for the much stronger semantic
  \kdnf resolution proof systems, $k \leq \kdnfboundsconst$, 
  it holds that any \mbox{\resknot-refutation} of $\fstd_n$ in formula space
  $\Littleoh{\sqrt[k+1]{\spacefuncupper(n)}}$
  must have superpolynomial length (or exponential length, correspondingly).
\end{iteMize}

\noindent This ``theorem template'' can be instantiated for a wide range of
space functions 
$\spacefunclower(n)$
and
$\spacefuncupper(n)$,
from constant space all the way up to nearly linear space,
\ifthenelse{\boolean{maybeLMCS}}
{using graph families with suitable trade-off properties,
  \eg, those in
  \cite{LT82AsymptoticallyTightBounds,Nordstrom11RelativeStrength}.
  Also, absolute lower bounds on black-white pebbling space, such as
  in~%
  \cite{GT78VariationsPebbleGameVerboseURL},
  yield corresponding lower bounds on clause space.}
{using graph families with suitable trade-off properties
  \ifthenelse{\boolean{maybeNOW}}
  {(\eg, those in Chapters~%
    \ref{sec:constant-space-trade-offs},
    \ref{sec:non-constant-space-trade-offs},
    \ref{sec:robust-trade-offs},
    and
    \ref{sec:exponential-trade-offs}).}
  {(\eg, those in Sections~%
    \ref{sec:constant-space-trade-offs},
    \ref{sec:non-constant-space-trade-offs},
    \ref{sec:robust-trade-offs},
    and
    \ref{sec:exponential-trade-offs}).}
  Also, absolute lower bounds on black-white pebbling space, such as in
  \refsec{sec:optimal-lower-bound},
  yield corresponding lower bounds on clause space.}

Moreover, these trade-offs are robust in that they are not
sensitive to small variations in either length or space. The way we
would like to think about this, with some handwaving intuition, is
that the trade-offs will not show up only for a SAT solver being
unlucky and picking just the wrong threshold when trying to hold down
the memory consumption. Instead, any resolution refutation having
length or space in the same general vicinity will be subject to the
same qualitative trade-off behaviour.

\subsubsectionNOW{Separations of Proof Systems}
\label{sec:pc-separations}


A number of restricted subsystems of resolution, often referred to as
\introduceterm{resolution refinements},
have been studied in the proof complexity literature. These refinements 
were introduced to model SAT solvers that try to make the
proof search more efficient by narrowing the search space, and they
are defined in terms of restrictions on the DAG
representations~$G_{\proofstd}$ of resolution
refutations~$\proofstd$.
An interesting question is how the strength of these refinements are
related to one another and to that of general, unrestricted
resolution, and pebbling has been used as a tool in several papers
investigating this.
We briefly discuss some of these restricted subsystems below, noting
that they are all known to be sound and complete.
We remark that more recently, a number of
different (but related) models for resolution with
\introduceterm{clause learning}  have also been proposed and studied
theoretically in 
\cite{%
  BKS04TowardsUnderstanding,
  BHJ09ResolutionTrees,
  BJ10LowerBoundsWidth,
  HBPV08ClauseLearning,
  PD11OnThePower,
  VanGelder05PoolResolution}
but going into details here is unfortunately outside the scope of
this survey.

A \introduceterm{regular resolution} refutation of a CNF
formula $\fstd$ is a refutation $\proofstd$ such that on any path in
$G_\proofstd$ from an axiom clause in $\fstd$ to the empty
clause~$\emptycl$, no variable is resolved over more than once.
We call a regular resolution refutation \introduceterm{ordered} if
in addition there exists an ordering of the variables such that every
sequence of variables labelling a path from an axiom to the empty
clause respects this ordering. Ordered resolution is also known as
\introduceterm{Davis-Putnam resolution}.
A \introduceterm{linear resolution} refutation  is a
refutation $\proofstd$ with the additional restriction that the
underlying DAG $G_\proofstd$ must be linear. That is, the proof
should consist of a sequence of clauses 
$\set{\clc_1, \clc_2, \ldots, \clc_m = \emptycl}$
such that for every
$i \in \intnfirst{m}$
it holds for the clause
$\clc_i$
that it is either an axiom clause of $\fstd$
or is derived from
$\clc_{i-1}$
and
$\clc_{j}$
for some $j < i$ (where $\clc_{j}$ can be an axiom clause).
Finally, as was already mentioned in
\refdef{def:sequential-refutation},
a
\introduceterm{tree-like} 
refutation is one in which the underlying DAG is a tree.
Tree-like resolution is also called 
\introduceterm{Davis-Logemann-Loveland}
or 
\introduceterm{DLL resolution}
in the literature. The reason for this is that tree-like resolution
refutations can be shown to correspond to refutations produced by the
proof search algorithm in 
\cite{DLL62MachineProgram},
known as DLL or DPLL,
that fixes one variable~$\varx$ in the formula~$\fstd$ to true or
false respectively, and then recursively tries to refute the two
formulas corresponding to the two values of~$\varx$ (after
simplifications, \ie removing satisfied clauses and shrinking clauses
with falsified literals).

It is known that tree-like resolution proofs can always be made
regular \wolog~\cite{U95Complexity}, and clearly ordered refutations
are regular by definition.
Alekhnovich \etal~\cite{AJPU07ExponentialSeparation}
established an exponential separation \wrt length between general and regular
resolution, improving a previous weaker separation by Goerdt~%
\cite{Goerdt93Regular},
and Bonet \etal~\cite{BEGJ00RelativeComplexity} 
showed that tree-like resolution can be exponentially weaker than
ordered resolution and some other resolution refinements.
%
%
Johannsen~\cite{Johannsen01ExponentialIncomparability}
exhibited formulas for which tree-like resolution is exponentially
stronger than ordered resolution, from which it follows that regular
resolution can also be exponentially stronger than ordered resolution
and that tree-like and ordered resolution are incomparable.
More separations for other
resolution refinements not mentioned above were presented in~%
\cite{BOP07Complexity},
but 
a detailed discussion of
these results are outside the scope of this survey. 

The construction in
\cite{AJPU07ExponentialSeparation}
uses an implicit encoding of the 
\ifthenelse{\boolean{maybeLMCS}}
{pebbling contradictions} 
{pebbling formulas} 
in
\refdef{def:pebbling-contradiction}
in the sense that they study formulas encoding that
each vertex in the DAG contains a pebble, identified by a unique
number. For every pebble, there is a variable encoding the colour of
this pebble---red or blue---where source vertices are known to have
red pebbles and the sink vertex should have a blue one. Finally, there are
clauses enforcing that if all predecessors of a vertex has red
pebbles, then the pebble on that vertex must be red. These formulas
can be refuted bottom-up in linear length just as our standard
pebbling contradictions, but such refutations are highly irregular. 
The paper~\cite{BEGJ00RelativeComplexity},
which also presents lower bounds for tree-like CP proofs for formulas
easy for resolution, uses another variant of pebbling contradictions
defined over pyramid graphs, but we omit the details. 
Later, \cite{BIW00Near-optimalSeparation} proved a stronger
exponential separation of general and tree-like resolution, improving
on the separation implied by~%
\cite{BEGJ00RelativeComplexity}, and this latter paper uses
substitution pebbling contradictions~$\pebcontrwithfunc[G]{}{\lor_2}$
and the 
$\bigomega{n / \log n}$
lower bound on black pebbling in~%
\ifthenelse{\boolean{maybeLMCS}}
{\cite{PTC76SpaceBounds}.}
{\cite{PTC76SpaceBounds}
  (see
  \refsec{sec:optimal-lower-bound}).}

Intriguingly, linear resolution is \emph{not} known to be weaker then
general resolution. 
%
%
%
%
The conventional wisdom seems to be that linear resolution should
indeed be weaker, but the difficulty is if so it can only be weaker on
a technicality.
Namely, it was shown in~%
\cite{BOP07Complexity}
that if a polynomial number of appropriately chosen tautological
clauses are added to any CNF formula, then linear resolution can
simulate general resolution by using these extra clauses.  Any
separation would therefore have to argue very ``syntactically.''

Esteban \etal~\cite{EGM04Complexity}
showed that tree-like $k$-DNF resolution proof systems form a strict
hierarchy \wrt proof length and proof space. The space separation they
obtain is for formulas requiring 
formula space
$\bigoh{1}$ in $\resknot[k+1]$
but formula space
$\bigomega{n / \log^2 n }$ in $\resknot[k]$.
Both of these separation results use a special flavour
$\pebcontrwithfunc[G]{}{\land_{l} \lor_{k}}$ 
of substitution pebbling formulas, again defined 
over the graphs 
in~%
\cite{PTC76SpaceBounds}
with black pebbling price
$\bigomega{n / \log n}$.
As was mentioned above, the space separation was strengthened to
general, unrestricted $\resknot$-systems in~%
\cite{BN11UnderstandingSpace},
but with worse parameters. This latter result is obtained using formulas
$\pebcontrwithfunc[G]{}{\xor_{k+1}}$
defined in terms of exclusive or of $k+1$ variables to get the
separation between
$\resknot[k+1]$
and~%
$\resknot[k]$,
as well as the stronger 
\emph{black-white} pebbling
price lower bound of
$\bigomega{n / \log n}$
in~\cite{GT78VariationsPebbleGameVerboseURL}.
%
%

Concluding our discussion of separation of resolution refinements, we
also want to mention that
Esteban and Torán~\cite{ET03CombinatorialCharacterization}
used substitution pebbling
contradictions~$\pebcontrwithfunc[G]{}{\lor_2}$ over complete binary
trees to prove that general resolution is strictly stronger than
tree-like resolution \wrt clause space. Expressed in terms of formula
size the separation one obtains is in the constant multiplicative
factor in front of the logarithmic space bound.%
\footnote{Such a constant-factor-only separation might not sound
too impressive, but recall that the space complexity it at most
linear in the number of variables and clauses, so it makes sense to
care about constant factors here. Also, it should be noted that this
paper had quite some impact in that the technique used to establish
the separation can be interpreted as a (limited) way of of simulating
black-white pebbling in resolution, and  this provided one of the key
insights for~%
\cite{Nordstrom09NarrowProofsSICOMP}
and the ensuing papers considered in
\refsubsec{sec:pc-time-vs-space}.}
This was recently improved to a logarithmic separation in
\cite{JMNZ12RelatingProofCplx},
obtained for XOR-pebbling contradictions over line graphs,
\ie graphs with vertex sets
$\set{v_1, \ldots, v_n}$
and edges
$(v_{i}, v_{i+1})$
for $i=1, \ldots, n-1$.

\subsubsectionNOW{Benchmark Formulas}
\label{sec:pc-benchmarks}

Pebbling contradictions have also been used as benchmark formulas for
evaluating and comparing different proof search heuristics. 
Ben-Sasson \etal~\cite{BIW00Near-optimalSeparation}
used the exponential lower bound discussed above for tree-like
resolution refutations of formulas
$\pebcontrwithfunc[G]{}{\lor_2}$ 
to show that a proof search heuristic that exhaustively searches for
resolution refutations in minimum width can sometimes be exponentially
faster than DLL-algorithms searching for tree-like resolutions,  while
it can never be too much slower.
Sabharwal \etal~\cite{SBK03UsingProblemStructure} 
also used pebbling formulas to evaluate heuristics for clause learning
algorithms. 
In a more theoretical work, 
Beame \etal~\cite{BIPS10FormulaCaching}
again used pebbling formulas
$\pebcontrwithfunc[G]{}{\lor_2}$ 
to compare and separate extensions of the resolution
proof system using ``formula caching,'' which is a
generalization of clause learning.

In view of the strong length-space trade-offs for resolution which
were hinted at in
\refsubsec{sec:pc-time-vs-space}
and will be examined in more detail below, a natural question is
whether these theoretical results also translate into trade-offs
between time and space in practice for state-of-the-art SAT solvers
using clause learning. Although the model in
\reftwodefs{def:sequential-refutation}{def:length-and-space}
for measuring time and space of resolution proofs is quite simple, it
still does not seem too unreasonable that it should be able to capture
the problem in clause learning concerning which of the learned clauses should
be kept in the clause database (which would roughly correspond to
configurations in our refutations). It would be interesting to
take graphs~$G$ as in 
\ifthenelse{\boolean{maybeLMCS}}
{\cite{LT82AsymptoticallyTightBounds,Nordstrom11RelativeStrength}}
{\reftwosecs{sec:constant-space-trade-offs}{sec:non-constant-space-trade-offs},
  or possibly as in 
  \reftwosecs{sec:robust-trade-offs}{sec:exponential-trade-offs}
  although these constructions are more complex and therefore perhaps
  not as good candidates,} 
and study formulas
$\pebcontrwithfunc[G]{}{\funcpebc[]}$ 
over these graphs for suitable substitution functions~%
$\funcpebc[]$.
If we for instance take
$\funcpebc[]$
to be exclusive or~$\xor_{d}$ for arity $d\geq 2$,
then
we have provable
length-space trade-offs in terms of pebbling trade-offs for the
corresponding DAGs (and although we cannot prove it, we strongly
suspect that the same should hold true also for formulas defined in
terms of the usual logical or of any arity), and the question is
whether one could observe similar trade-off phenomena also in practice.
%
%

\begin{openproblem}
  \label{openproblem:time-space-tradeoffs-SAT-solvers}
  Do pebbling contradictions
  $\pebcontrwithfunc[G]{}{\funcpebc[]}$
  for suitable $f$ (such as $\lor$ or $\xor$) 
  exhibit time-space trade-offs for 
  current state-of-the-art DPLL-based
  SAT solvers 
  similar to the pebbling trade-offs of the underlying DAGs~$G$?
\end{openproblem}

Let us try to present a very informal argument why the answer to this
question could be positive. On the one hand, 
all the length-space trade-offs that have been shown for pebbling
formulas hold for space in the sublinear regime (which is inherent,
since any pebbling formula can be refuted in simultaneous linear time
and linear space), 
and given that linear space is needed just to keep the formula in
memory such space bounds might not seem to relevant for real-life
applications.  On the other hand, suppose that we know for some
\cnfform~$\fstd$ that $\clspaceref{\fstd}$ is large. What this tells
us is that any algorithm, even a non-deterministic one making optimal
choices concerning which clauses to save or throw away at any given
point in time, will have to keep a fairly large number of ``active''
clauses in memory in order to carry out the refutation.
Since this is so, a real-life deterministic proof search algorithm,
which has no sure-fire way of knowing 
which clauses are the right ones to concentrate on 
at any given moment, might have to keep working on a
lot of extra clauses in order to be sure that the fairly large
critical set of clauses needed to find a refutation will be among the
``active'' clauses.

Intriguingly enough, in one sense one can argue that pebbling
contradictions have already been shown to be an example of this. 
We know that these formulas are very easy \wrt length and width,
having constant-width refutations that are essentially as short as the 
formulas themselves. 
But one way of interpreting the experimental results in~%
\cite{SBK03UsingProblemStructure},
is that one of the state-of-the-art SAT solvers at that time had
serious problems with even moderately large 
pebbling contradictions. Namely, the
``grid pebbling formulas'' 
in~%
\cite{SBK03UsingProblemStructure}
are precisely our
\orpebcontrtext{}s
$\pebcontrwithfunc[G]{}{\lor_2}$ 
over pyramids.
Although we are certainly not arguing that this is the whole
story---it was also shown 
in~\cite{SBK03UsingProblemStructure}
that the branching order is a critical factor, and that
given some extra structural information the algorithm can achieve an 
exponential speed-up---we wonder whether the high lower bound on
clause space can nevertheless be part of the explanation. 
It should be pointed out that pebbling contradictions are the only
formulas we know of that are really easy \wrt length and width but
hard for clause space. And if there is empirical data showing that for
these very formulas clause learning algorithms can have great
difficulties finding refutations, it might be worth investigating
whether this is just a coincidence or a sign of some deeper
connection.

\subsubsectionNOW{Complexity of Decision Problems}
\label{sec:pc-decision-problems}

A number of papers have also used pebble games to study how hard it is
to decide the complexity of a CNF formula~$\fstd$ with respect to some
proof complexity measure~$\measstd$. This is formalized in terms of
decision problems as follows: ``Given a CNF formula~$\fstd$ and a
parameter~$p$, is there a refutation $\proofstd$ of~$\fstd$ with
\mbox{$\measstdof{\proofstd} \leq p$?}''

The one proof complexity measure that is reasonably well understood is
proof length. It has been shown (using techniques not related to
pebbling) that the problem of finding a
shortest refutation of a CNF formula is \NP-hard~%
\cite{Iwama97ComplexityOfFindingShortResolutionProofs}
and remains hard even if we just want to approximate the minimum
refutation length~%
\cite{ABMP01MinimumPropositionalProofLength}.

With regard to proof space, 
Alex Hertel and Alasdair Urquhart~\cite{HU07Game} 
showed that tree-like resolution clause space is \PSPACE-complete,
using the exact combinatorial characterization of tree-like
resolution clause space given in
\cite{ET03CombinatorialCharacterization}
\ifthenelse{\boolean{maybeLMCS}}           
{and a generalization of the pebble game in
  \refdef{def:bw-pebble-game}
  introduced in~\cite{Lingas78PSPACE}.}
{and the generalized pebble game in~\cite{Lingas78PSPACE}
  mentioned in
  \refsubsec{sec:TLO-complexity}.}
They also proved (see
\cite[Chapter~6]{AlexHertel08Thesis})
that variable space in general resolution
is \PSPACE-hard, although this result requires CNF formulas of
unbounded width. Interestingly, variable space is \emph{not} known to
be in \PSPACE, and the best upper bound obtained in
\cite{AlexHertel08Thesis}
is that the problem is at least contained in \EXPSPACE.

Another very interesting space-related result is that of 
Philipp Hertel and Toni Pitassi
\cite{HP07ExponentialTimeSpaceSpeedupFOCS}, 
who presented a \PSPACE-completeness result for total space in
\ifthenelse{\boolean{maybeLMCS}}           
{resolution as well as some sharp trade-offs
for length \wrt total space,} 
{resolution as well as some sharp trade-offs 
  (in the sense of \refsubsec{sec:TLO-sharp-tradeoffs})
  for length \wrt total space,
  building on their \PSPACE-completeness result for black-white pebbling
  mentioned in
  \refsubsec{sec:TLO-complexity}
  and} 
using the original pebbling contradictions 
$\pebcontr[G]{}$
in 
\refdef{def:pebbling-contradiction}.
Their construction is highly nontrivial, and unfortunately a bug was
later found in the proofs leading to these results being withdrawn
in the journal version~\cite{HP10PspaceCompleteness}.
The trade-off results claimed in
\cite{HP07ExponentialTimeSpaceSpeedupFOCS}
were later subsumed by those in~%
\cite{Nordstrom09SimplifiedWay},
using other techniques not related to pebbling, 
but it remains open whether total space is \PSPACE-complete or not
(that this problem is in \PSPACE is fairly easy to show).

\begin{openproblem}
  \label{openproblem:total-space-complexity}
  Given a CNF formula~$\fstd$ (preferably of fixed width) and a
  parameter~$\spacestd$, is it \PSPACE-complete to determine whether
  $\fstd$ can be refuted 
  in the resolution proof system
  in total space at most~$\spacestd$?
\end{openproblem}

There are a number of other interesting open questions regarding the
hardness of proof complexity measures for resolution. 
An obvious question is whether the \PSPACE-completeness result for
tree-like resolution clause space in~%
\cite{HU07Game} 
can be extended to clause space in general resolution. (Again,
showing that clause space is in \PSPACE is relatively straightforward.) 

\begin{openproblem}
  \label{openproblem:clause-space-complexity}
  Given a CNF formula~$\fstd$ (preferably of fixed width) and a
  parameter~$\spacestd$, is it \PSPACE-complete to determine whether
  $\fstd$ can be refuted in resolution in clause space at most~$\spacestd$?
\end{openproblem}

A somewhat related question is whether it is possible to find a clean,
purely combinatorial characterization of clause space. This has been
done for resolution width~\cite{AD08CombinatoricalCharacterization}
and tree-like resolution clause
space~\cite{ET03CombinatorialCharacterization}, 
and this latter result was a key component in proving
the \PSPACE-completeness of tree-like space.
It would be very interesting to find similar characterizations of
clause space in general resolution and~$\resknot$.

\begin{openproblem}[\cite{ET03CombinatorialCharacterization,EGM04Complexity}]
  \label{openproblem:comb-charac-space}
  Is there a combinatorial characterization of refutation clause space for
  general, unrestricted resolution? For $k$-DNF resolution?
\end{openproblem}

The complexity of determining resolution width is also open.

\begin{openproblem}
  \label{openproblem:width-complexity}
  Given a \kcnfform~$\fstd$ and a
  parameter~$\widthstd$, is it \EXPTIME-complete to determine whether
  $\fstd$ can be refuted in resolution in width at most~$\widthstd$?%
  \footnote{%
    As the camera-ready version of this article was being prepared,
    a proof of the \EXPTIME-completeness of determining width
    complexity was announced in~%
    \cite{Berkholz12ComplexityNarrowProofs}.
    We refer to
    \refth{th:width-berkholz}
    below for more details on this result.}
\end{openproblem}

The width measure was conjectured to be 
\EXPTIME-complete by Moshe Vardi. As shown in~%
\cite{HU06ResolutionWidthEXPTIME},
using the combinatorial characterization of width in~%
\cite{AD08CombinatoricalCharacterization},
width is in \EXPTIME. The paper
\cite{HU06ResolutionWidthEXPTIME}
also claimed an  \EXPTIME-completeness result, but this was later
retracted in~%
\cite{HU09Comments}.
The conclusion that can be drawn from all of this is perhaps that
space is indeed a very tricky concept in proof complexity, and that we
do not really understand space-related measures very well, even for
such a simple proof system as resolution.

\subsectionNOW{Translating Time-Space Trade-offs from Pebbling to Resolution}
\label{sec:pc-tradeoffs}

So far, we have discussed in fairly non-technical terms how pebble
games have been used to prove different results in proof
complexity. In this section and the next, we want to elaborate on the
length-space trade-off results for resolution-based proof systems
mentioned in 
\refsubsec{sec:pc-separations}
and try to give a taste of how they are proven. Recall that the
general idea is to establish reductions between pebbling strategies
for DAGs on the one hand and refutations of corresponding pebbling
contradictions on the other. We start by describing the reductions
from pebblings to refutations in 
\refsubsec{sec:pc-upper-bounds},
and then examine how refutations can be translated to pebblings in
\refsubsec{sec:pc-lower-bounds}.

\subsubsectionNOW{Techniques for Upper Bounds on Resolution Trade-offs}
\label{sec:pc-upper-bounds}

Given any  black-only pebbling $\pebbling$ of a DAG~$G$ with bounded
fan-in~$\indegreedag$, it is straightforward to simulate this pebbling in resolution  
to refute the corresponding pebbling contradiction
$\pebcontrwithfunc[G]{}{\funcpebc[\pebdeg]}$
in length
$\Bigoh{\pebtime{\pebbling}}$
and space
$\Bigoh{\pebspace{\pebbling}}$.
This was perhaps first noted in
\cite{BIW00Near-optimalSeparation}
for the simple $\pebcontr{}$ formulas,
but the simulation extends readily to any formula
$\pebcontrwithfunc[G]{}{\funcpebc[\pebdeg]}$,
with the constants hidden in the asymptotic notation depending only on~%
$\funcpebc[\pebdeg]$
and~%
$\indegreedag$.
In view of the translations presented in
\cite{Ben-Sasson09SizeSpaceTradeoffs}
and subsequent works of resolution refutations to \emph{black-white}
pebblings, it is natural to ask whether this reduction goes both ways,
\ie whether resolution can simulate not only black pebblings but also
black-white ones. 

At first sight, it seems that resolution would have a  hard time
simulating black-white pebbling. To see why, let us start by
considering a black-only pebbling~$\pebbling$. 
We can easily mimic such a pebbling by a resolution refutation of 
$\pebcontrwithfunc[G]{}{\funcpebc[\pebdeg]}$
which derives that
$\funcpebc[\pebdeg] (v_1, \ldots, v_{\pebdeg})$ 
is true whenever the corresponding vertex $v$ in $G$
is black-pebbled.
If the pebbling strategy places a pebble on~$v$ at time~$t$, then we
know that all predecessors of~$v$ have pebbles at this point. By
induction, this implies that for all 
$w \in \prednode{v}$
we have clauses
$\substform{w}{\funcpebc[\pebdeg]}$ 
in the configuration~$\clsc_t$
encoding that all
$\funcpebc[\pebdeg] (w_1, \ldots, w_{\pebdeg})$  
are true, 
and if we download the pebbling axioms for~$v$ we can derive the
clauses 
$\substform{v}{\funcpebc[\pebdeg]}$ 
encoding that
$\funcpebc[\pebdeg] (v_1, \ldots, v_{\pebdeg})$  
is true by the implicational completeness of resolution. Furthermore,
this derivation can be carried out in  length and extra clause space 
$\bigoh{1}$, where the hidden constants depend only on 
$\indegreedag$ and~$\funcpebc[\pebdeg]$
as stated above.
We end up deriving that
$\funcpebc[\pebdeg] (z_1, \ldots, z_{\pebdeg})$ 
is true for the sink~$z$, at which point we can download the sink
axioms and derive a contradiction.

The intuition behind this translation is that a black pebble on~$v$ 
means that we know~$v$, which in resolution translates into truth
of~$v$. In the pebble game, having a white pebble on $v$ 
instead means that we need to assume~$v$. By duality, it seems
reasonable to let this correspond to falsity of~$v$ in resolution.
Focusing on the pyramid 
$\Pi_2$
in
\reffig{fig:pebbling-contradiction-for-Pi-2-graph},
and pebbling contradiction
$\pebcontrwithfunc[{\Pi_2}]{}{\lor_2}$ 
in
\reffig{fig:pebbling-contradiction-for-Pi-2-binary-or},
our intuitive understanding then becomes that white pebbles on $x$
and~$y$ and a black pebble on~$z$  should correspond to the 
set of  clauses
\begin{equation}
  \label{eq:x-y-white-z-black}
  \setdescr{\olnot{x}_i \lor \olnot{y}_j \lor z_1 \lor z_2}{i,j =  1,2}  
\end{equation}
which indeed encode that assuming
$x_1 \lor x_2$
and
$y_1 \lor y_2$,
we can deduce
$z_1 \lor z_2$.
See
\reffig{fig:examples-res-to-peb-a}
for an illustration of this.

\ifthenelse{\boolean{maybeLMCS}}           
{\begin{figure}[tp]
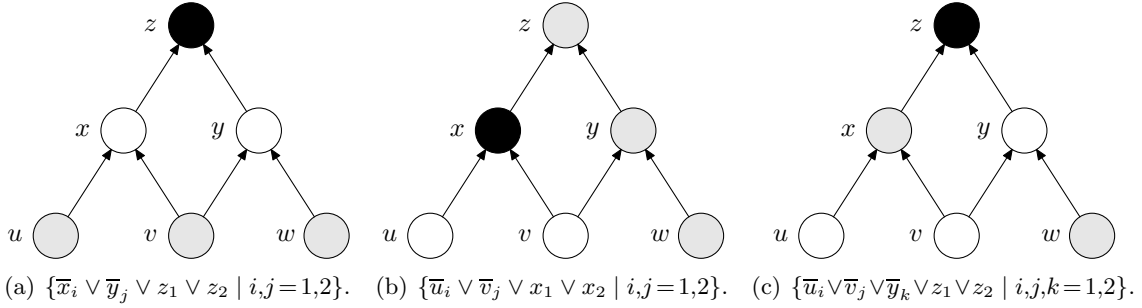

  \subfigure
  [$\setdescr{\olnot{x}_i \lor \olnot{y}_j \lor z_1 \lor z_2}{i,\!j \!=\!  1,\!2}$.]
  {
    \label{fig:examples-res-to-peb-a}
    \begin{minipage}[b]{0.30\linewidth}
      \centering
      \includegraphics{smallPyramidHeight2.2}%
    \end{minipage}
  }
  \hfill
  \subfigure
  [$\setdescr{\olnot{u}_i \lor \olnot{v}_j \lor x_1 \lor x_2}{i,\!j \!=\! 1,\!2}$.]
  {
    \label{fig:examples-res-to-peb-b}
    \begin{minipage}[b]{0.30\linewidth}
      \centering
      \includegraphics{smallPyramidHeight2.3}%
    \end{minipage}
  }
  \hfill
  \subfigure
  [$\setdescr{%
      \olnot{u}_i 
      \! \lor \! 
      \olnot{v}_j 
      \! \lor \! 
      \olnot{y}_k 
      \! \lor \! 
      z_1 
      \! \lor \! 
      z_2}
    {i,\!j,\!k \!=\! 1,\!2}$.]	    
  {
    \label{fig:examples-res-to-peb-c}
    \begin{minipage}[b]{0.33\linewidth}
      \centering
      \includegraphics{smallPyramidHeight2.4}%
    \end{minipage}
  }
  \caption{Black and white pebbles and (intuitively) corresponding 
    sets of clauses.}
  \label{fig:examples-res-to-peb}
\end{figure}

}
{\input{figExampleResToPeb.tex}}

If we now place white pebbles on $u$ and $v$, this allows us to remove
the white pebble from~$x$. Rephrasing this in terms of resolution,
we can say that $x$ follows if we assume $u$ and~$v$,
which is encoded as the set of clauses
\begin{equation}
  \label{eq:u-v-white-x-black}
  \setdescr{\olnot{u}_i \lor \olnot{v}_j \lor x_1 \lor x_2}{i,j = 1,2}  
\end{equation}
(see \reffig{fig:examples-res-to-peb-b}),
and indeed, from the clauses in
\refeq{eq:x-y-white-z-black}
and~\refeq{eq:u-v-white-x-black}
we can derive in resolution that $z$ is black-pebbled and $u$, $v$
and~$y$ are white pebbled, \ie the set of clauses
\begin{equation}
  \label{eq:u-v-y-white-z-black}
  \setdescr{\olnot{u}_i \lor \olnot{v}_j \lor \olnot{y}_k 
    \lor z_1 \lor z_2}{i,j,k = 1,2}  
\end{equation}
(see \reffig{fig:examples-res-to-peb-c}).
The above toy example indicates one of the problems one runs into when one
tries to simulate black-white pebbling in resolution: as the
number of white pebbles grows, there is an exponential blow-up in the
number of clauses. The clause set in~%
\refeq{eq:u-v-y-white-z-black}
is twice the size of those in
\refeq{eq:x-y-white-z-black}
and~\refeq{eq:u-v-white-x-black},
although it corresponds to only one more white pebble. This suggests
that as a pebbling starts to make heavy use of white pebbles,
resolution will not be able to mimic such a pebbling in a length-
and space-preserving manner.
This leads to the thought that perhaps black pebbling provides not
only upper but also lower bounds on resolution refutations of pebbling
contradictions.

However, it was shown in~\cite{Nordstrom11RelativeStrength} 
that at least in certain instances, resolution can in fact be strictly
better than black-only pebbling, both for time-space trade-offs and
\wrt space in absolute terms.
What is done in~%
\cite{Nordstrom11RelativeStrength} 
is to design a limited version of black-white pebbling,
tailor-made for resolution,
where one explicitly restricts the amount of nondeterminism, \ie white
pebbles, which a pebbling strategy can use. Such restricted pebblings use
``few white pebbles per black pebble'' (in a sense that will be made
formal below), and can therefore be simulated in a time- and
space-preserving fashion by resolution, avoiding the exponential
blow-up just discussed.
This game is essentially just a
formalization of the naive  simulation sketched above,
but before stating the formal definitions,
let us try to provide some intuition why the rules of this new game
look the way they do.  

First, if we want a game that can be mimicked by resolution, then
placements of isolated white vertices do not make much sense. What a
resolution derivation can do is to download axiom clauses, and
intuitively this corresponds to 
placing a black pebble on a vertex together with
white pebbles on its immediate predecessors, if the vertex has any.
Therefore, we adopt such aggregate moves as the only admissible way of
placing new pebbles. For instance, 
looking at  
\reffig{fig:examples-res-to-peb}
again, placing a black pebble on $z$ and white pebbles on $x$ and~$y$
corresponds to downloading the axiom clauses in~%
\refeq{eq:x-y-white-z-black}
for
$\pebcontrwithfunc[{\Pi_2}]{}{\lor_2}$.

Second, note that if we have
a black pebble on $z$ with white pebbles on $x$ and~$y$
corresponding to the clauses in~%
\refeq{eq:x-y-white-z-black}
and
a black pebble on $x$ with white pebbles on $u$ and~$v$
corresponding to the clauses in~%
\refeq{eq:u-v-white-x-black},
we can derive the clauses in~%
\refeq{eq:u-v-y-white-z-black}
corresponding to 
$z$ black-pebbled and  $u$, $v$ and~$y$ white-pebbled
but no pebble on $x$. This 
suggests   
that a natural rule for
white pebble removal is that a white pebble can be removed from a
vertex if a black pebble is placed \emph{on that same vertex} (and not on its
immediate predecessors).

Third, if we then just erase all clauses in~%
\refeq{eq:u-v-y-white-z-black},
this corresponds to all pebbles disappearing. On the face of it, this
is very much unlike the rule for white pebble removal in the standard
pebble game, where it is absolutely crucial that a white pebble can
only be removed when its predecessors are pebbled. However, the
important point here is that not only do the white pebbles
disappear, but the black pebble that has been placed on $z$ with the help
of these white pebbles disappears as well. 
What this means is that
we cannot treat black and white pebbles in isolation, but we have to
keep track of for each black pebble which white pebbles it depends on,
and make sure that the black pebble also is erased if any of the white
pebbles supporting it is erased. The way we do this is to label each
black pebble $v$ with its supporting white pebbles~$W$, and define the
pebble game in terms of moves of such labelled \introduceterm{pebble
subconfigurations} $\scnot{v}{W}$.
%
%
%
    
Formalizing the loose description above, our pebble game is then
defined as follows.

\begin{definition}[\Lpebblingtext{} \cite{Nordstrom11RelativeStrength}]
  \label{def:labelled-bw-pebble-game}
  For $v$ a vertex and $W$ a set of vertices 
  \st $v \notin W$,
  we say that
  $\scnot{v}{W}$
  is a 
  \introduceterm{pebble \subconftext{}}
  with a black pebble on~$v$ supported by white pebbles on all
  $w \in W$. 
  The black pebble on $v$  
  in $\scnot{v}{W}$   
  is   said to be
  \introduceterm{\blconditional{}} on   the white pebbles in
  its \introduceterm{support}~$W$.
  We refer to $\unconditionalblacknot{v}$ as 
  an \introduceterm{\blunconditional black pebble}.

  For $G$ any DAG with unique sink~$z$, 
  a (\pebcomplete{})
  \introduceterm{\lpebblingtext{}}
  of~$G$ is a sequence
  $\lpebbling
  =
  \setsmall{\lconf_0, 
    \ldots,
    \lconf_{\stoptime}}$
  of  
  labelled pebble configurations
  \st 
  ${\lconf_{0} = \emptyset}$,
  ${\lconf_{\stoptime} = \set{\unconditionalblacknot{z}}}$,
  and for all $t \in \intnfirst{\stoptime}$
  it holds that
  $\lconf_{t}$
  can be obtained from
  $\lconf_{t-1}  $
  by one of the following rules:
  \begin{enumerate}[\hbox to8 pt{\hfill}]

  \italicitem\noindent{\hskip-12 pt\bf\textit{Introduction:}}\
    $\lconf_{t} = \lconf_{t-1} \unionSP 
    \setsmall{\introsubconfnot{v}}$,
    where $\prednode{v}$ is the set of immediate predecessors of~$v$.

  \italicitem\noindent{\hskip-12 pt\bf\textit{Merger:}}\
    $\lconf_{t}  = 
    \lconf_{t-1} \unionSP
    \setcompact{\scnot{v}{(V \union W) \setminus \setsmall{w}}}$
    for 
    $\scnot{v}{V},  \scnot{w}{W} \in \lconf_{t-1}$
    such that
    $w \in V$
    (and
    $v \notin W$).       
    We denote this \subconftext
    $\spmerge{\scnot{v}{V}}{\scnot{w}{W}}$,
    and refer to it  as a
    \introduceterm{merger on~$w$}.

  \italicitem\noindent{\hskip-12 pt\bf\textit{Erasure:}}\
    $\lconf_{t} = \lconf_{t-1} \setminus \setsmall{\scnot{v}{V}}$
    for
    $\scnot{v}{V} \in \lconf_{t-1}$.

\end{enumerate}
Let
$\spinducedblack{\lconf_t}
=
\setdescrsmall{v}{\scnot{v}{W} \in \lconf_{t}}
$
denote the set of all black-pebbled vertices
in~$\lconf_{t}$
and
$\spinducedwhite{\lconf_t}
=
\Union
\setdescrsmall{W}{\scnot{v}{W} \in \lconf_{t}}
$
the set of all white-pebbled  vertices.
%
Then the  space of a \lpebblingtext
$\lpebbling = \setsmall{\lconf_0, \ldots, \lconf_{\stoptime}}$
is 
$\maxofsmall[\lconf \in \lpebbling]
{\setsize{\spinducedblack{\lconf}  \unionSP \spinducedwhite{\lconf}}}$
and
the time of 
$\lpebbling = \setsmall{\lconf_0, \ldots, \lconf_{\stoptime}}$
is
$\pebtime{\lpebbling} = \stoptime$.
\end{definition}

The game in 
\refdef{def:labelled-bw-pebble-game}
might look quite different from the standard black-white pebble game, but it
is not hard to show that \lpebblingtext{}s are essentially just a
restricted form of  black-white pebblings.
\ifthenelse{\boolean{maybeLMCS}}           
{}
{(We refer to~%
\cite{Nordstrom11RelativeStrength}
for formal proofs of this and all following claims in
\refsubsec{sec:pc-upper-bounds}).}

\begin{lemma}[\cite{Nordstrom11RelativeStrength}]
  \label{lem:bw-pebbling-simulates-l-pebbling}
  If $G$ is a \singlesinkdagtext 
  and
  $\lpebbling$
  is a \pebcomplete \lpebblingtext of~$G$,
  then there is a \pebcomplete black-white pebbling
  $\pebbling_{\lpebbling}$ of~$G$ with
  $
  \pebtime{\pebbling_{\lpebbling}} \leq
  \frac{4}{3}
  \pebtime{\lpebbling}
  $
  and
  $\pebspace{\pebbling_{\lpebbling}} \leq  \pebspace{\lpebbling}$.
\end{lemma}

However, the definition of space of \lpebblingtext{}s does not seem
quite right from the point of view of resolution. Not only does the
space measure fail to capture the exponential blow-up in the
number of white pebbles discussed above. We also have the problem that
if one white pebble is used to support many different black pebbles,
then in a resolution refutation simulating such a pebbling we have to
pay multiple times for this single white pebble, once for every black pebble
supported by it.  To get something that can be simulated by
resolution, we therefore need to restrict the \lpebblegame even
further.

\begin{definition}[Bounded \lpebblingtext{}s \cite{Nordstrom11RelativeStrength}]
    \label{def:bounded-l-pebblings}
    An
    \introduceterm{\nsboundedpebblingfull{\bpblackparam}{\bpwhiteparam}{}}
    is \alpebblingtext
    $\lpebbling = \setsmall{\lconf_0, \ldots, \lconf_{\stoptime}}$
    such that
    every 
    $\lconf_t$
    contains
    at most $\bpblackparam$ pebble \subconftext{}s
    $\scnot{v}{W}$
    and every
    $\scnot{v}{W}$
    has
    white support size
    $\setsize{W} \leq \bpwhiteparam$.
\end{definition}

Observe that if a graph $G$ with fan-in $\indegreedag$ has a
black-only pebbling strategy in time~$\stoptime$ and
space~$\spacestd$, then the \lpebblingtext simulating this strategy is
an  
\nsboundedpebbling{\spacestd+1}{\indegreedag} 
in time at most
$\stoptime (\indegreedag+1)$.%
\footnote{%
  Every black pebble in the black pebbling will 
  correspond to a \subconftext in the labelled pebbling plus that we need one extra
  \subconftext to remove the white pebbles in
  $\scnot{v}{\prednode{v}}$
  after an introduction move, so the total number of \subconftext{}s
  will be $\spacestd+1$.
  In a black pebbling every pebble except the one on the sink is both
  placed and removed. In a labelled pebbling, a black pebble placement
  on and removal from $v$ corresponds to one
  step for introducing
  $\scnot{v}{\prednode{v}}$,
  at most $\indegreedag$ steps for merging away all the white pebbles
  in $\prednode{v}$,
  at most $\indegreedag$ steps for erasing the intermediate
  \subconftext{}s, and one final step when
  $\scnot{v}{\emptyset}$ 
  is erased. Since the graph must contain at least one source vertex
  for which the $2 \indegreedag$ intermediate steps above are not needed,
  the stated time bound follows.}
Thus, the power of bounded \lpebblingtext is somewhere in between
black-only and black-white pebbling.

Note also that boundedness automatically implies low
space complexity,  since an
\nsboundedpebblingfull{\bpblackparam}{\bpwhiteparam}
$\lpebbling$ 
clearly satisfies
$\pebspace{\lpebbling} \leq \bpblackparam (\bpwhiteparam + 1)$.
%
And if we can design an
\nsboundedpebbling{\bpblackparam}{\bpwhiteparam}
for a graph~$G$, then such a pebbling can be used to refute
any pebbling contradiction 
$\genericpebcontr[G]{}$
in resolution by
constructing a refutation
mimicking~$\lpebbling$.

\begin{lemma}[\cite{Nordstrom11RelativeStrength}]
  \label{lem:resolution-simulates-bounded-pebblings}
  Suppose that
  $\lpebbling$
  is any \pebcomplete
  \nsboundedpebbling{\bpblackparam}{\bpwhiteparam}
  of   a DAG~$G$
  and that 
  \mbox{$\funcdescr{\funcpebc}{\zerooneset^{\pebdeg}}{\zerooneset}$}
  is any nonconstant Boolean function.
  Then there is a resolution refutation
  $\proofstd_{\lpebbling}$
  of
  $\genericpebcontr[G]{}$
  in
  length
  $
  \lengthofarg{\proofstd_{\lpebbling}}
  =
  \pebtime{\lpebbling}
  \cdot \exp \bigl(  \bigoh{\pebdeg \bpwhiteparam } \bigr)
  $
  and total space
  $
  \totspaceof{\proofstd_{\lpebbling}} =
  \bpblackparam \cdot \exp \bigl(  \bigoh{\pebdeg \bpwhiteparam } \bigr)
  $.
  In particular,
  fixing $\funcpebc$
  it holds that  resolution can simulate 
  \nsboundedpebbling{\bpblackparam}{\bigoh{1}}{}s
  in a time- and  space-preserving manner.    
\end{lemma}

The whole problem thus boils down to the question whether there are
graphs with (a)~asymptotically different properties for black
and black-white pebbling for which (b)~optimal black-white pebblings
can be carried out in the
\boundedpebblingfulltext
framework.
The answer to this question is positive, and using
\reflem{lem:resolution-simulates-bounded-pebblings}
one can prove that resolution can be strictly better than
black-only pebbling, both for time-space trade-offs and \wrt space in
absolute terms.
It turns out that  for all known separation results in the pebbling
literature where black-white pebbling does asymptotically better than
black-only pebbling, there are graphs exhibiting these separations for
which optimal black-white pebblings can be simulated by bounded
\lpebblingtext{}s. This means that resolution refutations of
pebbling contradictions over such DAGs can do asymptotically strictly
better than what is suggested by black-only pebbling, matching the
bounds in terms of black-white pebbling.

More precisely, such results can be obtained for 
(at least)
three families of graphs.
The first family are the 
so-called \introduceterm{bit reversal graphs},
for which Lengauer and Tarjan~%
\cite{LT82AsymptoticallyTightBounds}
established that black-white pebblings have quadratically better trade-offs
than black pebblings. More formally, they showed that
there are DAGs
$\set{G_n}_{n=1}^{\infty}$ 
of
size $\bigtheta{n}$
with black pebbling price 
$\pebblingprice{G_n} = 3$
such that
any optimal black pebbling $\pebbling_n$ of $G_n$ 
exhibits a trade-off
$
\pebtime{\pebbling_n} = 
\Bigtheta{n^2 / \pebspace{\pebbling_n} + n}
$
but optimal black-white pebblings
$\pebbling_n$ of $G_n$ 
achieve a    trade-off 
$
\pebtime{\pebbling_n} = 
\Bigtheta{(n / \pebspace{\pebbling_n})^2 + n}
$.

\begin{theorem}[\cite{Nordstrom11RelativeStrength}]
  \label{th:resolution-constant-trade-offs}
  Fix any non-constant Boolean function 
  $\funcpebc$
  and let
  $\genericpebcontr[G_n]{}$
  be pebbling contradictions over the 
  bit reversal graphs 
  $G_n$ of size~$\bigtheta{n}$
  in~%
  \cite{LT82AsymptoticallyTightBounds}.
  Then for any   monotonically nondecreasing function
  $s(n) = \bigoh{\sqrt{n}}$
  there are resolution refutations
  $\proofstd_n$
  of
  $\genericpebcontr[G_n]{}$
  in total space
  $\bigoh{s(n)}$
  and length
  $\Bigoh{n^2 / s(n)^2}$,
  beating the lower bound
  $\Bigomega{n^2 / s(n)}$
  for black-only pebblings of~$G_n$.
\end{theorem}

Let us next focus on
absolute bounds on space rather than time-space
trade-offs. Here  the best known separation between black and black-white
pebbling for polynomial-size graphs is the one shown by
Wilber~\cite{Wilber88WhitePebblesHelp}, who exhibited graphs
$\set{G(\spacestd)}_{\spacestd=1}^{\infty}$
of  size polynomial in~$\spacestd$
with black-white pebbling price
$\bwpebblingprice{G(\spacestd)} = \bigoh{\spacestd}$
and black pebbling price
$\pebblingprice{G(\spacestd)} =
\bigomega{\spacestd \log \spacestd / \log \log \spacestd}$.
For pebbling formulas over these graphs we do \emph{not} know how to
beat the black pebbling space bound---we return to this somewhat
intriguing problem below---but using instead the graphs with
essentially the same pebbling properties 
\ifthenelse{\boolean{maybeLMCS}}           
{constructed in~%
  \cite{KS91OnThePowerOfWhitePebbles},}
{constructed in~%
  \cite{KS91OnThePowerOfWhitePebbles}
  and covered in
  \refsec{sec:pebbling-space-separation},}
we can obtain the desired result.  

\begin{theorem}[\cite{Nordstrom11RelativeStrength}]
  \label{th:resolution-wilber}
  Fix any non-constant Boolean function 
  $\funcpebc$
  and let
  $\genericpebcontr[G(\spacestd)]{}$
  be pebbling contradictions over the graphs $G(\spacestd)$
  in~\cite{KS91OnThePowerOfWhitePebbles}
  with the same pebbling properties as
  in~\cite{Wilber88WhitePebblesHelp}.
  Then  there are resolution refutations
  $\proofstd_n$
  of
  $\genericpebcontr[G(\spacestd)]{}$
  in total space
  $\bigoh{s}$,
  beating the  lower bound
  $\bigomega{\spacestd \log \spacestd / \log \log \spacestd}$
  for black-only pebbling.
\end{theorem}

If we remove all restrictions on graph size,
there is a quadratic separation of black and black-white pebbling 
established by
Kalyanasundaram and Schnitger~\cite{KS91OnThePowerOfWhitePebbles}.
\ifthenelse{\boolean{maybeLMCS}}           
{They proved that}
{Recall again from 
  \refsec{sec:pebbling-space-separation}
  that they proved that} 
there are DAGs
$\set{G(\spacestd)}_{\spacestd=1}^{\infty}$
of  size
$\exp ( \bigtheta{\spacestd \log \spacestd} )$
such that
$\bwpebpricepersistent{G(\spacestd)} \leq 3 \spacestd + 1$
but
$\pebpricepersistent{G(\spacestd)} \geq \spacestd^2$.
For pebbling formulas over these graphs, resolution again
matches the black-white pebbling bounds.

\begin{theorem}[\cite{Nordstrom11RelativeStrength}]
  \label{th:resolution-ks}
  Fix any non-constant Boolean function 
  $\funcpebc$
  and let
  $\genericpebcontr[G(\spacestd)]{}$
  be pebbling contradictions over the graphs $G(\spacestd)$ in
  \cite{KS91OnThePowerOfWhitePebbles}
  exhibiting a quadratic separation of black and black-white pebbling.
  Then  there are resolution refutations
  $\proofstd_n$
  of
  $\genericpebcontr[G(\spacestd)]{}$
  in total space
  $\bigoh{s}$,
  beating the lower bound
  $\Bigomega{\spacestd^2}$
  for black-only pebbling.
\end{theorem}

Note that, in particular, this means that if we want to prove
\emph{lower bounds}
on resolution refutations of pebbling contradictions in terms of
pebble games, the best we can hope for in general are bounds expressed
in terms of black-white pebbling and not black-only pebbling.

Also, it should be noted that the best length-space separation that
could possibly be provided by pebbling contradictions are for formulas
of size~$\bigtheta{n}$ that are refutable in  length $\bigoh{n}$ 
but require clause space 
$\bigomega{n / \log n}$. 
\ifthenelse{\boolean{maybeLMCS}}
{This is so since
Hopcroft \etal~\cite{HPV77TimeVsSpace}
showed that} 
{This is so since as was discussed in
\refsec{sec:two-upper-bounds},
\cite{HPV77TimeVsSpace}~showed that} 
any graph of size~$n$ with bounded
maximal indegree has a black pebbling in space
$\bigoh{n / \log n}$. 
In fact, we can say more than that, namely that if any formula
$\fstd$ has a resolution refutation $\proofstd$ in length~$\lengthstd$, 
then it can be refuted in clause space
$\bigoh{\lengthstd / \log \lengthstd}$
(as was mentioned in 
\refsubsec{sec:pc-questions-relations-time-space}).
To see this, consider the graph representation $G_\proofstd$
of~$\proofstd$. 
By~\cite{HPV77TimeVsSpace}, 
this graph can be black-pebbled in space
$\bigoh{\lengthstd / \log \lengthstd}$.
It is not hard to see that we can construct another refutation that
simulates this pebbling  
$G_\proofstd$
by keeping exactly the clauses in memory that correspond to
black-pebbled vertices, and that this refutation will preserve the
pebbling space.%
\footnote{As a matter of fact, the original definition of the clause
  space of a resolution refutation in \cite{ET01SpaceBounds}
  was as the black pebbling price of the graph~$G_\proofstd$, 
  but (the equivalent)
  \refdef{def:length-and-space}
  as introduced by
  \cite{ABRW02SpaceComplexity}
  has turned out to be more convenient to work with for most purposes.}

In view of the results above, an intriguing open question is whether
resolution can \emph{always} simulate black-white pebblings, so that the
refutation space of pebbling contradictions is asymptotically equal to
the black-white pebbling price of the underlying graphs.

\begin{openproblem}[\cite{Nordstrom11RelativeStrength}]
  \label{open:bw-peb-achievable}
  Is it true for any DAG~$G$  with bounded vertex  indegree
  and any (fixed) Boolean function~$\funcpebc$
  that the pebbling contradiction
  $\genericpebcontr[G]{}$
  can be refuted in total space
  $\bigoh{\bwpebblingprice{G}}$?
\end{openproblem}

More specifically, one could ask---as a natural first line of attack
if one wants to investigate whether the answer to the above question
could be yes---if it holds that bounded labelled pebblings are in fact
as powerful as general black-white pebblings.  In a sense, this is
asking whether only a very limited form of nondeterminism is
sufficient to realize the full potential of black-white pebbling.

\begin{openproblem}[\cite{Nordstrom11RelativeStrength}]
  \label{open:bounded-pebbling-as-strong-as-bw-pebbling}
  Does it hold that any \pebcomplete black-white pebbling $\pebbling$
  of a \singlesinkdagtext~$G$ 
  with bounded vertex  indegree can be simulated by 
  a \nsboundedpebbling{\bigoh{\pebspace{\pebbling}}}{\bigoh{1}}
  $\lpebbling$?
\end{openproblem}

Note that a positive answer to this second question would immediately
imply a positive answer to the first question as well by
\reflem{lem:resolution-simulates-bounded-pebblings}.

We have no strong intuition either way regarding
Open Problem~\ref{open:bw-peb-achievable},
but as to 
Open Problem~\ref{open:bounded-pebbling-as-strong-as-bw-pebbling}
it would perhaps be somewhat surprising if bounded labelled pebblings
turned out to be as strong as general black-white pebblings.
Interestingly, although the optimal black-white pebblings of the
graphs in~%
\cite{KS91OnThePowerOfWhitePebbles}
can be simulated by 
\boundedpebblingtext{}s,
the same approach does \emph{not} work for
the original graphs separating black-white
from black-only pebbling in~%
\cite{Wilber88WhitePebblesHelp}.
Indeed, these latter graphs might be a candidate graph family 
for  answering 
Open Problem~\ref{open:bounded-pebbling-as-strong-as-bw-pebbling}
in the negative, \ie showing  that standard black-white pebblings can be
asymptotically stronger than bounded labelled pebblings.

\subsubsectionNOW{Techniques for Lower Bounds on Resolution Trade-offs}
\label{sec:pc-lower-bounds}

%
%

To prove lower bounds on resolution refutations in terms of  pebble games,
we need to construct a reduction from refutations to pebblings.
Let us again use formulas
$\substform{\pebcontr[G]{}}{\lor_2}$
to illustrate our reasoning.

For black pebbles, we can reuse the ideas above for transforming
pebblings into refutations and apply them in the other direction. That
is,  if the clause  
$v_1 \lor v_2$ 
is implied by the current content of the blackboard, 
we will let this correspond to a black pebble on $v$.  
A white pebble in a pebbling is a ``debt'' that has to be paid. It is
difficult to see how any clause could be a liability in the same way and 
therefore no set of clauses corresponds naturally to isolated white pebbles.
But if we think of white pebbles as assumptions that allow us to place
black pebbles higher up in the DAG, it makes sense to say that if the
content of the blackboard conditionally  implies
$v_1 \lor v_2$
given that we also assume the set of clauses
$\setdescr{w_1 \lor w_2}{w \in W}$ 
for some vertex set~$W$,
then this could be interpreted as a black pebble on~$v$ 
and white pebbles on the vertices in $W$.  
%
%
    
Using this intuitive correspondence, we can translate sets of clauses in a 
refutation of
$\substform{\pebcontr[G]{}}{\lor_2}$
into black and white pebbles in~$G$
as in 
\reffig{fig:intuition-pebbles-pyramid-height-2}.
To see this, note that
if we assume
$v_1 \lor v_2$ 
and
$w_1 \lor w_2$,
this assumption together with the clauses on the blackboard 
in \reffig{fig:intuition-pebbles-pyramid-height-2-a}
imply 
$y_1 \lor y_2$,
so $y$ should be black-pebbled and $v$ and~$w$ white-pebbled in
\reffig{fig:intuition-pebbles-pyramid-height-2-b}.
The vertex $x$ is also black since
$x_1 \lor x_2$ 
certainly is implied by the blackboard.
This translation from clauses to pebbles is arguably quite
straightforward, and furthermore it seems to yield well-behaved
black-white pebblings for all ``sensible'' resolution refutations of~%
$\substform{\pebcontr[G]{}}{\lor_2}$.
(What this actually means is that all refutations of pebbling
contradictions that we are able to come up with can be described as
simulations of \lpebblingtext{}s as defined in
\refdef{def:labelled-bw-pebble-game}, 
and for such refutations the reduction just sketched will essentially
give us back the pebbling we started with.)

\newlength{\clauseheight}
\settoheight{\clauseheight}{$\olnot{v}_2$}
\setlength{\clauseheight}{1.4\clauseheight}

\begin{figure}[t]
  \subfigure[Clauses on blackboard.]
  {
    \label{fig:intuition-pebbles-pyramid-height-2-a}
    \begin{minipage}[b]{.45\linewidth}
      \centering
      \begin{gather*}
        \left [
          \begin{array}{l}
            { x_1 \lor x_2 } 
            \rule{0pt}{\clauseheight}
            \\
            { \olnot{v}_1 \lor \olnot{w}_1 \lor y_1 \lor y_2 } 
            \rule{0pt}{\clauseheight}
            \\
            { \olnot{v}_1 \lor \olnot{w}_2 \lor y_1 \lor y_2 } 
            \rule{0pt}{\clauseheight}
            \\
            { \olnot{v}_2 \lor \olnot{w}_1 \lor y_1 \lor y_2 }
            \rule{0pt}{\clauseheight}
            \\
            { \olnot{v}_2 \lor \olnot{w}_2 \lor y_1 \lor y_2 }
            \rule{0pt}{\clauseheight}
          \end{array}
        \right ]
      \end{gather*}
    \end{minipage}%
  }
  \hfill
  \subfigure[Corresponding pebbles in the graph.]
  {
    \label{fig:intuition-pebbles-pyramid-height-2-b}
    \begin{minipage}[b]{.45\linewidth}
      \centering
      \includegraphics{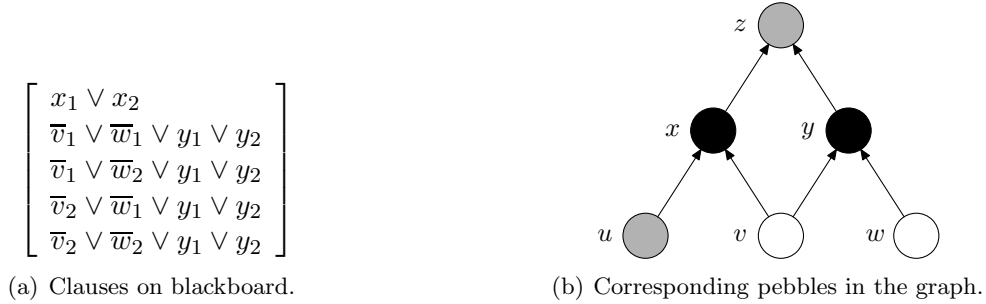}%
    \end{minipage}
  }
%
%
%
%
  \caption{Intuitive translation of clauses to black and white pebbles.}
  \label{fig:intuition-pebbles-pyramid-height-2}
\end{figure}

The problem, however, is that we have no guarantee that resolution
refutations will be ``sensible''.  Even though it might seem more or
less clear how an optimal refutation of a pebbling contradiction
should proceed, a particular refutation might contain unintuitive and
seemingly non-optimal derivation steps that do not make much sense
from a pebble game perspective. It can happen that clauses are derived
which cannot be translated, at least not in a natural way, to pebbles in the
fashion indicated above.

Some of these clauses we can afford to ignore. For example,
considering how axiom clauses can be used in derivations it seems
reasonable to expect that a derivation never writes an isolated axiom  
$\olnot{v}_i \lor \olnot{w}_j \lor y_1 \lor y_2$
on the blackboard. And in fact, if three of the four axioms for $v$ 
in
\reffig{fig:intuition-pebbles-pyramid-height-2}
are written on the blackboard but the fourth one
$\olnot{v}_2 \lor \olnot{w}_2 \lor
y_1 \lor y_2$
is missing, we will just discard these three clauses and there will be no
pebbles on $v$, $w$, and $y$ corresponding to them.

A more dangerous situation is when clauses are derived that are 
the disjunction of positive literals from different vertices.
For instance, a derivation starting from 
\reffig{fig:intuition-pebbles-pyramid-height-2-a}
could add the axioms
$\olnot{x}_1 \lor \olnot{y}_2  \lor z_1 \lor z_2$
and
$\olnot{x}_2 \lor \olnot{y}_2  \lor z_1 \lor z_2$
to the blackboard,
derive that the truth of
$v$ and~$w$ implies the truth of either $y$ or~$z$, \ie the clauses
$\olnot{v}_i \lor \olnot{w}_j \lor  y_1 \lor z_1 \lor z_2$
for $i,j = 1,2$,
and then erase
$x_1 \lor x_2$ 
to save space, resulting in the blackboard in
\reffig{fig:intuition-blobs-pyramid-height-2-a}.
As it stands, the content of this blackboard does not correspond to
any pebbles under our tentative translation. However, the clauses can
easily be used to derive something that does.
For instance, writing down
the axioms 
$\olnot{x}_i \lor \olnot{y}_j  \lor z_1 \lor z_2$, $i,j = 1, 2$,
on the blackboard, 
we get that the truth of 
$v$, $w$, and $x$ 
implies the truth of~$z$.
We have decided to interpret such a set of clauses as a black pebble
on $z$ and white pebbles on $v$, $w$, and $x$, but this pebble
configuration cannot arise out of nothing in an empty DAG.
Hence, the clauses in
\reffig{fig:intuition-blobs-pyramid-height-2-a}
have to correspond to some set of pebbles. But what pebbles?

%
%
%

\begin{figure}[t]
  \subfigure[New set of clauses on blackboard.]
  {
    \label{fig:intuition-blobs-pyramid-height-2-a}
    \begin{minipage}[b]{.5\linewidth}
      \centering
      \begin{gather*}
        \left [
          \begin{array}{l}
            { \olnot{v}_1 \lor \olnot{w}_1 \lor 
              y_1 \lor z_1 \lor z_2 } 
            \rule{0pt}{\clauseheight}
            \\
            { \olnot{v}_1 \lor \olnot{w}_2 \lor 
              y_1 \lor z_1 \lor z_2 } 
            \rule{0pt}{\clauseheight}
            \\
            { \olnot{v}_2 \lor \olnot{w}_1 \lor 
              y_1 \lor z_1 \lor z_2 }
            \rule{0pt}{\clauseheight}
            \\
            { \olnot{v}_2 \lor \olnot{w}_2 \lor 
              y_1 \lor z_1 \lor z_2 }
            \rule{0pt}{\clauseheight}
          \end{array}
        \right ]
      \end{gather*}
    \end{minipage}%
  }
  \hfill
  \subfigure[Corresponding blobs and pebbles.]
  {
    \label{fig:intuition-blobs-pyramid-height-2-b}
    \begin{minipage}[b]{.4\linewidth}
      \centering
      \includegraphics{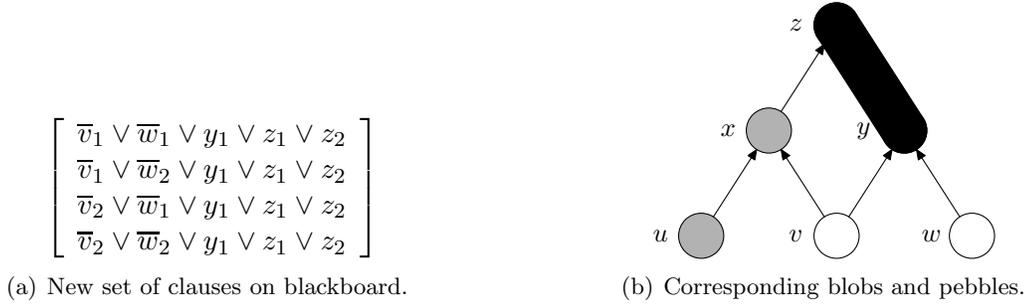}%
    \end{minipage}
  }
%
%
%
%
  \caption{Intepreting sets of clauses as black blobs and white pebbles.}
  \label{fig:intuition-blobs-pyramid-height-2}
\end{figure}

Although it is hard to motivate from such a small example, this turns
out to be a very serious problem. There appears to be no way that we
can interpret derivations as the one described above in terms of black
and white pebbles without making some component in the reduction from
resolution to pebbling break down.

So what can we do? Well, if  you can't beat 'em, join 'em!
In order to prove their results,
\cite{Nordstrom09NarrowProofsSICOMP, 
  NH13TowardsOptimalSeparation,
  BN08ShortProofs}
gave up the attempts to translate resolution refutations into
black-white pebblings and instead invented new pebble games (in three
different flavours). These pebble games are on the 
one hand somewhat similar to the black-white pebble game, but on the
other hand they have pebbling rules specifically designed
so that tricky clause sets such as that in
\reffig{fig:intuition-blobs-pyramid-height-2-a}
can be interpreted in a satisfying fashion. Once this has been taken care
of, one proceeds with the construction of the proof as outlined in
\refsubsec{sec:pc-time-vs-space},
but using the modified pebble games instead of standard
black-white pebbling. In what follows, we describe how this is done
employing the pebble game defined in
\cite{BN08ShortProofs}
(though using the more evocative terminology from
\cite{NH13TowardsOptimalSeparation}).
The games in
\cite{Nordstrom09NarrowProofsSICOMP, 
  NH13TowardsOptimalSeparation},
although somewhat different on the surface, can also be recast in the
framework presented below. 

The new pebble game in 
\cite{BN08ShortProofs}
is similar to the one in
\refdef{def:labelled-bw-pebble-game},
but with a crucial change in the definition of the
``\subconftext{}s.'' 
There are  white pebbles just as before, but the black pebbles are
generalized to \introduceterm{\blob{}s} that  can cover multiple
vertices instead of just a single vertex. A \blob on a vertex set $V$
can be thought of as truth of some vertex $v \in V$, unknown which.
The clauses in
\reffig{fig:intuition-blobs-pyramid-height-2-a}
are consequently translated into white pebbles on $v$
and~$w$, as before, and a black \blob covering both $y$ and~$z$ 
as in
\reffig{fig:intuition-blobs-pyramid-height-2-b}.
To  parse the formal definition of the game
given next, 
it might be helpful to 
study the examples in
\reffig{fig:blobpebblingmoves}.

\begin{figure}[t]   
  \centering
  \subfigure[Empty pyramid.]
  {
    \label{fig:blobpebblingmoves-a}
    \begin{minipage}[b]{.47\linewidth}
      \centering
      \includegraphics{blobpebblingmoves.2}%
    \end{minipage}
  }
  \hfill
  \subfigure[Introduction move.]
  {
    \label{fig:blobpebblingmoves-b}
    \begin{minipage}[b]{.47\linewidth}
      \centering
      \includegraphics{blobpebblingmoves.3}%
    \end{minipage}
  }

  \subfigure[Two \mpsctext{}s before merger.]
  {
    \label{fig:blobpebblingmoves-c}
    \begin{minipage}[b]{.47\linewidth}
      \centering
      \includegraphics{blobpebblingmoves.5}%
    \end{minipage}
  }
  \hfill
  \subfigure[The merged \mpsctext.]
  {
    \label{fig:blobpebblingmoves-d}
    \begin{minipage}[b]{.47\linewidth}
      \centering
      \includegraphics{blobpebblingmoves.6}%
    \end{minipage}
  }

  \subfigure[\Mpsctext before inflation.]
  {
    \label{fig:blobpebblingmoves-e}
    \begin{minipage}[b]{.47\linewidth}
      \centering
      \includegraphics{blobpebblingmoves.7}%
    \end{minipage}
  }
  \hfill
  \subfigure[\Mpsctext after inflation.]
  {
    \label{fig:blobpebblingmoves-f}
    \begin{minipage}[b]{.47\linewidth}
      \centering
      \includegraphics{blobpebblingmoves.8}%
    \end{minipage}
  }

%
%
%
%
  \caption{Examples of moves in the \multipebblegame.}
  \label{fig:blobpebblingmoves}
\end{figure}

%
%

\renewcommand{\rpconf}[1][P]{\pconf[#1]}
\renewcommand{\respebblingtext}{blob-pebbling\xspace}
\renewcommand{\Respebblingtext}{Blob-pebbling\xspace}
\renewcommand{\RESPEBBLINGTEXT}{Blob-Pebbling\xspace}
\renewcommand{\respebblinglongtext}{blob-pebbling\xspace}
\renewcommand{\Respebblinglongtext}{Blob-pebbling\xspace}
\renewcommand{\RESPEBBLINGLONGTEXT}{Blob-Pebbling\xspace}
\renewcommand{\respebbling}[1][P]{\pebbling[#1]}

%
%

\begin{definition}[Blob-pebble game \cite{BN08ShortProofs}]
  \label{def:res-pebbling-configuration}
  \label{def:res-pebbling-game}
  If $B$ and $W$ 
  are sets of vertices with 
  $B \neq \emptyset$,
  $B \intersectionSP W = \emptyset$,
  we say that
  $\mpscnot{B}{W}$
  is a 
  \introduceterm{blob \subconftext{}}
  with a black blob on $B$ and white pebbles on~$W$
  \introduceterm{supporting}~$B$.
%
%
  A \introduceterm{\respebblingtext{}}
  of a  DAG~$G$  with unique sink~$\sinkstd$ 
  is a sequence  
  $\respebbling =
  \Set{\rpconf_0, \ldots, \rpconf_{\stoptime}}$
  of blob \subconftext sets,   or 
  \introduceterm{\rpconftext{}s{}},
  \st
  $\rpconf_0 = \emptyset$,
  $\rpconf_\stoptime = \set{\mpscnot{\sinkstd}{\emptyset}}$,
  and   for all
  $t \in \intnfirst{\stoptime}$,
  $\rpconf_{t}$
  is obtained from
  $\rpconf_{t-1}$
  by one of the following rules:
  \begin{enumerate}[\hbox to8 pt{\hfill}]
    \italicitem\noindent{\hskip-12 pt\bf\textit{Introduction:}}\
    $\rpconf_{t} = \rpconf_{t-1} \unionSP
    \set{\rpscintro{v}}$.
    
    \italicitem\noindent{\hskip-12 pt\bf\textit{Merger:}}\
    $\rpconf_{t}  = 
    \rpconf_{t-1} \unionSP  
    \Set{\rpscnot{B_1 \unionSP B_2}{W_1 \unionSP W_2}}$
    \ifthenelse{\boolean{maybeLMCS}}
    {if there are blob subconfigurations\\}
    {if there are}
    $\rpscnot{B_1}{W_1 \unionSP \set{v}},
    \rpscnot{B_2 \unionSP \set{v}}{W_2}
    \in \rpconf_{t-1}$
    \st
    $B_1 \intersectionSP W_2 = \emptyset$.

    \italicitem\noindent{\hskip-12 pt\bf\textit{Inflation:}}\
    $\rpconf_{t} = \rpconf_{t-1} \unionSP
    \Set{\rpscnot{B \unionSP B'}{W \unionSP W'}}$
    if  $\rpscnotstd \in \rpconf_{t-1}$
    and
    $(B \unionSP B') \intersectionSP 
    \mbox{$(W \unionSP W')$} 
    = \emptyset$. 

    \italicitem\noindent{\hskip-12 pt\bf\textit{Erasure{}:}}\
    $\rpconf_{t} = \rpconf_{t-1} \setminus \Set{\rpscnotstd}$
    for
    $\rpscnotstd \in \rpconf_{t-1}$.
  \end{enumerate}
%
  \end{definition}

Let us now return to the proof outline in
\refsubsec{sec:pc-time-vs-space}.
The first step in our approach is to establish that any
resolution refutation of a pebbling contradiction can be interpreted
as a pebbling (but now in our modified game) of the DAG in terms of which this
pebbling contradiction is defined.  Intuitively, axiom downloads in
the refutation will be matched by introduction moves in the
blob-pebbling, erasures correspond to erasures, and seemingly
suboptimal derivation steps can be modelled by inflation moves in the
blob-pebbling. 
In all three papers
\cite{Nordstrom09NarrowProofsSICOMP, 
  NH13TowardsOptimalSeparation,
  BN08ShortProofs},
the formal definitions are set up so that a theorem along the
following lines can be proven.

\begin{theoremtemplate}
  \label{th:template-translation-derivations-to-pebblings}
  Consider a pebbling contradiction
  $\substform{\pebcontr[G]{}}{\funcpebc}$
  over any DAG~$G$. 
  Then there is a translation   function   from  sets of clauses 
  over $\Vars{\substform{\pebcontr[G]{}}{\funcpebc}}$ 
  to    sets of black blobs and white pebbles in~$G$ 
  that translates   any resolution   refutation   
  $\proofstd$
  of
  $\substform{\pebcontr[G]{}}{\funcpebc}$
  into a blob-pebbling~$\pebbling_{\proofstd}$
  of~$G$.
\end{theoremtemplate}

The next step is to show pebbling lower bounds.  Since the rules in
the blob-pebble game are different from those of the standard
black-white pebble game, known bounds on black-white pebbling price in
the literature no longer apply. But again, provided that we have got
the right definitions in place, we hope to be able to establish that
the blob-pebblings can do no better than standard black-white pebblings.

\begin{theoremtemplate}
  \label{th:template-lower-bound-pebbling-price}
  If there is a blob-pebbling of a DAG~$G$ in 
  time~$\stoptime$
  and space~$\spacestd$,
  then there is a standard black-white pebbling of~$G$ in 
  time~$\bigoh{\stoptime}$
  and space~$\bigoh{\spacestd}$.
\end{theoremtemplate}

Finally, we need to establish that the number of pebbles used in
$\pebbling_{\proofstd}$ in
\reftheoremtemplate{th:template-translation-derivations-to-pebblings}
is related to the space of the resolution refutation~$\proofstd$.
As we know from
\refsubsec{sec:pc-time-vs-space},
such a bound cannot be true for formulas
$\pebcontr[G]{}$
so this is where we need to do substitutions with some suitable
Boolean function $\funcpebc[\pebdeg]$ over 
$\pebdeg \geq 2$
variables and study
$\substform{\pebcontr[G]{}}{\funcpebc[\pebdeg]}$.

\begin{theoremtemplate}
  \label{th:template-pebbling-cost-bounded-by-space}
  If 
  $\proofstd$
  is a resolution refutation of a pebbling contradiction
  $\substform{\pebcontr[G]{}}{\funcpebc[\pebdeg]}$
  for some suitable Boolean function
  $\funcpebc[\pebdeg]$,
  then the time and space of the associated blob-pebbling
  $\pebbling_{\proofstd}$ of~$G$
  are upper bounded by $\proofstd$ by
  $
  \pebtime{\pebbling_{\proofstd}}
  =
  \bigoh{\lengthofarg{\proofstd}}
  $
  and
  $
  \pebspace{\pebbling_{\proofstd}}
  =
  \bigoh{\clspaceof{\proofstd}}
  $.
\end{theoremtemplate}

If we put these three theorems together, it is clear that we can
translate pebbling trade-offs to resolution trade-offs as described in the
``theorem template'' at the end of
\refsubsec{sec:pc-time-vs-space}.

There is a catch, however, which is why we have used the label
``\theoremtemplatename{}s'' above. 
It is reasonably straightforward to come up with natural definitions
that allow us to prove
\reftheoremtemplate{th:template-translation-derivations-to-pebblings}.
But this in itself does not yield any lower bounds. (Indeed, there is a
natural translation from refutations to pebbling even for
$\pebcontr[G]{}$, 
for which we know that the lower bounds we are after do \emph{not} hold!)
The lower bounds instead follow from the combination of
\reftwotheoremtemplates{th:template-lower-bound-pebbling-price}
{th:template-pebbling-cost-bounded-by-space},
but there is a tension between these two theorems.

The attentive reader might already have noted that two crucial details in
\refdef{def:res-pebbling-game}
are missing---we have not defined pebbling time and space for
blob-pebblings. And for a good reason, because this turns out to be
where the difficulty lies. On the one hand, we want the time and space
measures for blob-pebblings to be as strong as possible, so that we
can make
\reftheoremtemplate{th:template-lower-bound-pebbling-price}
hold, saying that blob-pebblings 
can do no better than
standard pebblings. On the other hand, we do not want the definitions
to be too  strong, for if so the bounds we need in
\reftheoremtemplate{th:template-pebbling-cost-bounded-by-space}
might break down. This turns out to be the major technical difficulty
in the construction

In the papers
\cite{Nordstrom09NarrowProofsSICOMP,NH13TowardsOptimalSeparation},
which study formulas
$\substform{\pebcontr[G]{}}{\lor_2}$
defined in terms of binary logical or,
we cannot make any connection between pebbling time and refutation
length in
\reftwotheoremtemplates{th:template-lower-bound-pebbling-price}
{th:template-pebbling-cost-bounded-by-space},
but instead have to focus on only clause space.
Also, the constructions work not for general DAGs but only for binary
trees in  
\cite{Nordstrom09NarrowProofsSICOMP},
and only for a somewhat more general class of graphs also including
pyramids in
\cite{NH13TowardsOptimalSeparation}.
The reason for this is that it is hard to charge for black blobs and
white pebbles. If we could charge for all vertices covered by blobs
and pebbles, or at least one space unit for every black blob and every
white pebble, we would be in good shape. But it appears hard to do so
without losing the connection to clause space that we want in
\reftheoremtemplate{th:template-pebbling-cost-bounded-by-space}.
Instead, for formulas
$\substform{\pebcontr[G]{}}{\lor_2}$
the best space measure that we can come up with is as follows.

\begin{definition}[\Blobpebblingtext price \wrt $\lor_2$]
  \label{def:blob-pebbling-price}
  Let
  $\pconf = \setdescr{\mpscnotstd[i]}{i=1, \ldots, n}$
  be a set of blob \subconftext{}s over some DAG~$G$.
 
  A
  \introduceterm{chargeable black blob collection}
  of $\pconf$ 
  is an ordered subset
  $\set{B_1, \ldots, B_m}$
  of black blobs in  $\pconf$
  \st for all $i \leq m$
  it holds that
  $B_i \setminus \Union_{j<i} B_j \neq \emptyset$
  (\ie the unions 
  $\Union_{j<i} B_j$
  are strictly expanding for $i=1, \ldots, m$). 
  We say that such a collection has 
  \introduceterm{black cost}~$m$.

  The set of \introduceterm{chargeable white pebbles}
  of a \subconftext $\mpscnotstd[i] \in \pconf$
  is the subset of vertices $w \in W_i$ that are below all
  $b \in B_i$ (where ``below'' means that  there is a path from $w$
  to~$b$ in~$G$). The
  \introduceterm{chargeable white pebble collection}
  of~$\pconf$ is the union of all such vertices for all
  $\mpscnotstd[i] \in \pconf$, and the 
  \introduceterm{white cost} is the size of this set.

  The \introduceterm{space} of a 
  \rpconftext $\pconf$
  is the largest black cost of a chargeable blob collection plus the
  cost of the chargeable white pebble collection, and the space of a
  blob-pebbling is the maximal space of any \rpconftext in it.
  The
  \blobpebblingtext price
  $\blobpebblingprice{G}$
  of a DAG~$G$  is the minimum space of any \pebcomplete{}
  \blobpebblingtext of~$G$. 
\end{definition}

Using the translation of clauses to blobs and pebbles in
\cite{BN08ShortProofs}
it can be verified that
\reftheoremtemplate{th:template-translation-derivations-to-pebblings}
as proven in that paper holds also for formulas
$\substform{\pebcontr[G]{}}{\lor_2}$.
Moreover, extending the proof techniques in 
\cite{Nordstrom09NarrowProofsSICOMP,NH13TowardsOptimalSeparation}
it is also not too hard to show the space bound in
\reftheoremtemplate{th:template-pebbling-cost-bounded-by-space}.%
\footnote{Although it is phrased in very different terms, what is shown in
  \cite{Nordstrom09NarrowProofsSICOMP,NH13TowardsOptimalSeparation}
  is essentially the somewhat more restricted result that if we charge
  only for the set of black vertices $V$ such that every $v \in V$ is the
  unique bottom black vertex in some \subconftext
  $\mpscnotstd$ 
  that have all vertices $b \in B$ topologically ordered (\ie the blob
  $B$ is a chain) and  only for supporting white pebbles 
  $w \in W$ 
  that are located below their bottom black vertex in such
  \subconftext{}s, then the space bound in 
  \reftheoremtemplate{th:template-pebbling-cost-bounded-by-space}.  
  holds. The proofs in
  \cite{Nordstrom09NarrowProofsSICOMP,NH13TowardsOptimalSeparation}
  extend to the more general definition of \blobpebblingtext space in
  \refdef{def:blob-pebbling-price},
  however.}
But we do not know how to establish the space part in
\reftheoremtemplate{th:template-lower-bound-pebbling-price}
for general DAGs. This is the part of the construction where
\cite{Nordstrom09NarrowProofsSICOMP}
works only for binary trees and
\cite{NH13TowardsOptimalSeparation}
can be made to work also for pyramids but not for general graphs.

The crucial new idea added in
\cite{BN08ShortProofs}
to make the approach outlined above work for general DAGs was to
switch formulas from
$\substform{\pebcontr[G]{}}{\lor_2}$
to 
$\substform{\pebcontr[G]{}}{\funcpebc}$
for other functions~$\funcpebc$ such as for instance 
binary exclusive or~$\xor_2$.
However, while this does make the analysis much simpler (and
stronger), it is not at all clear that the change of formulas should
be necessary. We find it an intriguing question whether the program in
Tentative Theorems~%
\ref{th:template-translation-derivations-to-pebblings},
\ref{th:template-lower-bound-pebbling-price},
and~%
\ref{th:template-pebbling-cost-bounded-by-space}
could in fact be carried out for formulas
$\pebcontrwithfunc[G]{}{\lor_2}$.

\begin{openproblem}[\cite{Nordstrom11RelativeStrength}]
  \label{open:or-pebbling-formulas}
  Is it true for any DAG~$G$ that
  any resolution refutation 
  $\proofstd$
  of
  $\pebcontrwithfunc[G]{}{\lor_2}$
  can be translated into a 
  black-white pebbling of $G$
  with time and space 
  upper-bounded in terms of
  the length and space
  of~$\proofstd$? 

  In particular, can we translate upper bounds in the blob-pebble game in
  \refdef{def:res-pebbling-game}
  with space defined as in
  \refdef{def:blob-pebbling-price}
  to upper bounds for standard black-white pebbling? (From which
  clause space lower bounds for
  $\pebcontrwithfunc[G]{}{\lor_2}$
  would immediately follow.)
\end{openproblem}

Our take on the results in
\cite{Nordstrom09NarrowProofsSICOMP,
  NH13TowardsOptimalSeparation}
is that they can be interpreted as  indicating that this should indeed
be the case. Although, as  noted above,
these results
apply only to limited classes of graphs, and only capture space lower
bounds and not time-space trade-offs, the problems arising in the
analysis seem to have to do more with artifacts in the proofs than
with any fundamental differences between formulas
$\pebcontrwithfunc[G]{}{\lor_2}$
and, say,
$\pebcontrwithfunc[G]{}{\xor_2}$.
We remark that the papers
\cite{BN08ShortProofs,BN11UnderstandingSpace}
do not shed any light on this question, as the techniques used there
inherently cannot work for formulas defined in terms of (non-exclusive)
logical or. 

If  
Open Problem~\ref{open:or-pebbling-formulas}
could be resolved in the positive, this could  potentially be useful
for settling the complexity of decision problems 
for resolution proof space, \ie the problem 
given a CNF formula~$\fstd$ and a space bound~$\spacestd$
to determine whether $\fstd$ has a resolution refutation
in space at most~$\spacestd$. Reducing from pebbling space 
by way of   
formulas $\pebcontrwithfunc[G]{}{\lor_2}$  would avoid the blow-up of the
gap between upper and lower bounds on pebbling space that cause
problems when using, for instance, exclusive or.

But let us return to the paper
\cite{BN08ShortProofs}
that resolves the problems identified in
\cite{Nordstrom09NarrowProofsSICOMP,
  NH13TowardsOptimalSeparation}.
The reason that we gain from switching from formulas
$\substform{\pebcontr[G]{}}{\lor_2}$
to, \eg, formulas
$\substform{\pebcontr[G]{}}{\xor_2}$
is that for the latter formulas we
can define a much stronger space measure for the
\blobpebblingtext{}s. In this case, it turns out that one can in fact
charge for all vertices covered by blobs or pebbles in the blob-pebble game,  
and then the space  bound in
\reftheoremtemplate{th:template-pebbling-cost-bounded-by-space}
follows for arbitrary DAGs. In the follow-up work
\cite{BN11UnderstandingSpace}
this result was improved to capture not only space but also the
connection between pebbling time and refutation length, thus realizing
the full program described in  
\refsubsec{sec:pc-time-vs-space}.

In this process, 
\cite{BN11UnderstandingSpace}~also presented a much cleaner way to 
argue more generally about how the refutation length and space  of a
CNF formula~$\fstd$ change when we do substitution with some Boolean
function~$\funcpebc$ to obtain~$\substform{\fstd}{\funcpebc}$. 
Since we believe that this is an interesting result in its own
right, we give an exposition of it in
\refsubsec{sec:pc-substitution-theorem}.
Before doing so, we want to conclude the current discussion by giving
some examples from
\cite{BN08ShortProofs,BN11UnderstandingSpace}
of the kind of results obtained by these techniques.

\subsubsectionNOW{Statement of Space Lower Bounds and Length-Space Trade-offs}
\label{sec:pc-statement-of-results}

Regarding the question of the relationship between length and space in
resolution, 
\cite{BN08ShortProofs}~showed that in contrast to the tight relation
between length versus width, length and space are as uncorrelated as
they can possibly be.

\begin{theorem}[Length-space separation for resolution 
    \cite{BN08ShortProofs}]
  \label{th:length-space-separation}
  There exist explicitly constructible families   of \kcnfform{}s 
  $\set{\fstd_n}_{n=1}^{\infty}$
  of size $\bigtheta{n}$ 
  that can be refuted in resolution in length $\bigoh{n}$ and width
  $\bigoh{1}$   simultaneously, but for which any  resolution
  refutation must have clause space
  $\bigomega{n / \log n}$.
\end{theorem}

An extension of this theorem to \kdnf resolution
in~\cite{BN11UnderstandingSpace}
showed that this family of proof systems does indeed form a strict
hierarchy \wrt space.

\begin{theorem}[$k$-DNF resolution space hierarchy
  \cite{BN11UnderstandingSpace}]
  \label{th:resk-space-hierarchy}
  For every $k\geq 1$ there exists an explicitly constructible family
  $\set{\fstd_n}_{n=1}^{\infty}$
  of CNF formulas of size $\bigtheta{n}$ and width $\bigoh{1}$
  such that
    there are $\reskone$-refutations
    $\refof{\proofstd_n}{\fstd_n}$ 
    in simultaneous length 
    $\lengthofarg{\proofstd_n}= \bigoh{n}$
    and formula space
    $\clspaceof{\proofstd_n} = \bigoh{1}$, but
    any   $\resknot$-refutation of $\fstd_n$ requires formula space
    $\Bigomega{\sqrt[k+1]{n/\log n}}$. 
  The constants hidden by the asymptotic notation depend only on $k$.
\end{theorem}

\ifthenelse{\boolean{maybeLMCS}}
{The formula families $\set{\fstd_n}_{n=1}^{\infty}$ in 
  \reftwoths{th:length-space-separation}{th:resk-space-hierarchy}
  are obtained by considering
  pebbling formulas defined in terms of the  graphs in
  \cite{GT78VariationsPebbleGameVerboseURL}}
{The formula families $\set{\fstd_n}_{n=1}^{\infty}$ in 
  \reftwoths{th:length-space-separation}{th:resk-space-hierarchy}
  are obtained by considering
  pebbling contradictions over the  graphs examined in
  \refsec{sec:optimal-lower-bound}{}}
requiring black-white pebbling space
$\bigtheta{n / \log n}$,
and substituting a $k$-non-authoritarian Boolean function
$\funcpebc$ of arity $k+1$, for instance XOR over $k+1$
variables, in these formulas.

The above theorems give absolute lower bounds on space for resolution
and~$\resknot$. 
Applying
the techniques in
\cite{BN11UnderstandingSpace} 
we can also derive length-space trade-offs for these proof systems.
In fact, we can obtain a multitude of such trade-offs, since for any graph
family with tight dual trade-offs for black and black-white pebbling,
or for which black-white pebblings can be cast in the framework of
\refsubsec{sec:pc-upper-bounds}
and simulated by resolution, we can obtain a corresponding trade-off
for resolution-based proof systems. Since a full catalogue listing all
of these trade-off results would be completely unreadable, we try to
focus on some of the more salient examples below.

From the point of view of space complexity, the easiest formulas are
those refutable in constant total space, \ie formulas having so
simple a structure that there are resolution refutations where we never
need to keep more than a constant number of symbols on the proof
blackboard. A priori, it is not even clear whether we should expect
that any trade-off phenomena could occur for such formulas, but it
turns out that there are quadratic length-space trade-offs.

\begin{theorem}[Quadratic trade-offs for constant space
    \cite{BN11UnderstandingSpace}]
  \label{th:tradeoffs-informal-quadratic}
  For any fixed positive integer
  $\kdnfboundsconst$ 
  there are explicitly constructible CNF formulas
  $\set{\fstd_n}_{n=1}^{\infty}$
  of size $\bigtheta{n}$ and width $\bigoh{1}$
  such that the following holds
  (where all multiplicative constants hidden in the asymptotic
  notation depend only on  $\kdnfboundsconst$):
  \begin{iteMize}{$\bullet$}
  \item 
    The formulas $\fstd_n$ are refutable in syntactic resolution
    in total space 
    $
    \mbox{$\totspaceref[\resnot]{\fstd_n}$} 
    \!=\! 
    \bigoh{1}$.
    
  \item 
    For any monotone function $\spacefuncupper(n) = \Bigoh{\sqrt{n}}$
    there are syntactic resolution refutations
    $\proofstd_n$ of~$\fstd_n$
    in simultaneous length
    $\lengthofarg{\proofstd_n} = \Bigoh{(n/\spacefuncupper(n))^2}$
    and total space 
    $\totspaceof{\proofstd_n} = \bigoh{\spacefuncupper(n)}$.

  \item 
    For any semantic resolution refutation
    $\refof{\proofstd_n}{\fstd_n}$
    in clause space
    $\clspaceof{\proofstd_n} \leq \spacefuncupper(n)$
    it holds that
    $\lengthofarg{\proofstd_n} =
    \Bigomega{(n/ \spacefuncupper(n))^2}$.
    
  \item 
    For any 
    $k \leq \kdnfboundsconst$,    
    any semantic 
    \kdnf resolution refutation
    $\proofstd_n$ of $\fstd_n$
    in formula space
    $\clspaceof{\proofstd_n} \leq \spacefuncupper(n)$
    must have length
    $\lengthofarg{\proofstd_n} = 
    \BIGOMEGA{\bigl( n / \spacefuncupper(n)^{k+1} \bigr)^{2}}
    $.
    In particular, any
    semantic
     constant-space \mbox{$\resknot$-refu}\-ta\-tion
    must also have quadratic length.
  \end{iteMize}
\end{theorem}

\noindent \Refth{th:tradeoffs-informal-quadratic} follows by combining the
techniques to be discussed in
\refsubsec{sec:pc-substitution-theorem}
\ifthenelse{\boolean{maybeLMCS}}           
{with the seminal work on pebbling trade-offs by  Lengauer and Tarjan~%
  \cite{LT82AsymptoticallyTightBounds}
  and}
{with the pebbling trade-offs 
  in
  \refsec{sec:constant-space-trade-offs}
  and}
the structural results on simulations of black-white pebblings by
resolution in
\refth{th:resolution-constant-trade-offs}.

\begin{remark}
  Notice that the trade-off applies to both formula space---\ie clause
  space for $\resknot[1]$---and total 
  space. This is because the upper bound is stated in terms of the
  larger of these two measures (total space) while the lower bound is in
  terms of the smaller one (formula space).  Note also that the upper
  bounds hold for the usual, syntactic versions of the proof systems,
  whereas the lower bounds hold for the much stronger semantic systems,
  and that for standard resolution the upper and lower bounds are tight
  up to constant factors.  These properties hold for all trade-offs stated below.
  Finally, we remark that we have to pick some arbitrary but fixed limit
  $\kdnfboundsconst$ for the size of the terms when stating the results
  for \kdnf resolution, since for any family of formulas we consider
  there will be very length- and space-efficient
  \mbox{$\resknot$-refutation} refutations if we allow terms of unbounded size.
\end{remark}

Our next example is based on the pebbling trade-off result 
\ifthenelse{\boolean{maybeLMCS}}                      
{in~\cite{Nordstrom11RelativeStrength}, 
  building on earlier work by Carlson and Savage~%
  \cite{CS80GraphPebbling, CS82ExtremeTimeSpaceTradeoffs}.}
{in \refsec{sec:non-constant-space-trade-offs}.}
Using this new result, we can derive among other things the rather striking
statement that for any \emph{arbitrarily slowly growing} 
non-constant function, there are explicit formulas of such
(arbitrarily small) space complexity that nevertheless exhibit
\emph{superpolynomial}
length-space trade-offs.

\begin{theorem}[Superpolynomial trade-offs for arbitrarily slowly
    growing space \cite{BN11UnderstandingSpace}]
  \label{th:tradeoffs-informal-super-constant}
  Let $\spacefunclower(n) = \littleomega{1}$ be any arbitrarily slowly growing function%
  \footnote{For technical reasons, we also assume that
    $\spacefunclower(n) = \Bigoh{n^{1/7}}$, 
    \ie that
    $\spacefunclower(n)$
    does not grow too quickly. This restriction is inconsequential
    since for faster-growing functions the trade-off results presented
    below yield much stronger bounds.}
  and fix any  $\epsilon >0$
  and positive integer
  $\kdnfboundsconst$.
  Then there are explicitly constructible CNF formulas
  $\set{\fstd_n}_{n=1}^{\infty}$
  of size $\bigtheta{n}$ and width $\bigoh{1}$
  such that the following holds:
  \begin{iteMize}{$\bullet$}
  \item 
    The formulas $\fstd_n$ are refutable in syntactic resolution
    in total space 
    $\mbox{$\totspaceref[\resnot]{\fstd_n}$}
    \!=\! \bigoh{\spacefunclower(n)}$.
    
  \item 
    There are syntactic resolution refutations
    $\proofstd_n$ of~$\fstd_n$
    in simultaneous length
    $\lengthofarg{\proofstd_n} = \bigoh{n}$ 
    and total space 
    $\totspaceof{\proofstd_n} 
    = \BIGOH{\bigl(n / \spacefunclower(n)^2\bigr)^{1/3}}$.
    
  \item 
    Any semantic resolution refutation of $\fstd_n$ in clause space
    $\BIGOH{\bigl( n / \spacefunclower(n)^2\bigr)^{1/3 - \epsilon}}$
    must have \superpoly length.
    
  \item 
    For any  $k \leq \kdnfboundsconst$,    
    \ifthenelse{\boolean{maybeLMCS}}                      
    {any semantic \kdnf resolution refutation}
    {any semantic $\resknot$-refutation}
    of $\fstd_n$
    in formula space
    $\BIGOH{\bigl( n / \spacefunclower(n)^2\bigr)^{1/(3(k+1)) - \epsilon}}$
    must have \superpoly length.
  \end{iteMize}
  All multiplicative constants hidden in the asymptotic
  notation depend only on  $\kdnfboundsconst$, $\epsilon$ and 
  $\spacefunclower$.

\end{theorem}

Observe the robust nature of this trade-off, which is displayed by the
long range of space complexity in standard resolution, from
$\littleomega{1}$ 
up to 
$\approx n^{1/3}$, 
which requires \superpoly length.  Note also that the trade-off result
for standard resolution is very nearly tight in the sense that the
superpolynomial lower bound on length in terms of space reaches up to
very close to where the linear upper bound kicks in.

The two theorems above focus on trade-offs for formulas of low space
complexity, and the lower bounds on length obtained in the trade-offs
are somewhat weak---the superpolynomial growth in
\refth{th:tradeoffs-informal-super-constant}
is something like $n^{\spacefunclower(n)}$.
We next present a theorem that has both a stronger superpolynomial
length lower bounds than 
\refth{th:tradeoffs-informal-super-constant}
and an even more robust trade-off covering a wider (although
non-overlapping) space interval. 
\ifthenelse{\boolean{maybeLMCS}}                      
{This theorem again follows by applying our tools to the  pebbling trade-offs in~%
\cite{LT82AsymptoticallyTightBounds}.}
{This theorem again follows by applying our tools to the  pebbling
  trade-offs in
  \refsec{sec:robust-trade-offs}
  established by~\cite{LT82AsymptoticallyTightBounds}.}

\begin{theorem}[Robust superpolynomial trade-off for 
    \ifthenelse{\boolean{maybeLMCS}}                      
    {medium}
    {medium-range}
    space \cite{BN11UnderstandingSpace}]
  \label{th:robust-trade-off-LT-informal}
  For any positive integer~$\kdnfboundsconst$, 
  there are explicitly constructible CNF formulas
  $\set{\fstd_n}_{n=1}^{\infty}$
  of size $\bigtheta{n}$ and width $\bigoh{1}$
  such that the following holds
  (where the hidden constants depend only on~$\kdnfboundsconst$):
  \begin{iteMize}{$\bullet$}
  \item 
    The formulas $\fstd_n$ are refutable in syntactic resolution
    in total space 
    $\mbox{$\totspaceref[\resnot]{\fstd_n}$}
    \!=\! \bigoh{\log^2 n}$.
    
  \item 
    There are syntactic resolution refutations
    of~$\fstd_n$
    in length
    $\bigoh{n}$ 
    and total space 
    $\bigoh{n / \log n}$.
    
  \item 
    Any semantic resolution refutation of $\fstd_n$ in clause space
    $\clspaceof{\proofstd_n} = \littleoh{n / \log n}$
    must have length
    $\lengthofsmall{\proofstd_n} =     n^{\bigomega{\log \log n}}$.

  \item 
    For any  $k \leq \kdnfboundsconst$,    
    any semantic \kdnf resolution refutation
    of~$\fstd_n$
    in formula space
    $\clspaceof{\proofstd_n} =
    \LITTLEOH{(n / \log n)^{1 / (k+1)}}$
    must have length
    $\lengthofsmall{\proofstd_n} =     n^{\bigomega{\log \log n}}$.
  \end{iteMize}
\end{theorem}

\noindent Having presented trade-off results in the low-space and medium-space
range, we conclude by presenting a result at the other end of the
space spectrum. Namely,  appealing one last time to yet another result
in~\cite{LT82AsymptoticallyTightBounds},
\ifthenelse{\boolean{maybeLMCS}}                  
{we can deduce}
{this time from
  \refsec{sec:exponential-trade-offs},
  we can deduce}
that there are formulas of nearly linear space complexity
(recall that any formula is refutable in linear formula space) that exhibit
not only superpolynomial but even exponential trade-offs.

We state this final theorem for standard resolution only since it is
not clear whether it makes sense for~$\resknot$. That is, we can certainly derive
formal trade-off bounds in terms of the $(k+1)$st square root
as in the theorems above, but we do not know whether
there actually exist $\resknot$-refutation in sufficiently small space so
that the trade-offs apply. Hence, such trade-off claims for~$\resknot$, 
although impressive-looking, might simply be vacuous.
It is possible to obtain other exponential trade-offs for $\resknot$
but they are not quite as strong as the result below for
resolution. We refer to 
\cite{BN11UnderstandingSpace}
for the details.

\begin{theorem}[Exponential trade-offs for nearly-linear space
    \cite{BN11UnderstandingSpace}]
  \label{th:tradeoffs-informal-exponential}
  Let $\largeconstant$ be any sufficiently large constant.
  Then   there are \cnfform{}s $\fstd_n$ of size
  $\bigtheta{n}$ and width $\bigoh{1}$
  and a constant 
  $\largeconstant' \ll \largeconstant$
  such that:
  \begin{iteMize}{$\bullet$}
  \item
    The formulas
    $\fstd_n$
    have syntactic resolution refutations in total space
    $\largeconstant' \cdot n/\log n$.
  \item
    $\fstd_n$ is also refutable in syntactic resolution in length
    $\bigoh{n}$
    and total space
    $\bigoh{n}$
    simultaneously.
    \item
      However, any semantic
      refutation
      of $\fstd_n$ in clause space at most
      $\largeconstant \cdot n/\log n$
      has   
      length
      $\exp\bigl(n^{\bigomega{1}}\bigr)$.
      
  \end{iteMize}
\end{theorem}

\noindent To get a feeling for this last trade-off result, note again that
the lower bound holds for proof systems with arbitrarily
strong derivation rules, as long as they operate with 
disjunctive clauses. 
In particular, it holds for proof systems that can in one step derive
anything that is semantically implied by the current content of the
blackboard. Recall that such a proof system can refute any
unsatisfiable \cnfform $\fstd$ with $n$~clauses in length
$n+1$ simply by writing down all clauses of $\fstd$ on the blackboard and then
concluding, in one single derivation step, the contradictory empty
clause implied by~$\fstd$.
In 
\refth{th:tradeoffs-informal-exponential}
the semantic resolution proof system has space nearly sufficient for
such an ultra-short refutation of the whole formula. But even so, when
we feed this proof system the formulas~$\fstd_n$ 
and restrict it to having at most
$\bigoh{n / \log n}$
clauses on the blackboard at any one given time, it will have to keep
going for an exponential number of steps before it is finished.

\subsectionNOW{Deriving Time-Space Trade-offs via the Substitution Theorem}
\label{sec:pc-substitution-theorem}

A paradigm that has turned out to be useful in many contexts in proof
complexity is to take a CNF formula family
$\set{\fstd_n}_{n=1}^{\infty}$
with interesting properties, tweak it by substituting some function
$\funcpebc(\varx_1, \ldots, \varx_\pebdeg)$ 
for each variable~$\varx$,
and then use this new formula family to prove the desired
result. Although this approach often is not made explicit in the
respective papers, most of the results reviewed in Sections~%
\ref{sec:pc-pebcontrs-and-proof-cplx}
and~%
\ref{sec:pc-tradeoffs}
can be viewed as applying variations on this theme to pebbling formulas.

Another example of this approach is the 
observation by
Alekhnovich and Razborov,
referenced (and used) in
\cite{Ben-Sasson09SizeSpaceTradeoffs,BOP07Complexity},
that if we take a CNF formula $\fstd$
and apply substitution with binary exclusive or~$\xor_2$, then the
length of refuting the substituted formula~%
$\substform{\fstd}{\xor_2}$
in resolution
is exponential in the
refutation width of the original formula, \ie
$
\lengthref{\substform{\fstd}{\xor_2}}
=
\exp \bigl( \bigomega{\widthref{\fstd}} \bigr)
$.
The proof is by applying a random restriction to the variables of
$\substform{\fstd}{\xor_2}$
by picking one variable~$\varx_i$ from each pair
$\varx_1, \varx_2$
and setting this variable to a random value. This restriction gives us
back the original formula~$\fstd$ (possibly after 
flipping polarity of 
literals), and also eliminates all wide clauses in a refutation with
high probability. Since restrictions preserve resolution refutations,
we can conclude that if there is no narrow refutation of~$\fstd$, then
there cannot be a short refutation of
$\substform{\fstd}{\xor_2}$.

%
%
%

If we take a closer look at the space lower bounds and length-space
trade-offs for substituted pebbling contradictions in
\refsubsec{sec:pc-statement-of-results},
it turns out that the only fact we need about the pebbling formulas
is that they have linear-length proofs
in small width
but that there is a weak
trade-off between length and variable space. The rest of the argument
can be seen to be totally oblivious to the fact that we are dealing
with pebbling formulas. The only property we use for these formulas is
the trade-off between length and variable space, and that substitution with
the right function~$\funcpebc$ can be used to lift these
trade-offs to length versus the much stronger measure clause space.

In this section, we want to give a clean exposition of how
substitution in CNF formulas can be used to amplify length-space
trade-offs for resolution-based proof systems. We believe that these
results are interesting in their own right, and that they can
potentially open the way for similar results for even stronger proof
systems.

\subsubsectionNOW{Preserving Upper Bounds for Substituted Formulas}
\label{sec:pc-subst-upper bounds}

If we want to use substitution to prove tight trade-offs, we need to
show that the substituted formulas become harder but not too hard,
since we want to be able to establish matching upper bounds.
%
It is straightforward to show that if
$\fstd$ is easy for resolution, then any substitution formula
$\substform{\fstd}{\funcpebc}$
is also easy in the following sense.

\begin{lemma}[\cite{BN11UnderstandingSpace}]
  \label{lem:from-original-formula-to-substitution-formula}
  Suppose that $\fstd$ is an unsatisfiable CNF formula and that 
  \ifthenelse{\boolean{maybeLMCS}}                  
  {$\funcpebc$ is a non-constant Boolean function of arity~$\pebdeg$.}
  {\mbox{$\funcdescr{\funcpebc}{\zerooneset^{\pebdeg}}{\zerooneset}$}
    is a non-constant Boolean function.}
  If there is a resolution
  refutation
  $\proofstd$ of~$\fstd$
  in length
  $\lengthofarg{\proofstd} = \lengthstd$,
  clause space
  $\clspaceof{\proofstd} = \spacestd$,
  and width
  $\widthofarg{\proofstd} = \widthstd$,
  then there is also a resolution refutation
  $\proofstd_{\funcpebc}$ of $\substform{\fstd}{\funcpebc}$
  in length
  $\Lengthofarg{\proofstd_{\funcpebc}} = \lengthstd 
  \cdot \exp ( \bigoh{\pebdeg \widthstd} )$,
  clause space
  $\Clspaceof{\proofstd_{\funcpebc}} = \spacestd
  \cdot \exp ( \bigoh{\pebdeg \widthstd} )$,
  and width
  $\Widthofarg{\proofstd_{\funcpebc}} = \bigoh{ \pebdeg \widthstd }$.
\end{lemma}

In particular, if the refutation
$\refof{\proofstd}{\fstd}$
has constant width, then it is possible to refute
$\substform{\fstd}{\funcpebc}$
with only a constant factor blow-up in length and space as compared
to~$\proofstd$ (where this constant depends on
$\widthofarg{\proofstd}$ and~$\funcpebc$). 
Of course, the same holds for any sequential proof
system $\psstd$ that can simulate resolution proofs 
efficiently line by line, such as, \eg, cutting planes or 
polynomial calculus resolution. 

The upper bounds for
$\substform{\fstd}{\funcpebc[]}$
in 
\reflem{lem:from-original-formula-to-substitution-formula}
are not hard to show. Briefly, the proof is as follows.
Given a resolution refutation $\proofstd$  of~$\fstd$,
we construct a refutation
$\refof{\proofstd_{\funcpebc}}{\substform{\fstd}{\funcpebc[]}}$
mimicking the derivation steps in~$\proofstd$.
When
$\proofstd$
downloads an axiom~$\clc$, we download all the axiom clauses in~%
$\substform{\clc}{\funcpebc[]}$, 
which is a set of at most
$
\exp
\bigl(
\bigoh{\pebdeg \cdot \widthofarg{\clc}}
\bigr)
$
clauses.
When
$\proofstd$
resolves
$\clc_1 \lor \varx$
and
$\clc_2 \lor \olnot{\varx}$
to derive
$\clc_1 \lor \clc_2$,
we use the fact that resolution is implicationally complete to derive
$\substform{(\clc_1 \lor \clc_2)}{\funcpebc[]}$
from
$\substform{(\clc_1 \lor \varx)}{\funcpebc[]}$
and
$\substform{(\clc_2 \lor \olnot{\varx})}{\funcpebc[]}$
in at most
$
\exp \bigl(\bigoh{\pebdeg \cdot \widthofarg{\clc_1 \lor \clc_2}} \bigl)
$
steps.
When a clause~$\clc$ is erased in~$\proofstd$, we erase all clauses
$\substform{\clc}{\funcpebc[]}$
in
$\proofstd_{\funcpebc[]}$.
We refer to
\cite{BN11UnderstandingSpace}
for the formal details.

A more interesting question is under which circumstances moderate
hardness results for~$\fstd$ can be amplified to stronger hardness
results for~$\substform{\fstd}{\funcpebc[]}$. In what follows, we will
describe a general framework in which weak space lower bounds and
length-space trade-offs for resolution can be lifted to strong lower
bounds and trade-offs for resolution or even more powerful proof
systems. This framework is just a straightforward generalization and
unification of what was done in
\cite{BN11UnderstandingSpace},
but we believe that the more general language below  helps us to
uncover the essence of the argument and makes it easier to see the
potential for (as well as the problems with) strengthening the results in
\cite{BN11UnderstandingSpace}
even further.

\subsubsectionNOW{Projections}
\label{sec:pc-projection}

The idea behind our approach is as follows: Start with a CNF formula~%
$\fstd$ which has some nice properties for resolution. Consider now
some proof system~$\psstd$ (which might be resolution, or some other,
stronger system) and study the substitution formula~%
$\substform{\fstd}{\funcpebc}$,
where we have chosen $\funcpebc$ to have the right properties
\wrt~$\psstd$.
Let 
$\proofstd_{\funcpebc}$
be any $\psstd$-refutation of 
$\substform{\fstd}{\funcpebc}$.
Intuitively, we want to argue that the best thing $\psstd$ can do is
to mimic a resolution refutation $\proofstd$ of~$\fstd$ as described
in the proof sketch for
\reflem{lem:from-original-formula-to-substitution-formula}
above. We then observe that in such simulations of
resolution refutations of~$\fstd$, the weak lower bounds for~$\fstd$ are
blown up to strong lower bounds for
$\substform{\fstd}{\funcpebc}$.

Formally speaking, we cannot really hope to prove this, since it is
hard to place restrictions on what $\psstd$ might or might not do when
refuting  
$\substform{\fstd}{\funcpebc}$.
However, what we can do is to argue that whatever a $\psstd$-refutation 
$\proofstd_{\funcpebc}$ 
of
$\substform{\fstd}{\funcpebc}$
looks like, we can
\emph{extract} from this~%
$\proofstd_{\funcpebc}$ 
a resolution refutation $\proofstd$ of~$\fstd$.
Our way of doing this is to define
\introduceterm{\projectionsubst{}s}
of arbitrary Boolean function over
$\varsdary{\pebdeg}{V}$ 
to clauses over~$V$, and to show that such projections translate
$\psstd$-refutations to resolution refutations.

Roughly, our intuition for \projectionsubst{}s is that if, \eg,
a
$\psstd$-configuration
$\clsd$ 
implies  
$\funcpebc(\varvecx) \lor \lnot \funcpebc(\varvecy)$,
then this should project the clause
$\varx \lor \olnot{\vary}$.
It will be useful for us, however, to relax this requirement a bit and
allow other definitions of \projectionsubst{}s as well as long as they
are ``in the same spirit.''
We specify next which formal properties a projection must satisfy
for  our approach to work. For this definition, it will be convenient
to have a compact notation for evaluating either $\funcpebc$ or the
complement of $\funcpebc$. Extending our notation for literals,  where
$\varx^1 = \varx$ 
and
$\varx^0 = \lnot \varx$,
we define
$\signedfunc[\funcpebc]{1}{\varvecx} = \funcpebc(\varvecx)$
and
$\signedfunc[\funcpebc]{0}{\varvecx} = \lnot \funcpebc(\varvecx)$.
Also for compactness, 
in what follows below
we will sometimes write
$\varx^{\truthval} \in \clc$
instead of 
$\varx^{\truthval} \in \lit{\clc}$
for a literal~$\varx^{\truthval}$ occuring in a clause~$\clc$.

\begin{definition}[Semantic \projectionsubst{}]   
  \label{def:f-projection}
  Let
  \mbox{$\funcdescr{\funcpebc}{\zerooneset^{\pebdeg}}{\zerooneset}$}
  be a fixed Boolean function. 
  Let $\psstd$ be a sequential proof system,
  and let $\clsd$ denote arbitrary sets of Boolean functions over
  $\varsdary{\pebdeg}{V}$ 
  of the form specified by~$\psstd$.
  Let $\clsc$ denote arbitrary sets of disjunctive clauses over~$V$. 
  Then the function
  $\projclnot[\funcpebc]$
  mapping set of Boolean functions $\clsd$ to clauses~$\clsc$
  is a
  \introduceterm{semantic \fprojsubst{}}, 
  or in what follows just an 
  \introduceterm{\fprojsubst{}}
  for short, if it is:
  \begin{enumerate}[\hbox to8 pt{\hfill}]     

   \item\noindent{\hskip-12 pt\bf Complete:}\ 
    If
    $
    \clsd
    \impl
    \Lor_{\varx^{\truthval} \in \clc} \signedfunc[\ff]{\truthval}{\varvecx}
    $
    then the clause $\clc$ either is in
    $\projcl[\funcpebc]{\clsd}$
    or is derivable from
    $\projcl[\funcpebc]{\clsd}$
    by weakening.

   \item\noindent{\hskip-12 pt\bf Nontrivial:}\ 
    If
    $\clsd = \emptyset$,
    then
    $\projcl[\funcpebc]{\clsd} = \emptyset$.

   \item\noindent{\hskip-12 pt\bf Monotone:}\ 
    If
    $\clsd' \impl \clsd$
    and
    $\clc \in \projcl[\funcpebc]{\clsd}$, 
    then either
    $\clc \in \projcl[\funcpebc]{\clsd'}$
    or $\clc$ is derivable from
    $\projcl[\funcpebc]{\clsd'}$
    by weakening.

   \item\noindent{\hskip-12 pt\bf Incrementally sound:}\ 
    Let $\cla$ be a clause over~$V$ and let
    $\linestd_{\cla}$
    be the encoding of some clause in
    $\substform{\cla}{\funcpebc}$
    as a Boolean function of the type prescribed by~$\psstd$.
    Then if
    $\clc \in \projcl[\funcpebc]{\clsd \unionSP
      \set{\linestd_{\cla}}}$,
    it holds for all literals
    $\lita \in \lit{A} \setminus \lit{\clc}$
    that 
    the clause
    $\olnot{\lita} \lor \clc$
    either is in
    $\projcl[\funcpebc]{\clsd}$
    or can be derived from
    $\projcl[\funcpebc]{\clsd}$
    by weakening.
  \end{enumerate}
\end{definition}

\noindent In what follows, for conciseness we will use the phrase that some
clause $\clc$ 
``can be derived from $\projcl[\funcpebc]{\clsd}$ by weakening''  
to mean that there is some clause 
$\clc' \subseteq \clc$ 
such that
$\clc' \in \projcl[\funcpebc]{\clsd}$,
\ie to cover both cases that  
$\clc \in \projcl[\funcpebc]{\clsd}$
or that the clause is derivable by weakening of some other clause in
$\projcl[\funcpebc]{\clsd}$.

A special kind of \projectionsubst{}s are those that look not only on
all of $\clsd$ ``globally,'' but measure the semantic content
of~$\clsd$ more precisely.

\begin{definition}[Local \projectionsubst{}]
  \label{def:local-projection}
  If
  $\projclnot[\funcpebc]$
  is \anfprojsubst, then its
  \introduceterm{localized version}
  $\localprojclnot[\funcpebc]$
  is defined to be
  $
  \localprojcl[\funcpebc]{\clsd}
  =
  \Union_{\clsd' \subseteq \clsd}
  \projcl[\funcpebc]{\clsd'}
  $.
  If
  $\projclnot[\funcpebc] = \localprojclnot[\funcpebc]$, 
  we say that 
  $\projclnot[\funcpebc]$
  is a \introduceterm{local projection}.
\end{definition}

It is easily verified that the localized version of a projection is
indeed itself a projection in the sense of
\refdef{def:f-projection}.

In order for our approach outlined above to work, the most
important property of a \projectionsubst is that it should somehow
preserve space when going from the proof system $\psstd$ to resolution.

\begin{definition}[\Spacerespecting \projectionsubst{}]
  \label{def:space-respecting-projection}
  Consider a sequential proof system $\psstd$  with space
  measure~$\clspaceof{\cdot}$.
  Let
  \mbox{$\funcdescr{\funcpebc}{\zerooneset^{\pebdeg}}{\zerooneset}$}
  be a fixed Boolean function, 
  and suppose that
  $\projclnot[\funcpebc]$
  is \anfprojsubst.
  Then we say that 
  $\projclnot[\funcpebc]$
  is \introduceterm{\spacerespecting of degree $\sprdeg$}
  \wrt~$\psstd$
  if there is a polynomial $Q$ of degree at most $\sprdeg$
  such that for any set of Boolean functions
  $\clsd$ 
  over
  $\varsdary{\pebdeg}{V}$ 
  on the form prescribed by~$\psstd$, it holds that
  $Q( \clspaceof{\clsd} ) \geq \Setsize{\vars{\projcl[\funcpebc]{\clsd}}}$.
  We say that 
  $\projclnot[\funcpebc]$
  is \introduceterm{linearly \spacerespecting{}}
  if $Q$ has degree~$1$, and that
  $\projclnot[\funcpebc]$
  is \introduceterm{exactly \spacerespecting{}}
  if we can choose $Q(x) = x$.
\end{definition}

The way 
\refdef{def:space-respecting-projection}
should be understood is that the smaller the degree, the tighter the
reduction between the proof system~$\psstd$ and resolution will be.

\subsubsectionNOW{The Substitution Theorem and Trade-off Lower Bounds}
\label{sec:pc-subst-thm-statement}

We now show that if we can design a  \projectionsubst in accordance with
\refdef{def:f-projection},
then this \projectionsubst can be used to extract resolution
refutations from 
refutations in 
sequential implicational
proof systems~$\psstd$
(as defined at the beginning of
\refsec{sec:pc-proof-systems}). 
Furthermore, if our
\projectionsubst is \spacerespecting, this extraction operation will
preserve length-space trade-off (with some loss in parameters
depending on how high the degree
$\sprdeg$
is).

\begin{lemma}
  \label{lem:projections-preserve-refutation}
  Let
  $\psstd$
  be 
  a sequential implicational
  proof system 
  and let
  \mbox{$\funcdescr{\funcpebc}{\zerooneset^{\pebdeg}}{\zerooneset}$}
  be a Boolean function, 
  and suppose that 
  $\projclnot$
  is \anfprojsubst.
  Then for any CNF formula~$\fstd$ it holds that if
  $\proofstd_{\funcpebc}
  =
  \set{\clsd_0, \clsd_1, \ldots, \clsd_\stoptime}
  $
  is a semantic $\psstd$-refutation of the substitution formula~%
  $\substform{\fstd}{\funcpebc}$,
  the sequence of sets of projected clauses
  $
  \Set{
    \projcl{\clsd_0}, \projcl{\clsd_1}, \ldots,
    \projcl{\clsd_\stoptime}}
  $
  forms the ``backbone'' of a resolution refutation $\proofstd$
  of~$\fstd$ in the following sense:
  \begin{enumerate}[\em(1)]
  \item 
    \label{part:ppr-1}
    $\projcl{\clsd_0} = \emptyset$.
    
  \item 
    \label{part:ppr-2}
    $\emptycl \in \projcl{\clsd_\stoptime}$.

  \item 
    \label{part:ppr-3}
    All transitions from
    $\projcl{\clsd_{t-1}}$
    to
    $\projcl{\clsd_{t}}$
    for $t \in \intnfirst{\stoptime}$ can be accomplished in syntactic
    resolution in such a fashion that
    $
    \varspaceof{\proofstd} =
    \Bigoh{
      \maxofexpr[\clsd \in \proofstd_{\funcpebc}]{\varspaceof{\projcl{\clsd}}}
    }
    $,
    or,
    if $\projclnot$ is a local projection, so that
    $
    \varspaceof{\proofstd} \leq
    \maxofexpr[\clsd \in \proofstd_{\funcpebc}]{\varspaceof{\projcl{\clsd}}}
    $.

  \item 
    \label{part:ppr-4}
    The length of $\proofstd$
    is upper-bounded by
    $\proofstd_{\funcpebc}$
    in the sense that the only time $\proofstd$ performs a download of an axiom
    $\clc \in \fstd$ is when $\proofstd_{\funcpebc}$ downloads some
    axiom
    $\cld \in \substform{\clc}{\funcpebc}$
    from~%
    $\substform{\fstd}{\funcpebc}$.
  \end{enumerate}
\end{lemma}

\noindent On the one hand,
\reflem{lem:projections-preserve-refutation}
is very strong in the sense that even
\emph{semantic} $\psstd$-refutations can be translated to
\emph{syntactic} resolution refutations. On the other hand, 
it would have been nice if the bound in
\refpart{part:ppr-4}
of
\reflem{lem:projections-preserve-refutation}
could have been made into a true upper bound in terms of 
the length of~$\proofstd_{\funcpebc}$,
but it is easy to see that this is \emph{not} possible. The
reason for this is precisely that the $\psstd$-proof refuting
$\substform{\fstd}{\funcpebc[]}$
is allowed to use any arbitrarily strong semantic inference
rules, and this can lead to exponential savings
compared to syntactic resolution. For a concrete example, just let
$\fstd$ be an encoding of the pigeonhole principle and let
$\proofstd_{\funcpebc}$
be the refutation that downloads all axioms of
$\substform{\fstd}{\funcpebc[]}$
and then derives contradiction in one step.
An interesting question is whether it would be possible to  circumvent
this problem by modifying
\refdef{def:f-projection}
to some kind of ``syntactic'' \projectionsubst instead, but we do not
know if and how this can be done. Also, it is not clear that it would help
much---for the applications we have in mind the bound in terms of
axiom downloads is enough, and allows us to obtain stronger bounds
that hold not only for syntactic but even semantic \mbox{$\psstd$-proofs}.

Before proving 
\reflem{lem:projections-preserve-refutation}
let us see how it can be used to prove trade-offs provided that we can
construct \spacerespecting \projectionsubst{}s.

\begin{theorem}
  \label{th:generic-subst-thm}
  Let $\psstd$ be a sequential proof system
  with space measure~$\clspaceof{\cdot}$.
  Suppose
  \mbox{$\funcdescr{\funcpebc}{\zerooneset^{\pebdeg}}{\zerooneset}$}
  is a Boolean function such that there exists \anfprojsubst which is
  \spacerespecting of degree~$\sprdeg$ \wrt~$\psstd$.
  Then if $\fstd$ is any unsatisfiable CNF formula and 
  $\proofstd_{\funcpebc}$
  is any semantic $\psstd$-refutation of the substitution formula~%
  $\substform{\fstd}{\funcpebc}$,
  there is a resolution refutation $\proofstd$ of~$\fstd$ such that:
  \begin{iteMize}{$\bullet$}
  \item 
    The length of $\proofstd$ is upper-bounded by
    $\proofstd_{\funcpebc}$
    in the sense that 
    $\proofstd$ makes at most as many axiom downloads as
    $\proofstd_{\funcpebc}$.
    
  \item 
    The space of $\proofstd$ is upper-bounded by
    $\proofstd_{\funcpebc}$
    in the sense that 
    $
    \varspaceof{\proofstd} = \Bigoh{\clspaceof{\proofstd_{\funcpebc}}^{\sprdeg}}
    $.
    
  \end{iteMize}
  In particular, if there is no syntactic resolution refutation
  of~$\fstd$ in 
  variable space~$\bigoh{\spacestd}$
  making $\bigoh{\lengthstd}$ axiom downloads, 
  then there is no semantic $\psstd$-refutation of 
  $\substform{\fstd}{\funcpebc}$
  in simultaneous length
  $\bigoh{\lengthstd}$
  and $\psstd$-space
  $\Bigoh{\sqrt[\sprdeg]{\spacestd}}$.
\end{theorem}

\begin{proof}[Proof of \refth{th:generic-subst-thm}]
  Let
  $\proofstd_{\funcpebc}$
  be a semantic $\psstd$-refutation of
  $\substform{\fstd}{\funcpebc}$,
  and let
  $\proofstd$
  be the resolution refutation we obtain by applying the the
  projection $\projclnot$ on   $\proofstd_{\funcpebc}$ as in
  \reflem{lem:projections-preserve-refutation}.
  By
  \refpart{part:ppr-4}
  of
  \reflem{lem:projections-preserve-refutation}
  we know that
  $\proofstd$ makes at most as many axiom downloads as
  $\proofstd_{\funcpebc}$.
  By 
  \refpart{part:ppr-3}
  of the lemma we have
  $
  \varspaceof{\proofstd} =
  \Bigoh{
    \maxofexpr[\clsd \in \proofstd_{\funcpebc}]{\varspaceof{\projcl{\clsd}}}
  }
  $.
  Fix some $\psstd$-configuration $\clsd$ maximizing the right-hand
  side of this expression. For this $\clsd$ we have
  $\varspaceof{\projcl{\clsd}} 
  = 
  \Bigoh{\clspaceof{\clsd}^{\sprdeg}}
  =
  \Bigoh{\clspaceof{\proofstd_{\funcpebc}}^{\sprdeg}}$
  according to 
  \refdef{def:space-respecting-projection}. The theorem follows.
\end{proof}

Clearly, the key to the proof of
\refth{th:generic-subst-thm}
is the claim that projections translate $\psstd$-refutations to
resolution refutations. Let us substantiate this claim.

\begin{proof}[Proof of \reflem{lem:projections-preserve-refutation}]
  Fix any sequential proof system
  $\psstd$,
  any \fprojsubst
  $\projclnot$,
  and any CNF formula~$\fstd$.
  Recall that we want to show that if
  $\proofstd_{\funcpebc}
  =
  \set{\clsd_0, \clsd_1, \ldots, \clsd_\stoptime}
  $
  is a semantic \mbox{$\psstd$-refu}\-tation of~%
  the substitution formula~%
  $\substform{\fstd}{\funcpebc}$,
  then the  sequence of projected clause sets
  $
  \Set{
    \projcl{\clsd_0}, \projcl{\clsd_1}, \ldots,
    \projcl{\clsd_\stoptime}}
  $
  is essentially a  resolution refutation $\proofstd$ except for
  some details that we might have to fill in when going from
  $\projcl{\clsd_{t-1}}$
  to
  $\projcl{\clsd_{t}}$
  in the derivation.

  Parts~\ref{part:ppr-1}  and~\ref{part:ppr-2}
  of
  \reflem{lem:projections-preserve-refutation}
  are immediate from
  \ifthenelse{\boolean{maybeLMCS}}           
  {the definition of projection,} 
  {\refdef{def:f-projection},}
  since we have
  $\projcl{\clsd_0} = \projcl{\emptyset} = \emptyset$
  by nontriviality
  and
  $\emptycl \in \projcl{\clsd_\stoptime}$
  by completeness
  (note that
  $\clsd_\stoptime \impl \emptycl 
  = \Lor_{\varx^\truthval \in \emptycl} \signedfunc[\ff]{\truthval}{\varvecx}$
  and the empty clause clearly cannot be derived by weakening).

  We want to show that a resolution refutation of~$\fstd$ can get from
  $\projcl{\clsd_{t-1}}$
  to
  $\projcl{\clsd_{t}}$
  as claimed in 
  \refpart{part:ppr-3}
  of the lemma.
  For brevity, let us write
  $\clsc_i = \projcl{\clsd_{i}}$ for all~$i$, and 
  consider the possible derivation steps at time~$t$.

  \ifthenelse{\boolean{maybeLMCS}}
  {\textbf{Inference:}}
  {\paragraph{Inference}}
  Suppose
  $\clsd_{t} = \clsd_{t-1} \unionSP \set{\linestd_t}$
  for some
  $\linestd_t$
  inferred from
  $\clsd_{t-1}$.
  Since $\psstd$ is sound we have
  $\clsd_{t-1} \impl \clsd_{t}$,
  and since the projection is monotone by definition we can conclude
  that all clauses in
  $\clsc_{t} \setminus \clsc_{t-1}$
  are derivable from
  $\clsc_{t-1}$  
  by weakening.
  We go from
  $\clsc_{t-1}$
  to  
  $\clsc_{t}$  
  in three steps. First, we erase all clauses
  $\clc \in \clsc_{t-1}$
  for which there are no clauses
  $\clc' \in \clsc_{t}$
  such that
  $\clc \subseteq \clc'$.
  Then, we derive all clauses in
  $\clsc_{t} \setminus \clsc_{t-1}$
  by weakening, noting that all clauses needed for weakening steps are
  still in the configuration. Finally, we erase the rest of
  $\clsc_{t} \setminus \clsc_{t-1}$.
  At all times during this transition from 
  $\clsc_{t-1}$
  to  
  $\clsc_{t-1}$,
  the variable space of the intermediate clause configurations is
  upper-bounded by
  $
  \maxofexpr{
    \varspaceof{\clsc_{t-1}},
    \varspaceof{\clsc_{t}}
  }
  $.

  \ifthenelse{\boolean{maybeLMCS}}
  {\textbf{Erasure:}}
  {\paragraph{Erasure}}
  Suppose
  $\clsd_{t} = \clsd_{t-1} \setminus \set{\linestd_{t-1}}$
  for some
  $\linestd_{t-1} \in \clsd_{t-1}$.
  Again we have that 
  $\clsd_{t-1} \impl \clsd_{t}$,
  and we can appeal to the monotonicity of the projection and proceed
  exactly as in the case of an inference above.

  \ifthenelse{\boolean{maybeLMCS}}
  {\textbf{Axiom download:}}
  {\paragraph{Axiom download}}
  So far, the only derivation rules used in the resolution refutation
  $\proofstd$ that we are constructing are weakening and erasure,
  which clearly does not help $\proofstd$ to make much progress
  towards proving a contradiction. Also, the only properties of the
  \fprojsubst that we have used are completeness, nontriviality, and
  monotonicity. Note, however, that a ``projection'' that sends
  $\emptyset$ to $\emptyset$ and all other configurations to~%
  $\set{\emptycl}$ also satisfies these conditions. 
  Hence, the axiom downloads are where we must expect the action to
  take place, and we can also expect that we will have to make crucial use
  of the incremental soundness of the projection.

  Assume that 
  $\clsd_{t} = \clsd_{t-1} \unionSP \set{\linestd_\cla}$ for
  a function $\linestd_\cla$ encoding some clause from the
  substitution clause set
  $\substform{\cla}{\funcpebc}$
  corresponding to an axiom $\cla \in \fstd$.
  We want to show that all clauses in
  $\clsc_{t} \setminus \clsc_{t-1}$
  can be derived in~$\proofstd$ by downloading~$\cla$, resolving (and
  possibly weakening) clauses, and then perhaps erasing~$\cla$, and
  that all this can be done without the variable space exceeding 
  $\varspaceof{\clsc_{t-1} \unionSP \clsc_{t}} \leq
  \varspaceof{\clsc_{t-1}} + \varspaceof{\clsc_{t}}$.

  We already know how to derive clauses by weakening, so consider a
  clause
  $\clc \in \clsc_{t} \setminus \clsc_{t-1}$
  that cannot be derived by weakening from~$\clsc_{t-1}$.
  By the incremental soundness of the projection, it holds for all
  literals
  $\lita \in \lit{\cla} \setminus \lit{\clc}$ 
  that the clauses
  $\olnot{\lita} \lor \clc$
  can be  derived from~$\clsc_{t-1}$ by weakening.
  Once we have these clauses, we can resolve them one by one with
  $\cla$ to derive~$\clc$.

  Some care is needed, though, to argue that we can stay within the
  variable space bound 
  $\varspaceof{\clsc_{t-1}} + \varspaceof{\clsc_{t}}$.
  Observe that what was just said implies that for all 
  $\lita \in \lit{\cla} \setminus \lit{\clc}$ 
  there are clauses
  $\olnot{\lita} \lor \clc_{\lita} \in \clsc_{t-1}$
  with
  $\clc_{\lita} \subseteq \clc$.
  In particular, we have
  $\olnot{\lita} \in \lit{\clsc_{t-1}}$
  for all
  $\lita \in \lit{\cla} \setminus \lit{\clc}$.
  This is so since by the incremental soundness there must exist some clause
  $\clc' \in \clsc_{t-1}$
  such that
  $\olnot{\lita} \lor \clc$
  is derivable by weakening from~$\clc'$,
  and if
  $\olnot{\lita} \notin \lit{\clc'}$
  we would have that
  $\clc$ is derivable by weakening from $\clc'$ as well, contrary to
  assumption. 
  Note furthermore that if the projection is local, then
  $\olnot{\lita} \lor \clc_{\lita} \in \clsc_{t-1}$
  implies that
  $\olnot{\lita} \lor \clc_{\lita} \in \clsc_{t}$
  as well, since no clauses can disappear from the projection when enlarging  
  $\clsd_{t-1}$
  to~%
  $\clsd_{t}$. 
  Thus, for local projections we have
  $\varspaceof{\clsc_{t-1} \unionSP \set{\cla}} 
  \subseteq
  \varspaceof{\clsc_{t}}$.

  If it happens that all clauses in
  $\clsc_{t} \setminus \clsc_{t-1}$
  can be derived by weakening, we act as in the cases of inference and
  erasure above. Otherwise,  to make the transition
  from
  $\clsc_{t-1}$
  to
  $\clsc_{t}$
  in a space-efficient fashion we proceed as follows.
  \begin{enumerate}[(1)]
  \item 
    Download the axiom clause $\cla$.

  \item 
    \label{step:ax-download-new-step-2}
    Infer all clauses in
    $\clsc_{t} \setminus \clsc_{t-1}$
    that can be derived by weakening from
    $\clsc_{t-1} \unionSP \set{\cla}$.

  \item 
    \label{step:ax-download-new-step-5}
    For all remaining clauses 
    $\clc \in \clsc_{t} \setminus \clsc_{t-1}$,
    derive   
    $\olnot{\lita} \lor \clc$
    for all literals
    $\lita \in \lit{\cla} \setminus \lit{\clc}$
    and then resolve these clauses with $\cla$ to obtain~$\clc$.

  \item 
    \label{step:ax-download-new-step-6}
    Erase all  clauses in the current configuration that are
    not present in~%
    $\clsc_{t}$,
    possibly including~$\cla$.
  \end{enumerate}
  Since it follows from what was argued above that 
  $
  \vars{\cla}
  \subseteq
  \vars{\clsc_{t-1} \unionSP \clsc_{t}}
  $,
  all the variables in all the intermediate configurations above will be
  contained in this set, meaning that the variable space never exceeds
  $\varspaceof{\clsc_{t-1}} + \varspaceof{\clsc_{t}}$.
  If in addition the projection is local, then
  $\varspaceof{\clsc_{t-1} \unionSP \set{\cla}} 
  \subseteq
  \varspaceof{\clsc_{t}}$
  and the variable space increases monotonically from
  $\clsc_{t-1}$
  to~%
  $\clsc_{t}$.

  Wrapping up the  proof, we have shown that
  no matter what $\psstd$-derivation step is
  made
  in the transition
  $\clcfgtransition{\clsd_{t-1}}{\clsd_t}$,
  it is possible to 
  perform the corresponding transition
  $\clcfgtransition{\clsc_{t-1}}{\clsc_t}$
  for our \projectedverb clause sets in resolution without
  the variable space going above
  $\varspaceof{\clsc_{t-1}} +  \varspaceof{\clsc_{t}}$
  (or even above  $\varspaceof{\clsc_{t}}$ for local projections).
  Also, the only time we need to download an axiom
  $\cla \in \fstd$
  in our \projectedverb refutation $\proofstd$ of~$\fstd$ is when
  $\proofstd_{\funcpebc}$
  downloads some axiom from~%
  $\substform{\cla}{\funcpebc[\pebdeg]}$.
  The lemma follows.
\end{proof}

\ifthenelse{\boolean{maybeLMCS}}
{\subsubsectionNOW{Resolution and 
    \texorpdfstring{\kdnf}{k-DNF} 
    Resolution Have
    \SPACERESPECTING \Projectionsubst{}s}}
{\subsubsectionNOW{Resolution and \kdnf Resolution Have
    \SPACERESPECTING \Projectionsubst{}s}}

\label{sec:pc-subst-thm-res-k}

Let us recall again what
\refth{th:generic-subst-thm}
says. Suppose we have a family of CNF formulas
with lower bounds for refutation variable space in resolution, or with
trade-offs between refutation length
and refutation variable space%
\footnote{Or, strictly speaking, between the \emph{number of axiom downloads}
  and variable space.}
(such as for instance pebbling contradictions over suitable
graphs). Then we can lift these lower bounds and trade-offs to
stronger measures in a potentially stronger proof system~$\psstd$,
provided that we can find a Boolean function
\mbox{$\funcdescr{\funcpebc}{\zerooneset^{\pebdeg}}{\zerooneset}$}
and \anfprojsubst
$\projclnot$
that is \spacerespecting \wrt~$\psstd$.

Thus, at this point we can in principle forget everything about proof
complexity. If we want to prove space lower bounds or time-space
trade-offs for a proof system~$\psstd$, we can focus on studying
Boolean functions of the form used by $\psstd$ and trying to devise
\spacerespecting projections for such functions. Below, we describe
how this can be done for resolution and $\resknot$-systems.

\begin{definition}[\Preciseimplsubst{} \cite{BN11UnderstandingSpace}]
  \label{def:precise-implication}
  Let $\funcpebc$ be a Boolean function of arity~$\pebdeg$, 
  let 
  $\clsd$
  be a set of Boolean functions over
  $\varsdary{\pebdeg}{V}$,
  and let $\clc$ be a disjunctive clause over~$V$.
  If
  \begin{subequations}
    \begin{align}
      \label{eq:precise-impl-condition-1}
      \clsd &\impl
      \Lor_{\varx^{\truthval} \in \clc} 
      \signedfunc[\funcpebc]{\truthval}{\varvecx}
      \\
      \intertext{but for all strict subclauses $\clc' \subsetneqq \clc$
        it holds that}
      \label{eq:IRP-precise-impl-condition-2}
      \clsd &\nimpl
      \Lor_{\varx^{\truthval} \in \clc'}
      \signedfunc[\funcpebc]{\truthval}{\varvecx}
      \eqcomma
    \end{align}
  \end{subequations}
  we say that the clause set $\clsd$ implies
  $\Lor_{\varx^{\truthval} \in \clc} 
  \signedfunc[\funcpebc]{\truthval}{\varvecx}$
  \introduceterm{\preciseimpladv{}}
  and write
  \begin{equation}
    \clsd \preciseimpl 
    \Lor_{\varx^{\truthval} \in \clc} 
    \signedfunc[\funcpebc]{\truthval}{\varvecx}
    \eqperiod
  \end{equation}
\end{definition}

\begin{definition}[Resolution \projectionsubst{}]
  \label{def:projected-clauses-res}
  Let $\funcpebc$ denote a Boolean function of arity~$\pebdeg$ 
  and let 
  $\clsd$
  be any set of Boolean functions over
  $\varsdary{\pebdeg}{V}$.
  Then we define
  \begin{equation}
    \rprojcl{\clsd} = 
    \setdescr{\clc}{\clsd \preciseimpl 
      \Lor_{\varx^{\truthval} \in \clc} 
      \signedfunc[\funcpebc]{\truthval}{\varvecx}}
  \end{equation}
  to be the \introduceterm{resolution \projectionsubst{}} of~$\clsd$.
  Also, we define
  $
  \localrprojcl{\clsd}
  =
  \Union_{\clsd' \subseteq \clsd}
  \rprojcl{\clsd'}
  $
  to be the 
  \introduceterm{local resolution \projectionsubst{}} of~$\clsd$.
\end{definition}

\begin{lemma}
  The mapping
  $\rprojclnot$
  is \anfprojsubst
  (for any sequential proof system~$\psstd$).  
\end{lemma}

\begin{proof}
  Suppose 
  $\clsd \impl
  \Lor_{\varx^{\truthval} \in \clc} 
  \signedfunc[\funcpebc]{\truthval}{\varvecx}$.
  Then we can remove literals from
  $\clc$ one by one until we have some minimal clause
  $\clc' \subseteq \clc$
  such that no more literal can be removed if the implication is to
  hold, and this clause $\clc'$ is projected by~$\clsd$ according to
  the definition. This proves
  both completeness and monotonicity for~%
  $\rprojclnot$. Nontriviality is obvious.
  
  For the incremental soundness,   if
  $\clc \in \rprojcl{\clsd \unionSP \set {\linestd_\cla}}$
  for an encoding
  $\linestd_\cla$
  of some clause in $\substform{\cla}{\funcpebc}$,
  then this means, in particular, that
  $\clsd \unionSP \set{\linestd_\cla}
  \impl
  \Lor_{\varx^{\truthval} \in \clc} 
  \signedfunc[\funcpebc]{\truthval}{\varvecx}$.
  Consider any \tva $\tvastd$ such that
  $\logeval{\clsd}{\tvastd} = 1$
  but
  $\Logeval{\Lor_{\varx^{\truthval} \in \clc} 
    \signedfunc[\funcpebc]{\truthval}{\varvecx}}
  {\tvastd} = 0$.
  By assumption,  
  $\logeval{\linestd_\cla}{\tvastd} = 0$.
  But this means that for all literals
  $\vary^{\truthvalalt} \in \lit{A}$
  we have
  $\Logeval{\signedfunc[\funcpebc]{1-\truthvalalt}{\varvecy}}{\tvastd} = 1$.
  Since this holds for any~$\tvastd$,
  it follows for all
  $\vary^{\truthvalalt} \in \lit{A}$
  that
  $\clsd
  \impl
  \Lor_{\varx^{\truthval} \in (\vary^{1-\truthvalalt} \lor \clc) } 
  \signedfunc[\funcpebc]{\truthval}{\varvecx}$,
  and we conclude by  the completeness of the projection that the
  clause
  $\vary^{1-\truthvalalt} \lor \clc$
  is derivable by weakening from
  $\rprojcl{\clsd}$.
\end{proof}

With this projection, and using 
\refth{th:generic-subst-thm},
the main technical result in
\cite{BN11UnderstandingSpace}
can now be rephrased as follows, 
where we recall the notion of \nonauthoritarian functions from
\refdef{def:non-authoritarian-func}.

\begin{theorem}[\cite{BN11UnderstandingSpace}]
  \label{th:projection-resolution}
%
  If
  $\funcpebc$ is a \nonauthoritarian
  Boolean function, then the projection
  $\localrprojclnot$
  is exactly \spacerespecting 
  \wrt the resolution proof system.
\end{theorem}

This result was later extended to \kdnf resolution, although with
slightly worse parameters.

\begin{theorem}[\cite{BN11UnderstandingSpace}]
  \label{th:projection-res-k}
%
  If
  $\funcpebc$ is a $(k+1)$-\nonauthoritarian
  function (for some fixed~$k$), then the projection
  $\rprojclnot$ 
  is
  \spacerespecting of degree $k+1$ \wrt \kdnf resolution.
\end{theorem}

It has subsequently been shown that the loss in the parameters in
\refth{th:projection-res-k}
as compared to 
\refth{th:projection-resolution}
is necessary, except perhaps for an additive constant~$1$ in the degree.

\begin{theorem}[\cite{NR11MinimalUnsatisfiability}]
  \label{th:projection-res-k-bound}
  If $\funcpebc$ is the exclusive or of $k+1$ variables, 
  then the projection
  $\rprojclnot$ 
  cannot be  \spacerespecting
  \wrt \kdnf resolution
  for any degree $\sprdeg < k$.
\end{theorem}

As an aside, it can be noted that
\refth{th:projection-resolution}
uses the local version of the projection, whereas
\refth{th:projection-res-k}
uses the non-local definition.
\Refth{th:projection-resolution}
can be made to work with either variant (if we are willing to settle
for a projection that is just linearly \spacerespecting instead of
exactly \spacerespecting), but for technical reasons the proof of
\refth{th:projection-res-k}
only seems to work with the ``global'' version of the projection.

Although this might not be immediately obvious,
\reftwoths{th:projection-resolution}{th:projection-res-k}
are remarkably powerful tools for understanding space in
resolution-based proof systems.
All the trade-off lower bounds in
\refsubsec{sec:pc-tradeoffs}
can be derived as immediate consequences of 
these two theorems.
Another interesting corollary of
\refth{th:projection-resolution}
is that that it yields optimal lower bounds on clause space for resolution.
Recall that Esteban and Torán~\cite{ET01SpaceBounds}
proved that the clause space of refuting $\fstd$
is upper-bounded by the formula size.
In the papers
\cite{ABRW02SpaceComplexity, BG03SpaceComplexity,ET01SpaceBounds}
it was shown, using quite elaborate arguments,  that there are
polynomial-size \kcnfform{}s with lower bounds on clause space
matching this upper bound up to constant factors. Using
\refth{th:projection-resolution}
we can get a different and  much shorter proof of this fact.

\begin{corollary}%
  [\cite{ABRW02SpaceComplexity, BG03SpaceComplexity,ET01SpaceBounds}]
  \label{cor:subst-space-thm-lower-bound-clause-space}
  There are 
  families of \kcnfform{}s
  $\set{\fstd}_{n=1}^{\infty}$
  with $\bigtheta{n}$ clauses over
  $\bigtheta{n}$ variables \st
  $\clspaceref[\resnot]{\fstd_n} = \bigtheta{n}$.
\end{corollary}

\begin{proof}
  Just pick any formula family
  for which it is shown that any refutation of
  $\fstd_n$
  must at some point in the refutation mention
  $\bigomega{n}$
  variables at the same time
  (for example, from   \cite{BW01ShortProofs}),
  and then apply
  \refth{th:projection-resolution}
  to,   
  say,
  $\substform{\fstd_n}{\xor_2}$.
\end{proof}

It should be noted, though, that to derive these linear lower bounds
we have to change the formula families by substitution, whereas the papers
\cite{ABRW02SpaceComplexity, BG03SpaceComplexity,ET01SpaceBounds}
prove their lower bounds for the original formulas. Moreover, 
there is another, and even more elegant way to derive
\refcor{cor:subst-space-thm-lower-bound-clause-space}
from~\cite{BW01ShortProofs}
\ifthenelse{\boolean{maybeLMCS}}
{without changing the formulas,}
{without having to modify the formulas under consideration,}
namely by using the lower bound on clause space in terms of width
in~\cite{AD08CombinatoricalCharacterization}.

As we indicated above, we believe that this projection framework can
potentially be extended to other proof systems than resolution
and~$\resknot$. In particular, it would be very interesting to see if
one could prove lower bounds for cutting planes, polynomial calculus,
or polynomial calculus resolution in this way. However, to do so
another projection 
(in the sense of
\refdef{def:f-projection})
than the one in
\refdef{def:projected-clauses-res}
would be needed.  We conclude this section by sketching the proofs of
\reftwoths{th:projection-resolution}{th:projection-res-k}, 
and then explaining why the same approach will not work for CP, PC, or
PCR.

\begin{proof}[Proof sketch for \refth{th:projection-resolution}]
  In the case of resolution, the set of Boolean functions
  $\clsd$
  just consists of disjunctive clauses over
  $\varsdary{\pebdeg}{V}$.
  Fix some clause set $\clsd$
  and let
  $V^* = \Vars{\localrprojcl{\clsd}}$.
  What we want to prove is that
  $
  \setsize{\clsd} \geq \setsize{V^*}$.

  To this end, consider the bipartite graph with the vertices on the
  left labelled by clauses $\cld \in \clsd$ and the vertices on the
  right labelled by variables $\varx \in V^*$. We draw an edge between
  $\cld$ and $\varx$ if some variable $\varx_i$ belonging to~$\varx$
  appears in~$\cld$. Let $\vneighbour{\clsd'}$ denote the neighbours
  on the right of a clause set~$\clsd'$. We claim without proof that
  $\vneighbour{\clsd} = V^*$, \ie that all $\varx \in V^*$ have
  incoming edges from $\clsd$ (this follows from the condition
  \refeq{eq:IRP-precise-impl-condition-2}  
  in
  \refdef{def:precise-implication}).

  Pick some
  $\clsd_1 \subseteq \clsd$
  of maximal size (possibly empty) with a set of neighbours
  $V^*_1 = \vneighbour{\clsd_1}$
  such that
  $\setsize{\clsd_1} \geq \setsize{V^*_1}$.
  If
  $\clsd_1 = \clsd$
  we are done, so let us suppose
  $\clsd_1 \neq \clsd$
  and argue by contradiction.
  Let
  $\clsd_2 = \clsd \setminus \clsd_1 \neq \emptyset$
  and
  $V^*_2 = V^* \setminus V^*_1$.
  For all 
  $\clsd' \subseteq \clsd_2$
  we must have
  $\setsize{\clsd'} \leq 
  \setsize{\vneighbour{\clsd'} \setminus V^*_1} = 
  \setsize{\vneighbour{\clsd'} \intersectionSP V^*_2}$,
  since otherwise $\clsd_1$ would not have been chosen of maximal
  size. This in turns implies by Hall's theorem that there is a
  matching $M$ from
  $\clsd_2$ into~$V^*_2$.

  Consider some clause 
  $\clc \in \localrprojcl{\clsd}$
  such that
  $\clsd_1$ is ``too weak'' to project~$\clc$ (we are fudging some
  details here again, but nothing important).
  Let $\clc_i$ be the part of $\clc$ that mentions variables from
  $V^*_i$ for $i=1,2$.
  Then by 
  \reftwodefs{def:precise-implication}{def:projected-clauses-res} it
  holds that
  $
  \clsd_1 \unionSP \clsd_2
  \impl
  \Lor_{\varx^{\truthval} \in \clc_1} 
  \signedfunc[\funcpebc]{\truthval}{\varvecx}
  \, \lor \,
  \Lor_{\vary^{\truthval} \in \clc_2} 
  \signedfunc[\funcpebc]{\truthval}{\varvecy}
  $
  but
  $
  \clsd_1
  \nimpl
  \Lor_{\varx^{\truthval} \in \clc_1} 
  \signedfunc[\funcpebc]{\truthval}{\varvecx}
  $.
  This means that there is a truth value assignment $\tvastd_1$ to
  $\varsdary{\pebdeg}{V^*_1}$
  satisfying  $\clsd_1$ but falsifying
  $\Lor_{\varx^{\truthval} \in \clc_1} 
  \signedfunc[\funcpebc]{\truthval}{\varvecx}$.
  Observe that     
  $\vars{\clsd_1} \subseteq \varsdary{\pebdeg}{V^*_1}$
  by construction.
%

  Using the matching~$M$, we can find another partial truth value
  assignment  $\tvastd_2$ to  $\varsdary{\pebdeg}{V^*_2}$
  that satisfies
  $\clsd_2$ by setting at most one variable
  $\varx_i$ for every $\varx \in V^*_2$.
%
  This
  $\tvastd_2$ 
  leaves the truth value of 
  $
  \Lor_{\vary^{\truthval} \in \clc_2} 
  \signedfunc[\funcpebc]{\truthval}{\varvecy}
  $
  undetermined
  since $\funcpebc$ is \nonauthoritarian, 
  and this means that we
  can extend $\tvastd_2$ to a full assignment over
  $\varsdary{\pebdeg}{V^*_2}$
  such that
  $
  \Lor_{\vary^{\truthval} \in \clc_2} 
  \signedfunc[\funcpebc]{\truthval}{\varvecy}
  $ 
  is falsified. But then
  $\tvastd_1 \unionSP \tvastd_2$
  is an assignment that satisfies
  $\clsd_1 \unionSP \clsd_2$
  but falsifies
  $
  \Lor_{\varx^{\truthval} \in \clc_1} 
  \signedfunc[\funcpebc]{\truthval}{\varvecx}
  \, \lor \,
  \Lor_{\vary^{\truthval} \in \clc_2} 
  \signedfunc[\funcpebc]{\truthval}{\varvecy}
  $,
  which is a contradiction.
\end{proof}

\begin{proof}[Proof sketch for \refth{th:projection-res-k} ]
  Let us restrict our attention to $2$-DNF resolution, since this
  already captures the   hardness of the general case. Also, we sweep
  quite a few technical details under the rug to focus on the main
  idea of the proof.

  Suppose that we have a set of \mbox{$2$-DNF} formulas $\clsd$ of
  size $\setsize{\clsd} = m$
  such that the set of projected variables
  $V^* = \vars{\rprojcl{\clsd}}$
  has size
  $\setsize{V^*} \geq K \cdot m^3$
  for some suitably large constant~$K$ of our choice. We want to
  derive a contradiction.

  As a first preprocessing step, let us prune all formulas
  $\cld \in \clsd$ one by one by shrinking any \mbox{$2$-term}
  $\lita \land \litb$
  in~$\cld$   to just
  $\lita$ or just~$\litb$,
  \ie making $\cld$ weaker,
  as long as this does not change the projection
  $\rprojcl{\clsd}$.
  This pruning step does not decrease the size (\ie the number of
  formulas) of~$\clsd$.

  By counting, there must exist some formula
  $\cld \in \clsd$
  containing literals belonging to at least
  $K \cdot m^2$
  different variables in~$V^*$.
  Consider some clause
  $\clc \in \rprojcl{\clsd}$
  such that $\clsd \setminus \set{\cld}$ is too weak to project it.
  This means that there is an assignment
  $\tvastd$
  such that
  $\logeval{\clsd \setminus \set{\cld}}{\tvastd} = 1$
  but
  $
  \Logeval{\Lor_{\varx^{\truthval} \in \clc} 
  \signedfunc[\funcpebc]{\truthval}{\varvecx}}{\tvastd}
  \neq 1 
  $,
  \ie $\tvastd$ either fixes 
  $\Logeval{\Lor_{\varx^{\truthval} \in \clc} 
    \signedfunc[\funcpebc]{\truthval}{\varvecx}}{\tvastd}$
  to false or leaves it undetermined.
  Let us pick such an $\tvastd$ assigning values to the minimal amount
  of variables. It is clear that the domain size of $\tvastd$ will
  then be at most
  $2(m-1)$
  since the assignment needs to fix only one \mbox{$2$-term} for every
  formula in $\clsd \setminus \set{\cld}$.
  But this means that the formula $\cld$ contains a huge number of
  unset variables. We would like to argue that somewhere in $\cld$
  there is a \mbox{$2$-term} that can be set to true without satisfying
  $\Lor_{\varx^{\truthval} \in \clc} 
    \signedfunc[\funcpebc]{\truthval}{\varvecx}$,
  which would lead to a contradiction.

  We note first that if $\cld$ contains $2m$ \mbox{$2$-terms}
  $\varx^{\truthval}_i \land \vary^{\truthvalalt}_j$
  with all literals in these terms belonging to pairwise disjoint
  variable sets for distinct terms
  (but where we can have $\varx=\vary$), 
  we immediately get a
  contradiction. Namely, if this is the case we can find at least one
  \mbox{$2$-term} 
  $\varx^{\truthval}_i \land \vary^{\truthvalalt}_j$
  such that $\tvastd$ does not assign values to any variables
  $\varx_{i'}, \vary_{j'}$.
  We can satisfy this 
  \mbox{$2$-term}, and hence all of $\clsd$,
  without satisfying
  $\Lor_{\varx^{\truthval} \in \clc} 
  \signedfunc[\funcpebc]{\truthval}{\varvecx}$
  since by assumption
  $\funcpebc$ is $3$-\nonauthoritarian
  (so any assignments to 
  $\varx_i$ and $\vary_j$
  can be repaired by setting other variables   
  $\varx_{i'}, \vary_{j'}$ 
  to appropriate values).

  But if $\cld$ does \emph{not} contain $2m$ such 
  \mbox{$2$-terms}
  over disjoint variables, then by counting (and adjusting our
  constant~$K$) there must exist some literal
  $\lita$
  that occurs in $\cld$ in at least $2m$ terms
  $\lita \land \varx^{\truthval}_i$
  with the $\varx_i$ belonging to different variables.
  Moreover, these 
  \mbox{$2$-terms}
  were not pruned in our preprocessing step, so they must all be
  necessary.
  Because of this, one can argue that there must exist some other
  assignment
  $\tvastd'$
  such that
  $\logeval{\clsd \setminus \set{\cld}}{\tvastd'} = 1$,
  $
  \Logeval{\Lor_{\varx^{\truthval} \in \clc} 
  \signedfunc[\funcpebc]{\truthval}{\varvecx}}{\tvastd'}
  \neq 1 
  $,
  and
  $\logeval{\lita}{\tvastd'} = 1$.
  Now at least one of the $2m$ companion variables of $\lita$ is
  untouched by $\tvastd'$ and can be set to true   without satisfying
  $\Lor_{\varx^{\truthval} \in \clc} 
  \signedfunc[\funcpebc]{\truthval}{\varvecx}$.
  This is a contradiction.
\end{proof}

To see why we cannot hope to prove lower bounds for 
cutting planes or polynomial calculus
in the same fashion, consider the following examples.

\begin{example}
  \label{ex:counterexample-cp}
  If we have variables
  $\varwindex{1}{}, \varwindex{2}{}, \varwindex{3}{}, \ldots$
  and make substitutions using binary exclusive or $\xor_2$ to get new
  variables
  $\varwindex{1}{1}, \varwindex{1}{2}, 
  \varwindex{2}{1}, \varwindex{2}{2}, 
  \varwindex{3}{1}, \varwindex{3}{2}, 
  \ldots$, 
  then the example
  \begin{equation}
    \sum_{i=1}^{k} (\varwindex{i}{1}  -\varwindex{i}{2} ) \geq k
  \end{equation}
  shows that just a single CP inequality can project an arbitrarily large conjunction
  $\varwindex{1}{} \land \varwindex{2}{} \land \formuladots \land \varwindex{k}{}$.
  Thus, here we have
  $\setsize{\clsd} = 1$
  while
  $\varspaceof{\rprojcl[\xor_2]{\clsd}}$ 
  goes to infinity.
\end{example}

\begin{example}
  \label{ex:counterexample-pc}
  Again using substitutions with $\xor_2$, for polynomial calculus we
  have the example 
  \begin{equation}
    -1 
    + \prod_{i=1}^{k} \varwindex{i}{1}  \varwindex{i}{2} 
  \end{equation}
  showing that just two monomials
  can project the 
  arbitrarily large conjunction 
  $\olnot{\varwindex{1}{}} \land \olnot{\varwindex{2}{}} \land
  \formuladots \land \olnot{\varwindex{k}{}}$
  if we use the projection in 
  \refdef{def:projected-clauses-res}.
\end{example}

Let us also give a slightly more involved example for polynomial
calculus resolution that uses the extra formal variables encoding negative
literals to obtain projected configurations with many positive
literals. Recall that in PCR we write
$\varx'$
to denote the formal variable that encodes the negation of~%
$\varx$.
  
\begin{example}
  \label{ex:counterexample-pcr}
  In PCR,
  three monomials
  \begin{equation}
    -1 
    + \prod_{i=1}^{k} \varwindex{i}{1}  \varwindex{i}{2}' 
    + \prod_{i=1}^{k} \varwindex{i}{1}'  \varwindex{i}{2} 
  \end{equation}
  can project the arbitrarily large conjunction 
  $\varwindex{1}{} \land \varwindex{2}{} \land \formuladots \land \varwindex{k}{}$
  if we use the projection in 
  \refdef{def:projected-clauses-res}.
\end{example}
  
Somehow, the reason 
that these counter-examples work
is that the projection in
\refdef{def:projected-clauses-res}
allows the Boolean functions in the implication to be far too
strong. These functions do not really imply just conjunctions of
exclusive ors, but something much stronger in that they actually fix
the variable assignments (to some particular assignment that happens
to satisfy exclusive ors). Note that formulas
$\substform{\fstd}{\xor_2}$
do not speak about fixed variable assignments for, say, 
$\varx_1$
or
$\varx_2$,
but only about the value of
$\varx_1 \xor \varx_2$.
Intuitively, therefore, the only way we can know something more about 
$\varx_1$
and
$\varx_2$
than the value of
$\varx_1 \xor \varx_2$
is if the refutation has already derived contradiction and is now
deriving all kinds of other interesting consequences from this. But
before this happens, we would like to argue that any refutation must
pass through a stage where all it can know about
$\varx_1$
and
$\varx_2$
is the  value of
$\varx_1 \xor \varx_2$
and nothing more. For this reason, we would like to find a more
``fine-grained'' definition of a projection that can capture only
these weaker implications and discard too strong implications.

\begin{openproblem}
  \label{openproblem:projections-for-CP-and-friends}
  Is it possible to prove 
  space lower bounds and/or 
  trade-offs between proof length/size and
  space for cutting planes, polynomial calculus, or  
  polynomial calculus resolution by designing smarter projections than in 
  \refdef{def:projected-clauses-res}
  that are \spacerespecting for these proof systems?%
  \footnote{%
    As the camera-ready version of this article was being prepared,
    some new results on
    time-space trade-offs 
    for PCR and CP were reported in 
    \cite{HN11Communicating,BNT12SomeTradeoffs}.
    We refer to
    Theorems~\ref{th:new-trade-off-PCR-and-CP},      
    \ref{th:tradeoffs-informal-super-constant-PCR},
    and~\ref{th:exp-trade-off-CS}
    below for detailed statements.}
\end{openproblem}

\subsectionNOW{Some Open Problems Regarding Space 
  Bounds and Trade-offs}
\label{sec:pc-open-problems}

Despite the progress made during the last few years on understanding
space in resolution and how it is related to other measures, there are
quite a few natural questions that still have not been resolved.

Perhaps one of the main open questions is how complex a
\kcnfform~$\fstd$ can be \wrt total space. 
If $\fstd$ has at most $n$ clauses or variables (which is the case if,
in particular, $\fstd$ has size~$n$) we know from~%
\cite{ET01SpaceBounds}
that
$\clspaceref[\resnot]{\fstd} \leq n + \bigoh{1}$. From this it
immediately follows that 
$\totspaceref[\resnot]{\fstd} = \Bigoh{n^2}$.
But is this upper bound tight?

\begin{openproblem}[\cite{ABRW02SpaceComplexity}]
  \label{openproblem:total-space}
  Are there  \kcnfform families $\set{\fstd_n}_{n=1}^{\infty}$ 
  of size~$\bigtheta{n}$ \st
  $\totspaceref{\fstd_n} = \Bigomega{n^2}$?
\end{openproblem}

As a first step towards resolving this question, 
Alekhnovich \etal
\cite{ABRW02SpaceComplexity}
posed the problem of finding \kcnfform{}s over $n$ variables and of size
polynomial in~$n$ \st 
$\totspaceref{\fstd} = \littleomega{n}$.
(There is a lower bound
$\totspaceref{\fstd} = \bigomega{n^2}$
proven in~\cite{ABRW02SpaceComplexity}, 
but it is for formulas of exponential size and linear width).
Alekhnovich \etal also conjectured that the answer to 
Open Problem~\ref{openproblem:total-space}
is yes, and suggested so-called Tseitin formulas defined over
\mbox{$3$-regular} expander graphs as candidates for
formulas $\fstd_n$ of size~$n$ with
$\totspaceref{\fstd_n} = \Bigomega{n^2}$.
%

Another natural open question is to separate 
polynomial calculus resolution from just resolution
\wrt space.

\begin{openproblem}
  \label{openproblem:separation-space-resolution-PCR}
  Are there  \kcnfform families $\set{\fstd_n}_{n=1}^{\infty}$ 
  such that
  $\clspaceref[\resnot]{\fstd_n} = 
  \Littleomega{\clspaceref[\pcrnot]{\fstd_n}}$?%
  \footnote{We note for completeness that
    \cite{ABRW02SpaceComplexity}
    proved a constant factor separation between resolution clause
    space and PCR monomial space, but this separation crucially
    depends on the definition of monomial space as not counting
    repetitions of monomials in different polynomial equations
    (and is in any case not strong enough to resolve this open question).
    For the arguably more natural concept of space in
    \refdef{def:PCR}
    nothing is known by way of separations from resolution.}
  
\end{openproblem}

The next two questions that we want to address concern upper bounds on
resolution length in terms of clause space. We know from
\cite{AD08CombinatoricalCharacterization}
that clause space is an upper bound for width, and width yields an
%
%
upper bound on length simply by counting. 
However, it would be more satisfying to find a more direct argument
that explains \emph{why} clause space upper-bounds
length.
Focusing on constant clause space for concreteness, the problem can be
formulated as follows.

\begin{openproblem}
  \label{openproblem:space-upper-bounds-length}
  For \kcnfform{}s $\fstd$ of size~$n$, we know that
  $\clspaceref{\fstd} = \bigoh{1}$ 
  implies
  $\lengthref{\fstd} = \polybound{n}$.
  Is there a direct proof of this fact, not going via
  \cite{AD08CombinatoricalCharacterization}?
\end{openproblem}

If we could understand this problem better, we could perhaps also find
out whether it is possible to derive stronger upper bounds on length
in terms of space. Esteban and Torán ask the following question.

\begin{openproblem}[\cite{ET01SpaceBounds}]
  \label{openproblem:stronger-bounds-on-length-from-space}
  Does it hold for \kcnfform{}s $\fstd$
  that
  $\clspaceref{\fstd} = \bigoh{\log n}$
  implies
  $\lengthref{\fstd} = \polybound{n}$?
\end{openproblem}

Turning to the relationship between width and length, 
recall that we  know from~\cite{BW01ShortProofs}   
that short resolution refutations imply the existence of narrow
refutations,  and that in view of this an appealing proof search
heuristic  is to search exhaustively for refutations in minimal width.
One serious drawback of this approach, however,  is that there is no
guarantee that the short and narrow refutations are the same one.
On the contrary, the narrow refutation constructed in the
proof in~\cite{BW01ShortProofs} is potentially exponentially longer
than the short refutation that one starts with.
However, we have no examples of formulas where the refutation in
minimum width is actually known to be substantially longer than the
minimum-length refutation. Therefore, it would be interesting to 
know whether this increase in length is necessary.
That is, is there a formula family which exhibits a length-width
trade-off in the sense that there are short refutations and narrow
refutations, but all narrow refutations have a length blow-up
(polynomial or superpolynomial)?
Or is the exponential blow-up in~\cite{BW01ShortProofs} just an
artifact of the proof?

\begin{openproblem}[\cite{NH13TowardsOptimalSeparation}]
  \label{openproblem:length-vs-width-BW}
  If $\fstd$ is a \kcnfform over $\nvar$ variables
  refutable in length~$\lengthstd$,   can one always find  a
  refutation $\proofstd$ of~$\fstd$ in width
  {$\widthofsmall{\proofstd} = \Ordocompact{\sqrt{\nvar \log \lengthstd}}$}
  with length no more than, say,
  {$\lengthofsmall{\proofstd} = \Ordosmall{\lengthstd}$}
  or at most
  {$\polyboundsmall{\lengthstd}$}?
  Or is there a formula family which necessarily exhibits a
  length-width trade-off in the sense that there are short
  refutations and narrow refutations,
  but all narrow refutations have a length blow-up
  (polynomial or superpolynomial)?
\end{openproblem}


As was mentioned above, for tree-like resolution 
Ben-Sasson~\cite{Ben-Sasson09SizeSpaceTradeoffs}
showed that there are formulas~$\fstd_n$ refutable in linear length
and also in constant width, but for which any refutation
$\proofstd_n$
must have
$\widthofsmall{\proofstd_n} \cdot 
\log \lengthofsmall{\proofstd_n} 
= \bigomega{n / \log n}$. 
This shows that the length blow-up in the proof of the tree-like
length-width relationships in  
\cite{BW01ShortProofs}
is necessary. 
That is, transforming a short tree-like proof into a
narrow proof might necessarily incur an exponential length blow-up.
But tree-like resolution is very different from unrestricted
resolution in that upper bounds on width do \emph{not} imply upper
bounds on length (as shown in~%
\cite{BW01ShortProofs}
using 
$\pebcontrwithfunc[G]{}{\lor_2}$-formulas),
so it is not clear that the result for tree-like resolution provides
any intuition for the general case.

A related question about trade-offs between length and width on a
finer scale, raised by Albert Atserias and Marc Thurley, is as follows.

\begin{openproblem}[\cite{AT09personalcommunication}]
  \label{openproblem:atserias-thurley}
  For 
  $\widthstd \geq 3$
  arbitrary but fixed, is there family of unsatisfiable $3$-CNF formulas 
  $\set{\fstd_n^{\widthstd}}_{n=1}^{\infty}$
  of size
  $\bigtheta{n}$
  that have resolution refutations of width~$\widthstd$
  but cannot be refuted in length
  $\Bigoh{n^{\widthstd -  c}}$
  for some small positive constant~$c$?
\end{openproblem}

This question was prompted by the paper~%
\cite{AFT11ClauseLearning}, 
where it was shown for a fairly general theoretical model of DPLL solvers
with clause learning that in this model contradictory formulas
$\fstd_n$
with
$\widthref{\fstd_n} = \widthstd$
are likely to be proven unsatisfiable in time
$n^{\bigoh{\widthstd}}$. 
%
%
It is natural to ask how much room for improvement there is for this
time bound. Since these algorithms are resolution-based, it would be
nice if one could prove a lower bound saying that there are formulas~%
$\fstd_n$
with
$\widthref{\fstd_n} = \widthstd$
that cannot be refuted by resolution in length~%
$n^{\littleoh{\widthstd}}$,
or even
$n^{\widthstd - \bigoh{1}}$.
As a step towards proving (or disproving) this, resolving special cases of
Open Problem~\ref{openproblem:atserias-thurley}
for concrete instantiations of the parameters, say 
$\widthstd=10$ and $\widthstd - c = 2$,
would also be of interest.

For resolution clause space, we know that there can be very strong trade-offs
with respect to length for space $\spacestd$ in the range
$\littleomega{1} = \spacestd = \littleoh{n / \log \log n}$,
but we do not know what holds for space outside of this
range. Consider first formulas refutable by proofs~$\proofstd$ in
constant space. When we run such a refutation through the proof in
\cite{AD08CombinatoricalCharacterization}
and obtain another narrow, and thus short, refutation $\proofstd'$ we
do not have any information about the space complexity of this
refutation. Is it possible to get a refutation in both short length
and small space simultaneously?

\begin{openproblem}[\cite{NH13TowardsOptimalSeparation}]
  \label{openproblem:space-vs-length}
  Suppose that
  $\setsmall{\fstd_n}_{n=1}^{\infty}$
  is a family of polynomial-size \kcnfform{}s 
  with refutation clause space
  $\mbox{$\clspaceref{\fstd_n}$} = \bigoh{1}$. 
  Does this imply that there are resolution refutations
%
%
  $\refof{\proofstd_n}{\fstd_n}$
  simultaneously in length
  {$\lengthofarg{\proofstd_n} = \polyboundsmall{n}$}
  and clause space
  {$\clspaceofsmall{\proofstd_n} = \bigoh{1}$}?
  Or can it be that restricting the clause space, we sometimes have to end up
  with really long refutations?
\end{openproblem}

Note that if we instead look at the total space measure (that also
counts the number of literals in each clause with repetitions), then
the answer to the above question is that we can obtain refutations
that are both short and space-efficient simultaneously, again by a
simple counting argument. But for clause space such a counting
argument does not seem to apply, and maybe strange things can
happen. (They certainly can in the sense that as soon as we go to
arbitrarily slowly growing non-constant space, there provably exist
strong space-length trade-offs.) 
%
Of course, one would expect here that
any insight regarding 
Open Problem~\ref{openproblem:space-upper-bounds-length}
should have bearing on this question as well.

Consider now space complexity at the other end of the range.
Note that all trade-offs for clause space proven so far
are in the regime where the space $\clspaceofsmall{\proofstd}$
is less than the number of clauses $\nclausesof{\fstd}$ in~$\fstd$.
On the one hand, this is quite natural, since 
the size of the formula is an upper bound on the
refutation clause space needed. On the other hand, it is not clear
that this should rule out length-space trade-offs for linear or
superlinear space, since the proof that any formula is refutable in
linear space constructs a resolution refutation that has exponential length.
Assume therefore that we have a \cnfform~$\fstd$ of size~$n$ 
refutable in length
$\lengthrefsmall{\fstd} = \lengthstd$ 
for $\lengthstd$ suitably large
(say, 
$\lengthstd = \polyboundsmall{n}$
or
$\lengthstd = n^{\log n}$
or so).
Suppose that we allow clause space more than the minimum
$n + \bigoh{1}$,
but less than the trivial upper bound
$\lengthstd / \log \lengthstd$.
Can we then find a  resolution refutation
using at most that much space
and achieving at most a polynomial increase in length compared to the
minimum? 

\begin{openproblem}[\cite{Ben-Sasson07personalcommunication,
      NH08TowardsOptimalSeparationSTOC}] 
  \label{openproblem:extra-clause-space-vs-length}
  Let $\fstd$ be any \kcnfform with
  $\nclausesof{\fstd} = n$ clauses.
  Suppose that
  $\lengthrefsmall{\fstd} = \lengthstd = \polyboundsmall{n}$.
  Does this imply that there is  a resolution refutation
  $\refof{\proofstd}{\fstd}$
  in clause space
  $\clspaceofsmall{\proofstd} = \bigoh{n}$
  and length
  $\lengthofarg{\proofstd}=\polyboundsmall{\lengthstd}$?
  Or are there formulas with trade-offs in the range space $\geq$ formula
  size?%
  \footnote{%
    As the camera-ready version of this article was being prepared,
    a solution of this problem was
    reported in 
    \cite{BBI12TimeSpace},
    and this result was then further strengthened in
    \cite{BNT12SomeTradeoffs}.
    The answer is that there are (very) strong trade-offs even
    in the superlinear space regime.
    See
    \reftwoths{th:new-superlinear-trade-off-res}       
              {th:BBI-tradeoff-for-PCR}       
    for details.}
\end{openproblem}

%

Finally, 
a slightly curious aspect of the space lower bounds and length-space
trade-offs surveyed above is that the results in
\cite{Nordstrom09NarrowProofsSICOMP,NH13TowardsOptimalSeparation}
only work for
\kcnfform{}s
of width
$\clwidth \geq 4$, and in
\cite{BN08ShortProofs,BN11UnderstandingSpace}
we even have to choose
$\clwidth \geq 6$
to find \kcnfform families that optimally separate space and length
and exhibit time-space trade-offs. We know from~%
\cite{ET01SpaceBounds}
that any
\xcnfform{2}{}
is refutable in constant clause space, 
but should there not be
\xcnfform{3}{}s 
for which we could prove similar separations and trade-offs?

Given any \cnfform $\fstd$, we can transform it to a
\xcnfform{3}{} 
by rewriting every   clause 
$\clc = \lita_1 \lor \ldots \lor \lita_m$
in $\fstd$
with $m>3$
as a  conjunction of  \xclause{3}{}s
\begin{equation}
  \label{eq:extended-version-of-clause}
  \olnot{\vary}_0 \, \land \,
  \Land_{1 \leq i \leq m} 
  (\vary_{i-1} \lor \lita_i \lor \olnot{\vary}_i)
  \, \land \,
  \vary_n,
\end{equation}
for some new auxiliary variables
$\vary_{0},  \vary_{1}, \ldots, \vary_{m}$ 
unique for this clause~$\clc$.
\label{page:extended-version-discussed-here}
Let us write
$\extendedversion{\fstd}$
to denote the \xcnfform{3}{} obtained from $\fstd$ in this way.
It is easy to see that
$\extendedversion{\fstd}$
is unsatisfiable \ifaoif
$\fstd$ is unsatisfiable.
Also, it is straightforward to verify that
$\mbox{$\Lengthref{\extendedversion{\fstd}}$} \leq
\lengthref{\fstd} + \bigoh{\widthofarg{\fstd} \cdot \lengthofarg{\fstd}}
$
and
$\Clspaceref{\extendedversion{\fstd}} \leq
\clspaceref{\fstd} + \Ordocompact{1}$.
(Just note that each clause of $\fstd$ can be derived from
$\extendedversion{\fstd}$ 
in length $2 \cdot \widthofsmall{\fstd} + 3$ and 
space~$3$, 
and then  use
this   together with an optimal refutation of $\fstd$.)

It seems like a natural idea to rewrite pebbling contradictions
$\genericpebcontr[G]{}$
for suitable functions $\funcpebc$
as \xcnfform{3}{}s
$\extendedpebcontrwithfunc[G]{}{\funcpebc}$
and study length-space trade-offs for such formulas. For this to work,
we would need \emph{lower} bounds on the refutation clause space of 
$\extendedversion{\fstd}$
in terms of  the refutation clause space of 
$\fstd$,  
however.

\begin{openproblem}
  \label{openproblem:clause-space-of-extended-version}
  Is it true that
  $
  \Clspaceref{\extendedpebcontrwithfunc[G]{}{\xor}}
  \geq
  \bwpebblingprice{G}
  $?
  In general, can we prove lower bounds on
  $\Clspaceref{\extendedversion{\fstd}}$
  in terms of
  $\clspaceref{\fstd}$,%
  \footnote{Some limited results along these lines were reported in~%
    \cite{FLNTZ12SpaceCplx}
    but the general question still remains wide open.
    We refer to
    \refth{th:flntr-weight-constrained}
    below for more details.}
  or are there counter-examples where the two measures differ asymptotically?
\end{openproblem}

This final open problem is of course of a somewhat technical
nature. However, we still find it interesting in the sense that if it
could be shown to hold in general that
$\Clspaceref{\extendedversion{\fstd}} = \Bigtheta{\clspaceref{\fstd}}$
or even
$\Clspaceref{\extendedversion{\fstd}} = \clspaceref{\fstd} + \bigoh{1}$,
then we would get all space lower bounds, and maybe
also the length-space trade-offs,  for free for
\mbox{$3$-CNF} formulas. It would be aesthetically satisfying not
having to insist on using \mbox{$6$-CNF} formulas to obtain these
bounds. 
If, on the other hand, such an equality does \emph{not} hold, it would further
strengthen the argument that space should only be studied for formulas
of fixed width (as was discussed above).

\section{Concluding Remarks}
\label{sec:concluding-remarks}

In this survey, we have tried to give an overview of how pebble games
have been used as a tool to derive many strong results in proof
complexity. Our main focus has been on explaining how CNF formulas
encoding pebble games played on graphs can be shown to inherit
trade-off properties for proof length (size) and space from pebbling time-space
trade-offs for these graphs. While these connections have turned out
to be very useful, the reductions are far from tight. It would be
interesting to clarify the true relationship between pebble games and
pebbling formulas for resolution-based proof systems.

As 
discussed in
\refsec{sec:pc-substitution-theorem},
all the length-space trade-offs for resolution and \kdnf resolution
can be described as starting with simple CNF formulas possessing
useful moderate hardness properties (namely, the pebbling
contradictions with just one variable per vertex exhibiting weak
length-space trade-offs) and then studying how variable substitutions
in these formulas can amplify the hardness properties. We find it an
intriguing open question whether this approach can be made to work for
stronger proof systems such as cutting planes or variants of
polynomial calculus.

Finally, throughout the survey, and in particular in
\refsec{sec:pc-open-problems},
we have tried to highlight a number of open problems, pebbling-related
or non-pebbling-related, which we believe merit further study.


\newcommand{\mspacestd}{{\spacestd}}
\newcommand{\mspaceof}[2][]{\genericformsmall{Sp_{#1}}{#2}}
\newcommand{\Mspaceof}[2][]{\genericformcompact{Sp_{#1}}{#2}}
\newcommand{\mspaceref}[2][]{\genericrefsmall{Sp}{#1}{#2}}
\newcommand{\Mspaceref}[2][]{\genericrefcompact{Sp}{#1}{#2}}
\newcommand{\mspacederiv}[3][]{\genericderivsmall{Sp}{#1}{#2}{#3}}
\newcommand{\Mspacederiv}[3][]{\genericderivcompact{Sp}{#1}{#2}{#3}}


\section*{Note Added in Proof}

While this survey paper was being reviewed and revised, there were a
number of new developments regarding some of the open problems
discussed in the paper. We attempt to give an overview of (some of)
these recent 
results below.

As we have tried to describe in this survey, there is a a wealth of
results on space lower bounds and time-space trade-offs for
resolution-based proof systems. For polynomial calculus and cutting
planes the situation has been very different in that almost nothing
has been known.  As far as we are aware, the first trade-off results
to be reported for these latter proof systems are those in~%
\cite{HN11Communicating}.
%
%

\begin{theorem}[\cite{HN11Communicating}]
  \label{th:new-trade-off-PCR-and-CP}
  There are $k$-CNF formulas 
  $\set{\fstd_n}_{n=1}^{\infty}$
  of size $\bigtheta{n}$ that can be refuted
  in length $\bigoh{n}$
  in resolution, polynomial calculus (and hence also PCR) and
  cutting planes, but for which the following holds:
  \begin{iteMize}{$\bullet$}
  \item 
    Any PCR refutation of $\fstd_n$ in length~$L$ and monomial
    space~$s$ must satisfy
    $$
      s \log L = \Bigomega{\sqrt[4]{n}}.
      $$
  \item
    Any CP refutation
    of $\fstd_n$ in length~$L$ and line space%
    \footnote{That is, charging one unit of space for each linear inequality.}
    $s$
    must satisfy
    $$
      s \log L\log(s\log L) = \BIGOMEGA{\sqrt[4]{n} 
        \, 
        / \log^{2} n}.
      $$

%
  \end{iteMize}
\end{theorem}

\noindent The formulas used in this theorem are a particular flavour of pebbling
formulas over pyramid graphs (not obtained by substitution but by a
related operation of \introduceterm{lifting} as defined in~%
\cite{BHP10HardnessAmplificationSTOC}), but the results were obtained
by other techniques than the \spacerespecting projections discussed in
this survey; namely, by tools from communication complexity.

It should be pointed out that it is not clear whether
\refth{th:new-trade-off-PCR-and-CP} provides ``true trade-offs,''
however.  The issue is that to be able to speak about a trade-off
between proof length and proof space in a strict sense, we would also
like the formulas to have space complexity smaller than the space
bound at where the trade-off kicks in. This is not the case for
pebbling formulas over pyramid graphs, since we do not know how to
refute them in space less than~$\bigoh{\sqrt{n}}$. And indeed, it can
be argued that the formulas seem more likely to have space
complexity~$\bigomega{\sqrt{n}}$, as is known to be the case for
similar formulas \wrt resolution.

Interestingly, the statements in 
\refth{th:new-trade-off-PCR-and-CP}
closely parallel the result obtained by
Ben-Sasson in 2002 (journal version in
\cite{Ben-Sasson09SizeSpaceTradeoffs}),
when he proved for 
pebbling formulas obtained by XOR substition
that a trade-off on the form
$s \log L =  \bigomega{n / \log n}$
must hold in resolution for any refutation in length~$L$ and
clause space~$s$. Six years later, the factor $\log L$ was shown to be an
artifact of the proof technique and was removed in
\cite{BN08ShortProofs}
to obtain an unconditional space lower bound
$\bigomega{n / \log n}$
as stated in 
\refth{th:length-space-separation}.
It is very tempting to conjecture that the situation should be the
same for PCR and CP, so that 
\refth{th:new-trade-off-PCR-and-CP}
should be understood as providing ``conditional space lower bounds''
from which we should strive to remove the 
$\log L$ factors. 

%
%
In another line of research, Beame \etal~\cite{BBI12TimeSpace}
made progress on Open
Problem~\ref{openproblem:extra-clause-space-vs-length} by showing that
there are non-trivial trade-offs between proof length and proof space
in resolution even for superlinear space.

\begin{theorem}[\cite{BBI12TimeSpace}]
  \label{th:new-superlinear-trade-off-res}
  There is a family of explicitly constructible CNF formulas $\fstd_n$ 
  of size~$\bigtheta{n}$
  and  width $\bigtheta{\log n}$
  such that the following holds:
  \begin{iteMize}{$\bullet$}
  \item 
    The formulas $F_n$
    have resolution refutations in simultaneous length and clause
    space 
    $n^{\bigoh{\log n}}$.

  \item 
    The formulas $F_n$ can also be refuted in 
    clause space~$\bigoh{n}$
    (since the refutation clause space is always upper-bounded by
    the formula size).  

  \item 
    Any resolution refutation of $F_n$ 
    in length~$L$
    and
    clause space~$s$
    must satisfy
    the inequalitity
    \mbox{$
    L
    \geq
    \bigl( n^{0.58 \log n} / s \bigr)^
         {\bigomega{\log \log n / \log \log \log n}} 
    $.} 
  \end{iteMize}
  In particular, resolution refutations of $F_n$ in linear clause
  space must have length superpolynomial in  
  $n^{\log n}$.
\end{theorem}

For the restricted system of regular resolution
(see
\refsec{sec:pc-separations}),
the paper \cite{BBI12TimeSpace} contains more dramatic trade-offs that
hold for formulas of constant width.
The proof of
\refth{th:new-superlinear-trade-off-res},
however, crucially requires CNF formulas of unbounded width.
In the very recent paper~%
\cite{BNT12SomeTradeoffs}, 
the analysis in
\cite{BBI12TimeSpace}
was simplified and strengthened to produce similar time-space
trade-offs for CNF formulas of constant width, completely resolving
Open Problem~\ref{openproblem:extra-clause-space-vs-length},
and these trade-off results were then further extended to apply also
to PCR. 

\begin{theorem}[\cite{BNT12SomeTradeoffs}]
  \label{th:BBI-tradeoff-for-PCR}       
  Let $\mathbb{F}$ be a field of odd characteristic.
  There are explicitly constructible
  families of $8$-CNF formulas
  $\set{\fstd_{n,w}}$, 
  with
  $w = w(n)$ satisfying 
 $1 \leq w\leq n^{1/4}$, 
  which are
  of size $\bigtheta{n}$
  and have the following properties:
  %
  \begin{iteMize}{$\bullet$}
  \item
    \label{item:BBI-refut-1}
    The formulas $\fstd_n$ have  resolution 
    refutations $\proofstd_n$
    in (short) length $\lengthofarg{\proofstd_n} \leq n^{O(1)} 2^w$
    and clause space $\clspaceof{\proofstd_n} \leq 2^w + n^{O(1)}$.

  \item
    \label{item:BBI-refut-2}
    They also have
    resolution refutations
    $\proofstd'_n$ 
    in (small) clause space $\clspaceof{\proofstd'_n} = \Bigoh{w \log n}$
    and length $\lengthofarg{\proofstd'_n} \leq 2^{\bigoh{w \log n}}$. 

    
  \item
    \label{item:BBI-part-3}
    For any PCR refutation $\proofstd_n$ of $\fstd_n$ over $\mathbb{F}$,
    the proof  size is bounded  by\\
    $\sizeofarg{\proofstd_n} 
    = \left( 
    \frac{2^{\Omega(w)}}{\mspaceof{\proofstd_n}} \right)^{ \Omega \left( \frac{
        \log \log n }{ \log \log \log n } \right)}
    $.
  \end{iteMize}

\end{theorem}

\noindent As a side note, we remark that the formulas in 
\reftwoths{th:new-superlinear-trade-off-res}{th:BBI-tradeoff-for-PCR}       
are not pebbling formulas---not surprisingly, since as discussed above
such formulas do not exhibit trade-offs in the superlinear space
regime---but so-called \introduceterm{Tseitin formulas} over grid-like
graphs. 

The paper \cite{BNT12SomeTradeoffs} also lifts the time-space
trade-offs for pebbling formulas in
\cite{BN11UnderstandingSpace}
from resolution to PCR. This is achieved by using \spacerespecting
projections, but combined with a random restriction argument that
leads to some loss in the parameters as compared to corresponding
results for resolution. Two examples of the results obtained in this
way are as follows.

\begin{theorem}[\cite{BNT12SomeTradeoffs}]
  \label{th:tradeoffs-informal-super-constant-PCR}
  Let $\spacefunclower(n) = \littleomega{1}$ be any arbitrarily slowly
  growing function%
  \footnote{As in 
    \refth{th:tradeoffs-informal-super-constant},
    for technical reasons we also need the assumption
    $\spacefunclower(n) = \Bigoh{n^{1/7}}$ 
    but again this restriction is not important.}
  and fix any  $\varepsilon >0$.
  Then there are explicitly constructible $6$-CNF formulas
  $\set{\fstd_n}_{n=1}^{\infty}$
  of size $\bigtheta{n}$ 
  such that the following holds:
  \begin{iteMize}{$\bullet$}
  \item
    There are 
    resolution refutations and 
    polynomial calculus refutations
    $\proofstd_n$ of~$\fstd_n$
    in total space
    $\totspaceof{\proofstd_n}    = 
    \bigoh{\spacefunclower(n)}$.
    
  \item
    There are 
    resolution refutations and 
    PC refutations
    $\proofstd_n$ of~$\fstd_n$
    in simultaneous size 
    $\sizeofarg{\proofstd_n} = \bigoh{n}$
    and total space
    $\totspaceof{\proofstd_n}
    = \BIGOH{\bigl(n / \spacefunclower(n)^2\bigr)^{1/3}}$.

  \item
    Any PCR refutation of $\fstd_n$ in monomial space   
    $\BIGOH{
      \bigl( n / ( \spacefunclower(n)^3 \log n )\bigr)^{1/3 - \varepsilon}}$
    must have superpolynomial size.

  \end{iteMize}

\end{theorem}

\begin{theorem}[\cite{BNT12SomeTradeoffs}]
  \label{th:exp-trade-off-CS}
  There is a family of
  explicitly constructible $6$-CNF formulas
  $\set{\fstd_n}_{n=1}^{\infty}$
  of size $\bigtheta{n}$ 
  such that the following holds:
  \begin{iteMize}{$\bullet$}
  \item
    The formulas $\fstd_n$ are refutable in
    resolution and
    PC
    in total space
    $\Bigoh{n^{1/11}}$.
    
  \item
    There are 
    resolution refutations and 
    PC refutations
    $\proofstd_n$ of~$\fstd_n$
    in 
    size
    $\sizeofarg{\proofstd_n} = \bigoh{n}$
    and total space
    $\totspaceof{\proofstd_n} = \Bigoh{n^{3/11}}$.

  \item
    Any PCR refutation of $\fstd_n$ in monomial space at most
    $n^{2/11}/(10 \log n)$ must have size at least 
    $\bigl( n^{1/11} \bigr)!\,$.

  \end{iteMize}
\end{theorem}

\noindent The trade-off results obtained in
\cite{BNT12SomeTradeoffs}
subsume those from
\cite{HN11Communicating}
for polynomial calculus and PCR in 
\refth{th:new-trade-off-PCR-and-CP},
but for cutting planes 
\cite{HN11Communicating}
still remains state-of-the-art to the best of our knowledge.

Intriguingly, what
Theorems~\ref{th:new-trade-off-PCR-and-CP},
\ref{th:tradeoffs-informal-super-constant-PCR},
and~\ref{th:exp-trade-off-CS}
seem to say is that the pebbling properties of graphs are
preserved even when these graphs are translated to CNF formulas and
one uses polynomial calculus/PCR and cutting planes to reason about
such CNF formulas. A priori, it was not clear at
all whether any such connections should exist or not. 
We still
cannot prove that the correspondence is as tight as for
resolution---since we lose a logarithmic factor in the proof
length/pebbling time, and in addition can only prove the
correspondence for a fairly limited set of graphs when it comes to
cutting planes---but as already hinted to above we believe that this is
due to limitations in our current techniques.

The attentive reader will have noticed that we need CNF formulas of
width~$6$ or higher in the theorems above. Somewhat annoyingly, we
still do not know of any time-space trade-offs for \mbox{$3$-CNF}
formulas, but some limited progress on 
Open Problem~\ref{openproblem:clause-space-of-extended-version} 
has been reported in~%
\cite{FLNTZ12SpaceCplx}.

\begin{theorem}[\cite{FLNTZ12SpaceCplx}]
  \label{th:flntr-weight-constrained}
  Let us say that a CNF formula $F$ is
  \introduceterm{weight-constrained}
  if it has the property that for every clause 
  $\lita_{1} \lor \lita_{2} \lor \formuladots \lor \lita_{w}$ 
  in $F$ of width $w \geq 4$,
  $F$ also contains the clauses 
  $\olnot{\lita}_{i} \lor \olnot{\lita}_{j}$ 
  for all $1\leq i < j \leq w$.
  Then if $F$ is a weight-constrained CNF and $\extendedversion{F}$ is
  the corresponding \mbox{$3$-CNF} formula as defined in~%
  \refeq{eq:extended-version-of-clause},
  it holds that
  \begin{equation*}
    \clspaceref[\resnot]{F} = \Clspaceref[\resnot]{\extendedversion{F}}
    + \bigoh{1}
    \eqperiod
  \end{equation*}
\end{theorem}

\noindent Last but not least, we want to mention that in a technical
tour-de-force, 
Berkholz~\cite{Berkholz12ComplexityNarrowProofs} 
recently managed to show that resolution width is 
\EXPTIME-complete. In fact, he obtained the slightly stronger result
stated next.

\begin{theorem}[\cite{Berkholz12ComplexityNarrowProofs}]
  \label{th:width-berkholz}
  Let $F$ be a CNF formula and let
  $w \geq 15$ be an integer.
  Then the question of whether $F$ has a resolution refutation in
  width at most~$w$ cannot be decided in time
  $\setsize{F}^{(w-3)/12}$
  on a multi-tape Turing machine. In particular, the problem of
  deciding resolution refutation width is
  \EXPTIME-complete.
\end{theorem}

Using similar techniques, Berkholz also proved a limited form of
trade-off between length and width, but unfortunately nothing that
seems to shed light on
Open Problems~\ref{openproblem:length-vs-width-BW}
or~\ref{openproblem:atserias-thurley}. And the problem of proving (or
disproving) the \PSPACE-completeness of resolution clause space  
(Open Problem \ref{openproblem:clause-space-complexity})
remains stubbornly out of reach.

\section*{Acknowledgements}

I would like to thank 
\mbox{Eli Ben-Sasson},
\mbox{Johan Håstad},
and 
\mbox{Alexander Razborov},
with whom I have co-authored some of the papers on which this survey is based,  
for many interesting and fruitful discussions.
I also want to thank my more recent  co-auhors
\mbox{Chris Beck},
Yuval Filmus,
Trinh Huynh,
Massimo Lauria,
Mladen Mik\v{s}a,
Noga Ron-Zewi,
Bangsheng Tang,
Neil Thapen,
and
Marc Vinyals,
with whom I have written papers on related subjects, although 
these papers are mostly too recent to have made it properly into this survey.
I am grateful to   
Joshua Buresh-Oppenheim,
\mbox{Nicola Galesi},
\mbox{Jan Johannsen},
and
Pavel Pudlák
for helping me 
sort out some proof complexity related questions
that arose while working on this article.
Thanks also go to 
Paul Beame and Russell Impagliazzo for
keeping me posted about the line of work that led to the 
time-space trade-off results for resolution in the superlinear space regime.
Last but not least, I would like to thank the anonymous \emph{LMCS} referees 
for their extraordinary work in reviewing the paper---their comments
were truly invaluable and helped to improve the 
presentation
substantially---and
Mladen Mik\v{s}a and
Marc Vinyals for helping me 
catch some bugs and typos in the 
final, revised version of the paper.
Needless to say, any remaining errors are solely the 
responsibility of the author.

The author did part of this work while at 
the Massachusetts Institute of Technology,
supported by
the Royal Swedish Academy of Sciences,
the Ericsson Research Foundation,
the Sweden-America Foundation,
the Foundation Olle Engkvist Byggmästare, and
the Foundation Blanceflor Boncompagni-Ludovisi, née Bildt.
He is currently supported by
the European Research Council under the European Union's Seventh Framework
Programme \mbox{(FP7/2007--2013) /} ERC grant agreement no.~279611
and by the Swedish Research Council grants 
\mbox{621-2010-4797}
and
\mbox{621-2012-5645}.

%
%

\bibliography{refArticles,refBooks,refOther}

%
%
%
%
%
%

\bibliographystyle{alpha}   

%
%

 \proofcomment{Last ``proof comment''.
   (This comment added to know how many there are in total.)}
 \editcomment{Last ``edit comment''.
   (This comment added to know how many there are in total.)}
 
\end{document}